\PassOptionsToPackage{pdfversion=1.7,hyperfootnotes=false}{hyperref}
\documentclass[fleqn,usenatbib]{mnras}

\usepackage[T1]{fontenc}
\usepackage{ae,aecompl}
\usepackage{amsmath}
\usepackage{amssymb}
\usepackage{newtxtext}
\usepackage[varvw]{newtxmath}
\usepackage{graphicx}
\usepackage{booktabs}
\usepackage{multirow}
\usepackage{hyperref}
\usepackage{orcidlink}

\usepackage{xspace}
\makeatletter
\newcommand{\Bonetti}{%
  \@ifundefined{cited@Bonetti}{%
    \citet{Bonetti2018}%
    \expandafter\gdef\csname cited@Bonetti\endcsname{}%
  }{%
    \hyperlink{cite.Bonetti2018}{Bonetti}%
  }%
  \xspace
}
\makeatother

\usepackage{xcolor}
\hypersetup{hidelinks}
\usepackage{etoolbox}
\makeatletter
\patchcmd{\caption}{\refstepcounter}{\H@refstepcounter}{}{}
\makeatother

\title[RAMCOAL-3body]{Set them free: extending RAMCOAL to model massive black hole triplets in hydrodynamical simulations of galaxies}

\author[Kunyang Li et al.]{
Kunyang Li$^{1,2}$\orcidlink{0000-0002-0867-8946}\thanks{E-mail: lli@flatironinstitute.org},
Ricarda S. Beckmann$^{3}$\orcidlink{0000-0002-2850-0192},
Yohan Dubois$^{2}$\orcidlink{0000-0003-0225-6387},
and Marta Volonteri$^{2}$\orcidlink{0000-0002-3216-1322}
\\
$^{1}$Center for Computational Astrophysics, Flatiron Institute, 162 5th Avenue, New York, NY 10010 \\
$^{2}$Institut d'astrophysique de Paris, UMR 7095, CNRS, Sorbonne Universit\'e, 98 bis boulevard Arago, 75014 Paris, France\\
$^{3}$Institute for Astronomy, University of Edinburgh, Royal Observatory, Edinburgh EH9 3HJ, UK
}

\date{Accepted 2026 November 05. Received 2026 October 30; in original form 2026 July 07}
\pubyear{2026}

\begin{document}

\maketitle

\begin{abstract}
Massive black hole binaries (MBHBs) and the higher-order multiples produced by repeated galaxy mergers evolve through regimes that cosmological simulations cannot resolve. These regimes determine merger delays, spins, recoils, and host galaxies at coalescence. We extend the RAMCOAL subgrid model to follow massive black hole triplets within hydrodynamical galaxy simulations. In the original three-stage binary model, black holes evolve as sink particles, undergo dynamical friction, and form bound binaries that harden through stellar scattering, gas torques, circumbinary-disc coupling, and gravitational-wave emission. When a hierarchical triplet becomes chaotic, RAMCOAL samples a library of outcomes from direct $N$-body experiments and follows the resulting mergers, exchanges, and ejections alongside the accretion and spin evolution of each black hole. Isolated-galaxy tests with different triplet geometries show that orbital configuration alone can change which pair merges and its coalescence time. For the first time, we follow an MBH triplet to coalescence within a hydrodynamical simulation of gas-rich galaxies, coupling triplet-driven mergers to host-galaxy evolution. Subgrid integration costs $\lesssim1$\,ms of single-core time per MBH system and time step, independently of particle resolution. The resulting coalescence properties reflect the environment encountered during subgrid evolution. RAMCOAL thus provides a step towards merger catalogues linking gravitational-wave signatures to the galaxies in which the black holes evolve.

\end{abstract}


\section{Introduction}

Massive black holes (MBHs) are closely tied to galaxy formation. Their growth is linked to gas inflow, star formation, feedback, and galaxy mergers across cosmic time, and their coalescences trace hierarchical structure formation \citep{Begelman1980,Volonteri2010}. In galaxy mergers, the central MBHs may form massive black hole binaries (MBHBs), which then evolve through dynamical friction, stellar scattering, gas-disc coupling, and gravitational-wave (GW) emission before coalescence. This sequence is well established, but tracking it directly in galaxy simulations remains difficult because the relevant scales range from kiloparsec separations to sub-parsec separations and, eventually, to the relativistic regime.

Classical MBHB theory still provides the conceptual basis for much of the field. Dynamical friction brings initially unbound MBHs towards the remnant nucleus. Stellar three-body scattering can then harden a bound binary, gas discs may exchange angular momentum with the binary, and GW emission dominates once the separation becomes small enough \citep{Begelman1980,Quinlan1996,Peters1964,SesanaKhan2015}. In an idealized spherical stellar background, however, the binary can deplete the stars that intersect its orbit. It may then harden too slowly to merge within a Hubble time, which is known as the ``final parsec'' problem \citep{MilosavljevicMerritt2003}. Later work has shown that non-spherical stellar potentials, gas inflow, massive perturbers, and repeated galaxy mergers can help refill the loss cone or otherwise drive the binary inward \citep{Sesana2011,Gualandris2022,Bonetti2018}. MBHB coalescence appears plausible in many galaxies, but its timescale still depends on the structure of the host galaxy on both resolved and unresolved scales.

This uncertainty is now central to low-frequency GW astrophysics. At nanohertz frequencies, pulsar timing arrays (PTAs) are probing the stochastic background expected from the cosmic MBHB population. In June 2023, the NANOGrav, EPTA/InPTA, PPTA, and CPTA collaborations reported evidence for a low-frequency GW signal that has been interpreted as broadly consistent with an inspiralling supermassive-black-hole-binary background \citep{Agazie2023,EPTA2023,Reardon2023,Xu2023}. NANOGrav analyses indicate that astrophysically motivated MBHB population models can reproduce the current background spectrum, although its amplitude is higher than predicted by most models, thus emphasizing that a better understanding of binary evolution is needed to interpret the signal \citep{Agazie2023SMBHB}. Individual continuous-wave sources have not yet been firmly identified by any of the PTAs \citep{Agazie2023Individual}. Current PTA observations are therefore compatible with MBHB populations, but cannot by themselves characterize the underlying binary demographics, hardening channels, or environmental coupling without guidance from theoretical models. 

At millihertz frequencies, the Laser Interferometer Space Antenna (LISA) is planned to open a complementary window onto individual MBHB inspirals and mergers. LISA is designed to detect coalescing systems with  masses of roughly $10^4$--$10^7\,{\rm M_\odot}$ across a large fraction of cosmic history. LISA is capable of measuring merging MBH masses and spins, and, in favourable cases electromagnetic counterparts, including the host galaxies \citep{AmaroSeoane2017,Klein2016}. The mission was formally adopted by ESA on 25 January 2024, with launch planned for 2035 \citep{ESA2024}. Another space-based detector, the TianQin mission, was proposed in 2014 \citep{Luo2016TianQin}. TianQin is designed to probe MBHBs in a frequency band similar to that of LISA, providing complementary constraints on masses, merger rates, and parameter-estimation accuracy \citep{Wang2019TianQin}. A third mission, Taiji, is also being proposed in the same frequency band \citep{Hu2017Taiji}. If LISA operates concurrently with TianQin or Taiji, the resulting detector network would sharpen the sky localization of MBHB mergers and improve the prospects for identifying electromagnetic counterparts \citep{Ruan2021LISATaiji}. PTA, LISA, TianQin, and Taiji science would all benefit from models that predict whether MBHs merge, when they merge, in what environments, and with what masses, spins, eccentricities, and recoil velocities.

Electromagnetic observations probe another part of the same evolutionary chain. Dual AGNs trace systems in which two accreting MBHs remain spatially separated on scales from several parsecs to tens of kiloparsecs. Binary AGNs instead probe smaller separations, at which the MBHs may be gravitationally bound \citep{DeRosa2020}. Such systems connect galaxy mergers, MBH accretion, and possible precursors of low-frequency GW sources, although the connection is indirect. Semi-analytic modelling based on TNG50-3 galaxy mergers indicates that dynamical-friction delays can last several gigayears, and that the probability of eventual coalescence depends on redshift, separation, host-galaxy properties, mass ratio, and feedback assumptions \citep{Li2022TNG50,Li2023TNG50DualAGN}. Romulus-based analyses likewise indicate that dual AGNs include rapidly evolving systems, slowly evolving systems, and systems in which the MBHs do not merge efficiently \citep{Saeedzadeh2024}. Recent JWST/NIRSpec results have raised the possibility of a high dual-AGN fraction at redshift $z\sim3$, which has been interpreted as a potential constraint on whether MBHs are simultaneously fuelled and on whether subsequent binaries stall or undergo rapid gas-driven inspiral \citep{Perna2025,PadmanabhanLoeb2024}. JWST/NIRSpec surveys have further revealed a diverse population of accreting MBHs at $4<z<11$, including candidate merging systems and MBHs that appear overmassive relative to their host galaxies \citep{Maiolino2024}. The surveys have also revealed an offset AGN identified only $\sim740$\,Myr after the Big Bang \citep{Uebler2024}. These observational links are promising, but they also expose the gap between observable dual activity and the unresolved dynamics that determine whether and when MBHs actually coalesce.

Theoretical predictions for MBH merger rates have developed along several partly complementary paths. Semi-analytic and semi-empirical models combine halo or galaxy merger trees with prescriptions for seeding, accretion, feedback, dynamical friction, stellar hardening, gas-driven migration, and GW emission \citep{Barausse2012,Klein2016,Barausse2020}.  Early semi-analytic merger-tree models already followed, for example, the cosmic evolution of the MBH spin distribution through accretion and mergers \citep{Volonteri2005}. They also followed the dynamical evolution and observable signatures of intermediate-mass black holes in the nearby Universe \citep{VolonteriPerna2005}. A more recent example applies the L-Galaxies semi-analytic model to characterize the host galaxies of coalescing MBHBs detectable by LISA \citep{IzquierdoVillalba2023}. The main advantage of these models is that they can explore large parameter spaces and connect assumptions about MBH formation and binary evolution to PTA and LISA observables. This usually comes at the cost of reducing the environmental history of each binary to a small set of analytical host-galaxy parameters.

Hydrodynamical simulations instead model the host galaxy explicitly. They can follow gas inflow, star formation, feedback, galaxy mergers, and MBH growth self-consistently on resolved scales \citep{Chapon2013,Tremmel2017Romulus,Pfister2019,Massonneau2023,Beckmann2025}. These simulations are therefore important for connecting MBH dynamics to the evolving host-galaxy environment, even though the final inspiral remains far below the spatial resolution of large-volume simulations. The treatment of MBH repositioning, dynamical friction, and numerical merger criteria can therefore affect inferred merger rates and the predicted population of wandering or stalled MBHs \citep{Chen2022DF,Genina2024DF}. The central modelling issue is not whether the host environment matters, but how much of that environment can be retained once the binary enters a subgrid regime. A recent LISA Astrophysics Working Group comparison of about 20 semi-analytic models and cosmological simulations found that predicted MBH merger and LISA detection rates remain sensitive to seeding, accretion, feedback, resolution, and dynamical-delay prescriptions \citep{IzquierdoVillalba2026}. In particular, dynamical delays generally suppress predicted rates, with strong effects at high redshift and for low-mass MBHs \citep{IzquierdoVillalba2026}.

At the high-accuracy end, direct $N$-body simulation can follow close MBH--star interactions, eccentricity evolution, and post-Newtonian binary dynamics much more accurately than cosmological simulations \citep{Sesana2011,Varisco2021,Gualandris2022,Mannerkoski2023}. KETJU is especially relevant because it embeds a regularized treatment of close MBH dynamics within a galaxy-simulation framework \citep{Mannerkoski2023}. Such simulations are powerful calibration tools, although their computational cost limits their direct use in large cosmological samples. Many such simulations are also purely collisionless and omit gas and star formation. They are therefore best suited to gas-poor systems rather than to the gas-rich nuclei where gas-driven hardening, circumbinary torques, and accretion are expected to be important. Scattering libraries and recoil formulae provide another link between small and large scales. Three-body scattering experiments quantify merger, exchange, stalled, and ejection outcomes \citep{Bonetti2018}, while numerical-relativity fits connect remnant properties to GW recoil \citep{Lousto2013}. Cosmological simulations indicate that a non-negligible fraction of MBHBs can encounter a third MBH before coalescence, especially when binary inspiral is stalled  \citep{Sayeb2024Triples}, building on early dynamical studies of triplet systems \citep{Volonteri2003,Hoffman2007,Bonetti2018,Koehn2023}. These results motivate models that include MBH triplets, although they do not by themselves prescribe how such interactions should be coupled to gas, feedback, and host-galaxy evolution in a hydrodynamical simulation.

Large-volume models can sample MBH populations, but often describe the unresolved environment with analytical prescriptions. High-resolution simulations can resolve selected nuclei, but cannot easily cover cosmological volumes. Standard hydrodynamical simulations follow the evolving galaxy environment, but usually merge MBHs before they would actually coalesce. The first RAMCOAL paper introduced a three-stage subgrid model for MBHB evolution in RAMSES \citep{RAMCOAL2024}. RAMCOAL does not let MBHs coalesce immediately once they satisfy a numerical merger condition. Instead, it follows close pairs with a subgrid model that accounts for stellar and gaseous dynamical friction, stellar hardening, circumbinary-disc torques, and eventual GW emission. During the subgrid evolution, the binary remains embedded in the simulated galaxy. Local gas and stellar properties continue to be measured on the fly, accretion and feedback remain coupled to the parent simulation, and the subgrid orbit responds to the environment of the host galaxy.

The present paper extends RAMCOAL to include subgrid MBH triplets. Such systems may arise when a later galaxy merger brings an incoming MBH into a nucleus that already hosts a close pair. Triplets can accelerate or delay coalescence, trigger exchanges and ejections, change the spins of the remnants and the distribution of recoil velocities, and modify the accretion history of MBHs \citep{Bonetti2018,Sayeb2024Triples}. Whether a third MBH matters depends on the ordering of two timescales. One is $t_{\rm coal,in}$, the time the first pair needs to coalesce once it has formed. The other is $t_{\rm DF,out}$, the time the incoming MBH needs to reach the nucleus, which is set by the interval between successive galaxy mergers and by dynamical friction. If $t_{\rm coal,in}\ll t_{\rm DF,out}$, or if the intruder settles on a wide orbit that remains dynamically stable with respect to the inner pair \citep{MardlingAarseth2001,Myllari2018}, the system can safely be treated as two successive binaries. The first pair may instead stall, for instance at parsec separations in gas-poor, nearly spherical nuclei \citep{MilosavljevicMerritt2003}. In that case, the intruder arrives before the first coalescence and a three-body interaction follows. Neglecting such interactions can bias the results in several ways. The coalescing pair need not be the original binary, and the merger time can be shortened by orders of magnitude. The lightest MBH can be ejected altogether, and the remnant masses, spins, eccentricities, and recoils are altered \citep{Hoffman2007,Bonetti2018,Bonetti2019}. Cosmological models indicate that these configurations are not rare. With merger-tree models, \citet{Volonteri2003} already found that MBHs in massive haloes experience tens of triple interactions over their lifetime. In the Illustris-based analysis of \citet{Sayeb2024Triples}, $22\%$ of all MBH binaries encounter a third MBH before merging, and $71\%$ of these triple interactions involve binaries that would not otherwise merge by $z=0$. \citet{Kulkarni2012} estimate that about $30 (60) \%$ of galaxies in haloes with present-day masses of $\sim10^{14}$ ($\gtrsim10^{15}$)\,M$_\odot$ host more than two MBHs at $2\lesssim z\lesssim6$. \citet{Hoffman2023ASTRID} identify 588 MBH triplets and 174 quadruplets with maximum separations below 200\,kpc in the ASTRID volume at $z\sim2$. Triplet-induced coalescences also contribute to the nanohertz background and to the LISA event rate even when binaries would otherwise stall \citep{Bonetti2018b,Bonetti2019}. 

RAMCOAL combines elements of several of these approaches. Like hydrodynamical simulations, it follows MBHs in a self-consistently evolving galaxy, and like semi-analytic population models, it uses prescriptions for the unresolved scales. In a pure post-processing delay model, dynamical evolution is applied analytically to MBH mergers drawn from a completed cosmological simulation \citep{Kelley2017}. RAMCOAL instead updates the subgrid orbit during the simulation, taking into account how the local gas, stellar, and feedback properties evolve. This makes the model well suited to PTA and LISA science, to the interpretation of dual AGNs, and to multimessenger studies. The present paper has two aims. The first is to introduce a triplet extension of RAMCOAL that identifies when a third MBH becomes bound to an MBH binary. The extension then maps the ensuing chaotic interaction onto the scattering survey of \citet{Bonetti2018} and updates the subgrid system accordingly. The second is to revise the subgrid treatment of MBHB accretion and spin evolution, including preferential accretion across the circumbinary cavity, the suppression of mini-disc accretion by feedback, the recoil of merger remnants, and stage-dependent spin alignment. In addition, we quantify the computational cost of the extended model in a cosmological simulation of the COSCOAL suite, the type of application for which RAMCOAL is designed (Section~\ref{sec:cosmo}). These additions bring RAMCOAL closer to a general subgrid model for multiple MBH systems and their coalescence in cosmological simulations of galaxy formation.

\section{General RAMSES Framework}
\label{sec:ramses}

RAMCOAL is implemented within the adaptive mesh refinement code RAMSES~\citep{Teyssier2002}. RAMSES evolves collisionless components with a particle-mesh gravity solver and integrates gas dynamics with a second-order Godunov scheme on the adaptive mesh. %

The RAMSES configuration relevant here includes radiative cooling and heating, star formation, stellar feedback from core-collapse supernovae, and MBH sink particles \citep{Dubois2012}. The adaptive mesh supplies the dynamical range needed to model galaxy dynamics. Even with refinement, however, it does not resolve the full inspiral of MBH pairs and higher-order multiples down to the separations relevant for hardening, circumbinary-disc interaction, and GW-driven coalescence. RAMCOAL addresses this subgrid component of the problem. RAMSES models galactic scales by explicitly following the gas, stellar, and DM dynamics that drive MBH evolution. The final unresolved dynamical stages of close pairs and triplets are instead modelled semi-analytically below the grid scale.

Within RAMSES, individual MBHs are represented by collisionless sink particles that move under gravity, accrete gas, and inject AGN feedback into the surrounding gas \citep{Dubois2012,Dubois2014b,Dubois2021,Beckmann2025}. This model provides the numerical environment in which RAMCOAL operates. In particular, it supplies the sink-particle dynamics and the local gas and stellar properties measured around each BH. It also supplies the accretion and feedback channels that regulate MBH growth, and the spin variables that later enter the binary and triplet extensions.

For the purposes of this paper, RAMCOAL should be understood as an extension of the standard RAMSES MBH implementation rather than as a replacement for it. Isolated BHs are still evolved with the standard RAMSES treatment of sink-particle motion, mass and spin evolution, and feedback. RAMCOAL intervenes only when two or more BHs enter a regime that is unresolved by the grid, at which point their subsequent orbital evolution is transferred to a dedicated subgrid model. The triplet extension introduced here is therefore the third layer in a hierarchy consisting of the parent RAMSES simulation, the original RAMCOAL MBHB subgrid model, and the new treatment of three-body interactions.

RAMCOAL is not tied to RAMSES or to its adaptive-mesh solver. Besides the RAMSES implementation used in this paper, it is available as a standalone library, CCOAL, a C version of RAMCOAL that includes the triplet module presented here. A C++ interface makes the library easy to call from other simulation codes. Through this interface, CCOAL has already been coupled to the Santa Cruz semi-analytic model of galaxy formation (SantaCoal, Li \& Somerville, in preparation). RAMCOAL can therefore be used as a subgrid model both in hydrodynamical simulations and in semi-analytic models. Implementing it in a hydrodynamical code requires a set of inputs, including kernel-averaged local quantities around each MBH, namely the gas density and sound speed, and the densities and velocity dispersions of stars and dark matter within the resolution sphere. The accretion rate and feedback state are also needed. Particle-based and moving-mesh codes (e.g. GADGET-4, AREPO, GIZMO, or ChaNGa) provide equivalent quantities, because they estimate the gas properties around black-hole particles with neighbour kernels. RAMCOAL can therefore be implemented in such codes with the same inputs, without changing the physics of the subgrid model. Only the numerical definition of the resolution sphere would have to be tied to the corresponding resolution scale of the host code. Here, $R_{\rm sp}=4\Delta x$ is set by the cell size. In other codes, the corresponding scale could be the gravitational softening or the black-hole kernel radius.

The subgrid prescriptions described below are embedded in the RAMSES model for BH growth and feedback. Following the RAMSES model for super-Eddington accretion and feedback \citep{Massonneau2023,Massonneau2023b}, MBHs grow through Bondi--Hoyle--Lyttleton accretion \citep{Hoyle1939,Bondi1952},
\begin{equation}
\dot M_{\rm BHL}=
\frac{4\pi G^2 M_{\rm BH}^2 \bar{\rho}}
{\left(\bar{c}_{\rm s}^2+\bar{v}_{\rm rel}^2\right)^{3/2}},
\end{equation}
where the local density $\bar \rho$, sound speed $\bar c_{\rm s}$, and BH--gas relative velocity $\bar v_{\rm rel}$ are kernel-weighted averages measured around the sink (see~\citealp{Dubois2012} for details). An MBH sink carries only the mass of the MBH (or, in a subgrid system, of the MBHs) that it hosts. The stellar and dark-matter properties entering the subgrid dynamics are measured from the resolved star and dark-matter particles inside the resolution sphere. When the sphere satisfies the star-formation criterion (Section~\ref{sec:stage1}), these measurements are supplemented by the analytic cored-isothermal gas and stellar profiles of paper~I. In the super-Eddington RAMSES model, the accretion rate is not capped at the Eddington limit by default\footnote{However, the MBH mass accretion rate $\dot M_{\rm BH}$ can differ from $\dot M_{\rm BHL}$ when accretion is limited by the gas supply. See \cite{Dubois2012} for details.}. Instead, the Eddington ratio,
\begin{equation}
f_{\rm Edd} \equiv \frac{L}{L_{\rm Edd}},
\end{equation}
where $L$ is the bolometric luminosity and $L_{\rm Edd}$ is the Eddington luminosity, is used to determine the active feedback regime using the fit from \cite{Madau2014}.

The radiative efficiency is set by the MBH spin through the radius of the innermost stable circular orbit, $r_{\rm ISCO}(a_\ast)$ in units of $GM_{\rm BH}/c^2$ \citep{Bardeen1972},
\begin{equation}
\varepsilon_{\rm r}=1-\sqrt{1-\frac{2}{3\,r_{\rm ISCO}}},
\label{equ:epsr}
\end{equation}
which increases from $\varepsilon_{\rm r}\simeq0.038$ for a maximally spinning MBH accreting from a retrograde disc, through $0.057$ for $a_\ast=0$, to $0.32$ for $a_\ast=0.998$. The Eddington luminosity is $L_{\rm Edd}=4\pi GM_{\rm BH}m_{\rm p}c/\sigma_{\rm T}$, with $m_{\rm p}$  the proton mass, $\sigma_{\rm T}$ the Thomson cross-section, and the Eddington accretion rate is $\dot M_{\rm Edd}=L_{\rm Edd}/(\varepsilon_{\rm r}c^2)$. For sub-Eddington, radiatively efficient accretion $L=\varepsilon_{\rm r}\dot M_{\rm BH}c^2$, so that $f_{\rm Edd}=\dot M_{\rm BH}/\dot M_{\rm Edd}$. Below $f_{\rm Edd}=0.01$ the radiative efficiency is further reduced in proportion to $f_{\rm Edd}/0.01$ to represent a radiatively inefficient flow \citep{BensonBabul2009}. The RAMSES accretion model allows for super-Eddington accretion \citep{Massonneau2023}, but, for simplicity, we enforce the Eddington limit in the isolated-galaxy simulations presented here.

RAMSES distinguishes three feedback modes. For radiatively inefficient accretion with $f_{\rm Edd}\leq 0.01$, the MBH is in the radio mode and injects kinetic jet power
\begin{equation}
\dot E_{\rm jet}=\eta_{\rm jet}\dot M_{\rm BH}c^2,
\end{equation}
with $c$ the speed of light, and with the jet efficiency written in the magnetically arrested disc (MAD)-inspired form
\begin{equation}
\eta_{\rm jet}=1.3\,a_\ast^2 f_{\rm MAD}^2,
\label{equ:etajet}
\end{equation}
following  \cite{Tchekhovskoy2015} and \citet{Sadowski2016}, where $a_\ast$ is the MBH spin and $0\leq f_{\rm MAD}\leq1$ is the ``MAD-ness'' state of the disc (i.e.~a proxy for the strength of the poloidal magnetic field). Throughout, $a_\ast\equiv cJ_{\rm BH}/(GM_{\rm BH}^2)$ is the dimensionless spin parameter, where $J_{\rm BH}$ is the MBH angular momentum. Its magnitude is limited to $|a_\ast|\le0.998$ \citep{Thorne1974}, and a negative value denotes an MBH accreting from a counter-rotating (retrograde) disc. The NewHorizon cosmological simulation~\citep{Dubois2021} models the spin evolution of MBHs in a cosmological context. In this simulation, coherent gas accretion spins most MBHs up to high values ($|a_\ast|\approx0.4$--0.99, with near-maximal spins common). MBH mergers broaden the distribution, and jet extraction in the radio mode spins some massive MBHs down over Gyr timescales \citep{Beckmann2025}. In the tests presented here all MBHs are initialized with spins close to zero, so that their spin evolution is driven entirely by the accretion histories discussed in Section~\ref{sec:testcases}. The radio-mode efficiency is taken from a fourth-order polynomial fit to the MAD simulations of \citet{McKinney2012} (jet plus wind). This fit, introduced in \cite{Dubois2021}, also increases steeply with $|a_\ast|$.

The jet is modelled as a mass-loaded bipolar outflow launched along the MBH spin axis. Mass, momentum, and kinetic energy of the jet are deposited together in a cylinder of radius $r_{\rm jet}=2\Delta x$ and height $2r_{\rm jet}$ centred on the sink \citep{Dubois2010,Dubois2012}. The jet power is $\dot E_{\rm jet}=\eta_{\rm jet}\dot M_{\rm BH}c^2$, and the gas is injected with a velocity $v_{\rm jet}=0.1c$ and a mass-loading factor $\beta_{\rm jet}$. The jet therefore carries a mass flux $\beta_{\rm jet}\dot M_{\rm BH}$ and a momentum flux $\dot p_{\rm jet}=\beta_{\rm jet}\dot M_{\rm BH}v_{\rm jet}$. In the Eddington-limited configuration used for all the simulations of this paper, the mass loading is fixed to $\beta_{\rm jet}=100$,  Following \citet{Dubois2012}, the jet velocity is determined by the kinetic power and mass loading,
$v_{\rm jet}=c\sqrt{2\eta_{\rm jet}/\beta_{\rm jet}}$.
The corresponding momentum flux is
$\dot p_{\rm jet}=\sqrt{2\eta_{\rm jet}\beta_{\rm jet}}\,\dot M_{\rm BH}c
=\sqrt{2\beta_{\rm jet}/\eta_{\rm jet}}\,\dot E_{\rm jet}/c$.
For $\beta_{\rm jet}=100$ and the nearly non-spinning MBHs in our tests, for which $\eta_{\rm jet}\simeq0.04$, this gives
$\dot p_{\rm jet}\simeq2.8\,\dot M_{\rm BH}c
\simeq71\,\dot E_{\rm jet}/c$.
The injected kinetic power is therefore normalized to $\dot E_{\rm jet}$. At intermediate accretion rates, $0.01<f_{\rm Edd}\leq 1$, the MBH enters the quasar mode, in which the feedback is primarily thermal to mimic radiatively-driven winds,
\begin{equation}
\dot E_{\rm thm}=\eta_{\rm thm}\dot M_{\rm BH}c^2,
\label{equ:ethm}
\end{equation}
where $\eta_{\rm thm}$ is the spin-dependent thermal efficiency of the disc, $\eta_{\rm thm}=\varepsilon_{\rm f}\,\varepsilon_{\rm r}(a_\ast)$. Here $\varepsilon_{\rm r}$ is the radiative efficiency of equation~(\ref{equ:epsr}), and $\varepsilon_{\rm f}$ is the fraction of the radiated energy that couples thermally to the surrounding gas \citep{Dubois2012}.

For super-critical accretion, $f_{\rm Edd}>1$, RAMSES activates a mixed mode in which both kinetic and thermal channels operate simultaneously. The total injected power is
\begin{equation}
\dot E_{\rm sEdd}=
\left(\eta_{\rm jet}+0.5\,\eta_{\rm thm}\right)\dot M_{\rm BH}c^2.
\label{equ:esedd}
\end{equation}
This regime is designed to approximate the coexistence of powerful outflows and radiative losses in thick discs \citep{Madau2014,Sadowski2016,Massonneau2023b}. The jet efficiency in equation~(\ref{equ:esedd}) is that of equation~(\ref{equ:etajet}), and $\eta_{\rm thm}$ is the thermal efficiency of equation~(\ref{equ:ethm}), both evaluated at the instantaneous spin. The factor 0.5 on the thermal term is the value adopted in the RAMSES model of \citet{Massonneau2023}. The coupling is relevant for RAMCOAL because the MBH spin, radiative efficiency, and AGN mode evolve together. The spin state tracked by the subgrid model also enters the RAMSES feedback efficiencies and therefore changes the partition of released energy between jets and thermal heating.

\section{The RAMCOAL Subgrid Model for MBHBs}

Before introducing the triplet extension, we summarize the RAMCOAL model for MBHBs as presented in paper~I \citep{RAMCOAL2024}. In the parent RAMSES simulation, two MBH sink particles satisfy the standard numerical merger condition once their separation falls below the resolution sphere, $R_{\rm sp}=4\Delta x$. RAMCOAL intercepts that event, stores the subgrid pair, and follows the subsequent inspiral using local gas, stellar, and dark-matter quantities measured on the fly from the hydrodynamical simulation.

\begin{figure*}
    \centering
    \includegraphics[width=\textwidth]{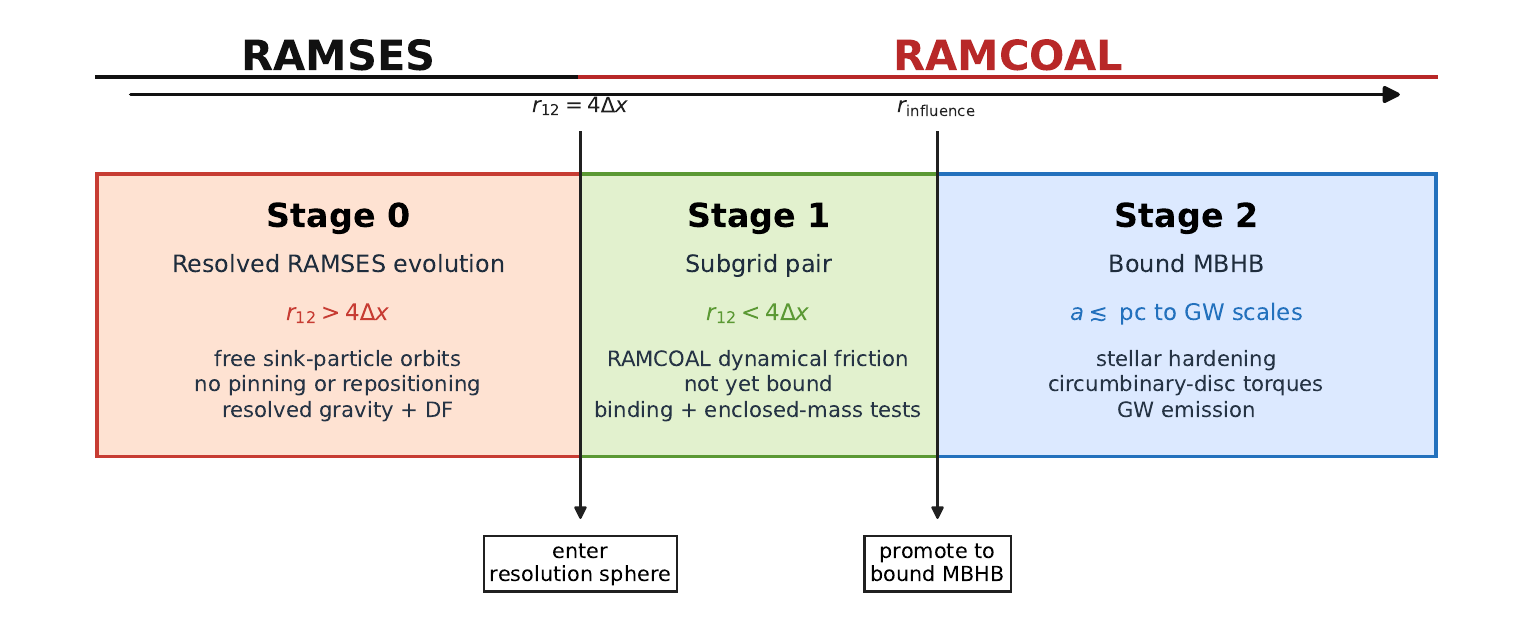}
    \caption{Three-stage RAMSES--RAMCOAL treatment of MBH-pair evolution, progressing from left to right. The stages (Stage~0 $\rightarrow$ Stage~1 $\rightarrow$ Stage~2) follow a single pair as its separation shrinks towards coalescence. In stage~0 (left), outside the resolution sphere, the MBHs are standard RAMSES sink particles whose positions are not pinned to potential minima. They move freely under resolved gravity plus the RAMSES dynamical-friction correction. Once their separation reaches $4\Delta x$ and they enter the resolution sphere, RAMCOAL replaces the numerical sink merger by the stage~1 subgrid dynamical-friction evolution of a not-yet-bound pair. The pair is promoted to a stage~2 bound MBHB (right) only after the binding and enclosed-mass criteria are satisfied, and then hardens through stellar scattering, circumbinary-disc torques, and gravitational-wave emission.}
    \label{fig:ramcoalstages}
\end{figure*}

The overall RAMSES--RAMCOAL treatment can be organized into three stages, illustrated schematically in Fig.~\ref{fig:ramcoalstages}. Stage~0 is the resolved RAMSES evolution outside the resolution sphere. Stage~1 describes the subgrid, not-yet-bound pair inside $4\Delta x$, where RAMCOAL models dynamical friction from gas, stars, and dark matter. Stage~2 begins only after the pair is gravitationally bound and the enclosed gas-plus-stellar mass falls below twice the pair mass (the binding and enclosed-mass criteria given below). The subgrid state is then a bound MBHB hardened by stellar scattering, circumbinary-disc torques, and gravitational-wave emission.

\subsection{Stage 0 (resolved RAMSES sink-particle evolution)}

Before RAMCOAL is activated, each MBH is a standard RAMSES collisionless sink particle. Accretion and AGN feedback are computed through the standard RAMSES MBH prescriptions described above. The MBHs are not pinned to the galaxy centre but evolve freely as massive particles in the resolved gravitational field. This preserves off-centre or wandering MBHs and allows the orbital decay time to be set by the simulated merger remnant rather than by an imposed repositioning rule \citep{Wurster2013, Ni2022, Bahe2022, Damiano2025}.

At this stage RAMSES supplements the resolved particle-mesh force with subgrid dynamical friction when the gravitational wake is not resolved. The gas contribution is evaluated from kernel-averaged quantities inside a sphere of radius $4\Delta x$ around the MBH,
\begin{equation}
F_{\rm DF,gas}^{\rm RAMSES}
= f_{\rm gas}({\cal M})\,4\pi\,\bar{\rho}
\left(\frac{G M_{\rm BH}}{\bar{c}_{\rm s}}\right)^2 ,
\end{equation}
where ${\cal M}$ is the Mach number, and $f_{\rm gas}$ is given by the Ostriker drag formula with the finite Coulomb logarithm adopted in RAMSES \citep{Ostriker1999,Chapon2013,Lescaudron23,RAMCOAL2024}. For the collisionless component, RAMSES adopts the approach from~\cite{Pfister2019} in which stars and dark matter are treated separately and the local velocity distribution is measured from particles inside the same $4\Delta x$ sphere. In the notation of the first RAMCOAL paper,
\begin{equation}
4\pi v^2 f(v)
=
\frac{3}{256\pi \Delta x^3}
\sum_{i\in S} m_i\,\delta(v_i-v),
\end{equation}
where $S$ is the $4\Delta x$ resolution sphere around the sink, $m_i$ and $v_i$ are the particle masses and velocities, and the Coulomb logarithm is $\ln(4\Delta x/r_{\rm def})$ with $r_{\rm def}\simeq GM_{\rm BH}/v_{\rm BH}^2$. The collisionless subgrid correction is turned off when the 90-degree deflection scale is already resolved, so that resolved gravity is not double counted \citep{Pfister2019,RAMCOAL2024}. Stage~0 ends when two MBHs enter each other's resolution sphere and would otherwise be numerically merged by RAMSES.

\subsection{Stage 1 (subgrid dynamical-friction inspiral)}
\label{sec:stage1}

RAMCOAL stage~1 begins at the numerical merger boundary, $r_{12}<4\Delta x$. The pair is not immediately treated as a bound Keplerian binary. Instead, the two MBHs remain separate subgrid components associated with the numerically merged RAMSES sink, and their relative orbit is integrated until the criteria for a bound binary are satisfied. The orbital equations are sub-cycled relative to the hydrodynamical time step, with an adaptive time step chosen to limit abrupt changes in the subgrid orbital elements.

The stage~1 drag is computed from the local RAMSES quantities (gas or stars) but with subgrid assumptions appropriate to the unresolved region inside the resolution sphere. For stars and dark matter, RAMCOAL adopts the \cite{AntoniniMerritt2012} form used in the first paper, retaining both slow and fast collisionless particles,
\begin{equation}
\boldsymbol{F}_\star =
\boldsymbol{F}_\star^{(v_\star<v)}
+
\boldsymbol{F}_\star^{(v_\star>v)} ,
\end{equation}
where the subgrid velocity distribution is approximated by a Maxwellian distribution function, $v_\star$ is the velocity of the collisionless particles, and $v$ is the velocity of the MBH.
\begin{equation}
f(v_\star)=
\frac{1}{(2\pi\sigma_\star^2)^{3/2}}
\exp\left(-\frac{v_\star^2}{2\sigma_\star^2}\right),
\quad
\sigma_\star=
\left[\frac{G(M_{\rm sp}+\widetilde{M}_{12})}{R_{\rm sp}}\right]^{1/2}.
\label{equ:sigmastar}
\end{equation}
Here $M_{\rm sp}$ is the total mass of gas, stars, and dark matter measured in the resolution sphere, and $\widetilde{M}_{12}=M_1M_2/(M_1+M_2)$ is the reduced mass~\citep{AntoniniMerritt2012,DosopoulouAntonini2017,RAMCOAL2024}. The maximum impact parameter is set by $4\Delta x$, and the minimum impact parameter is the larger of two scales. One is the gravitational-deflection scale, given by the $90^\circ$-deflection radius $b_{90}=GM_{\rm BH}/v_{\rm rel}^2$, where $v_{\rm rel}$ is the MBH velocity relative to the collisionless background. The other is the gravitational radius $R_{\rm g}=GM_{\rm BH}/c^2$. These define $b_{\rm min}=\max(b_{90},R_{\rm g})$, which enters the Coulomb logarithm $\ln\Lambda=\ln(4\Delta x/b_{\rm min})$ and keeps the stellar and dark-matter drag from diverging at small impact parameters.

The gas drag in stage~1 is analogous to the RAMSES gas drag. However, RAMCOAL uses the \cite{KimKim2007} formulation for an MBH orbiting through a gaseous disc and therefore includes the radial and azimuthal components of the spiral wake,
\begin{equation}
\boldsymbol{F}_{\rm gas}
=
-\frac{4\pi(GM_{\rm BH})^2\bar{\rho}_{\rm sp}}
{\Delta\bar{v}^{\,2}}
\left(I_r\boldsymbol{e}_r+I_\phi\boldsymbol{e}_\phi\right),
\end{equation}
where $\bar{\rho}_{\rm sp}$ is the mean gas density in the resolution sphere, $\Delta\bar{v}$ is the MBH--gas relative velocity, and $I_r$ and $I_\phi$ are dimensionless Mach-number-dependent wake functions \citep{KimKim2007}. To reduce the explicit resolution dependence, paper~I also introduced subgrid gas and stellar density profiles when the entire resolution sphere satisfies the star-formation criterion. The resulting cored-isothermal profile approximates the dense molecular-cloud and nuclear-star-cluster environment that cannot be resolved directly.

Stage~1 ends only when the pair satisfies the criteria for treatment as a bound MBHB. RAMCOAL requires the orbital kinetic energy to be smaller than the magnitude of the potential energy and the gas-plus-stellar mass enclosed by the binary orbit to be less than twice the total MBH-pair mass,
\begin{equation}
E_{\rm kin}<|E_{\rm pot}|,
\qquad
M_{\rm enc,gas+\star}<2(M_1+M_2),
\label{equ:boundcrit}
\end{equation}
respectively.
These criteria prevent a transient close passage inside $4\Delta x$ from being promoted prematurely to a hard MBHB.

The two criteria test different aspects of the dynamics. The binding criterion uses only the MBH masses, so it tests whether the two MBHs are bound to each other by their own gravity. The enclosed-mass criterion tests whether the pair lies inside its own sphere of influence. By convention, this is the sphere that contains a enclosed mass equal to twice the binary mass \citep{Merritt2013}.

Before the MBHs are bound in a binary, the orbit shrinks only when the MBHs lose energy through dynamical friction, the drag from the wake that each MBH creates in the surrounding gas, stars and dark matter. Stage~1 computes this friction, and the surrounding mass enters it through the velocity dispersion (equation~\ref{equ:sigmastar}) and the densities in the drag forces. After the MBHs are bound, dynamical friction becomes inefficient in shrinking the binary further \citep{Begelman1980}. Therefore stage~2 treats the pair as a Keplerian binary that hardens through loss-cone scattering and circumbinary-disc torques. This treatment is valid only inside the sphere of influence, where the stellar-hardening rates of stage~2 are calibrated \citep{SesanaKhan2015}.

\subsection{Stage 2 (bound MBHB hardening and coalescence)}

Once the stage~1 binding and enclosed-mass criteria are met, RAMCOAL promotes the system to a bound MBHB. The subgrid state is then described by an inner semi-major axis $a_{\rm in}$, eccentricity $e_{\rm in}$, component masses, spins, accretion rates, and local environmental quantities. The orbital evolution is written as a sum of the principal subgrid hardening channels,
\begin{equation}
\frac{{\rm d}a_{\rm in}}{{\rm d}t}
=
\left(\frac{{\rm d}a_{\rm in}}{{\rm d}t}\right)_{\rm LC}
+
\left(\frac{{\rm d}a_{\rm in}}{{\rm d}t}\right)_{\rm CBD}
+
\left(\frac{{\rm d}a_{\rm in}}{{\rm d}t}\right)_{\rm GW},
\end{equation}
\begin{equation}
\frac{{\rm d}e_{\rm in}}{{\rm d}t}
=
\left(\frac{{\rm d}e_{\rm in}}{{\rm d}t}\right)_{\rm LC}
+
\left(\frac{{\rm d}e_{\rm in}}{{\rm d}t}\right)_{\rm CBD}
+
\left(\frac{{\rm d}e_{\rm in}}{{\rm d}t}\right)_{\rm GW}.
\end{equation}
The loss-cone term (LC) follows the standard stellar-scattering hardening form, with the hardening rate $s\equiv {\rm d}(1/a)/{\rm d}t\simeq H\,G\rho_{\rm inf}/\sigma_{\rm inf}$. Here $H$ is the dimensionless hardening rate, and $\rho_{\rm inf}$ and $\sigma_{\rm inf}$ are the stellar density and velocity dispersion at the binary's influence radius. The LC term also includes an eccentricity source calibrated to scattering experiments \citep{Quinlan1996,SesanaKhan2015}. In gas-rich systems, RAMCOAL includes circumbinary-disc (CBD) migration and viscous torques, using the local RAMSES accretion and feedback state to regulate whether a coherent disc can persist. This feedback-regulated circumbinary accretion is described in detail in the CBD accretion model below. At small enough separations, the \cite{Peters1964} gravitational-wave (GW) terms dominate and drive the final coalescence. The stage~2 model therefore links the evolving galactic environment to the properties of the subgrid binary as a GW source. The GW terms themselves depend only on the component masses, semi-major axis, and eccentricity of the binary, but the galaxy still affects the GW-driven phase indirectly in three ways. First, the separation and eccentricity at which GW emission takes over are set by the preceding environmental hardening. This hardening rate depends on the stellar density and velocity dispersion around the binary, and on the gas inflow measured on the fly. In a denser or more gas-rich nucleus, the binary enters the GW-driven regime at a different separation and eccentricity, which changes the remaining GW-driven inspiral time, which scales steeply with both, $t_{\rm GW}\propto a^4(1-e^2)^{7/2}$ \citep{Peters1964}. Second, the component masses and spins that enter the waveform, the remnant spin and the recoil are shaped by accretion from the surrounding gas during stages~1 and~2. Third, the outcome feeds back on the host. The GW recoil and the orbit of the remnant determine where the MBH accretes and deposits its feedback afterwards, and a large enough kick removes it from the nucleus (Section~\ref{sec:recoil}). The GW-driven phase itself is integrated with the vacuum rates of \citet{Peters1964}.

\section{Merger classification of two subgrid systems}
\label{sec:mergerclass}

The updated model also changes how RAMCOAL identifies and combines two subgrid merger candidates before advancing the triplet evolution. Here a \emph{system} is the subgrid object carried by one sink particle. It can be a single MBH, a bound MBHB, or a triplet (the seven states of Table~\ref{tab:mergerfinderstates}). The individual MBHs contained in such a subgrid system are called its \emph{members}. The merger finder runs when two sink particles fall within the numerical merger radius, and its task is to combine the two systems they carry into a single updated subgrid system.

Each of the two systems of an accepted candidate pair is mapped to one of seven possible physical states, labelled 0--6 and summarized in Table~\ref{tab:mergerfinderstates}. These states encode the physical configuration of the subgrid system. The system can be a single MBH, a stage-1 pair, a bound binary, or three stage-1 MBHs. It can also be a hierarchical triplet, an active triplet, or a triplet in which the ejection of one MBH has already been scheduled. The classification gives $7\times 7=49$ possible input combinations, which are then all handled by a single reduction path.

\begin{table*}
\centering
\caption{Physical states used by the merger-classification scheme. The same 0--6 classification is applied to both systems of a candidate encounter, which gives the 49 combinations handled by the reduction path.}
\label{tab:mergerfinderstates}
\begin{tabular}{clp{0.34\textwidth}p{0.36\textwidth}}
\toprule
State & Short name & Physical meaning & Role in the reduction step \\
\midrule
0 & Single MBH & An isolated MBH with no active subgrid companion. & May form a new pair or join an existing subgrid binary or triplet. \\
1 & Stage-1 pair & Two MBHs inside the RAMCOAL resolution sphere but not yet promoted to a bound MBHB. & Evolves by subgrid dynamical friction and can be combined with another incoming MBH system. \\
2 & Bound binary & A two-body MBHB evolving in stage~2, with no third component stored. & Preserves the existing inner binary unless a higher-order encounter requires reduction. \\
3 & Three stage-1 MBHs & Three MBHs present, but no bound inner pair has formed yet. & Represents a subgrid triplet whose evolution is still dominated by dynamical friction. \\
4 & Hierarchical triplet & Three MBHs with an inner bound pair and an outer MBH that is still in stage~1 and has not entered the chaotic phase. & Retains the inner binary until the outer MBH is captured. \\
5 & Active triplet & Chaotic three-body phase active, with both the inner and the outer orbits in stage~2. & Allows the \Bonetti outcome to be drawn and applied, which can produce mergers, exchanges, stalled systems, and surviving binaries. \\
6 & Pending ejection & Active triplet whose interaction outcome has scheduled a three-body ejection. & Ensures that the ejected MBH is released before the remaining system is reduced or evolved further. \\
\bottomrule
\end{tabular}
\end{table*}

The reduction that follows is straightforward. When two systems are combined, RAMCOAL places them in a common centre-of-mass frame. It then coalesces any bound inner pair that must merge, using the remnant-spin and recoil prescription below. Finally, it reduces the remaining members to at most three MBHs, which are ordered by mass and re-classified into one of the seven states. 

\section{Triplet Extension and Revised Subgrid Prescriptions}

A subgrid triplet can form when two subgrid systems are combined through the merger classification described above. The treatment of triplets extends the 3-stage binary RAMCOAL model (Fig.~\ref{fig:ramcoalstages}). It reuses the same three stages, environmental quantities, and orbital, accretion, feedback, and spin variables introduced for binaries in paper~I. It also updates the subgrid models for MBHB accretion in the CBD, for feedback-regulated mini-disc accretion, and for spin evolution. The updated CBD accretion model includes the fraction of the inflow that is stored in the ``lump'' at the inner edge of the disc. 

\subsection{Entering the three-body regime}

The extension described here adds an on-the-fly prescription for MBH triplets to the binary model presented in paper~I. The triplet model applies when all three MBHs are closer to their common centre of mass than the resolution sphere radius, $R_{\rm sp}=4\Delta x$.  RAMCOAL then determines whether the system remains a hierarchical triplet or enters a chaotic three-body phase. Once the triplet enters this phase, the model does not integrate the full resonant interaction directly. Instead, it maps the masses and mass ratios onto the scattering survey of \cite{Bonetti2018} (\Bonetti hereafter) and draws one of seven interaction outcomes. It then translates that outcome into the corresponding merger or ejection of the subgrid triplet.

\begin{figure*}
    \centering
    \includegraphics[width=\textwidth,height=0.90\textheight,keepaspectratio]{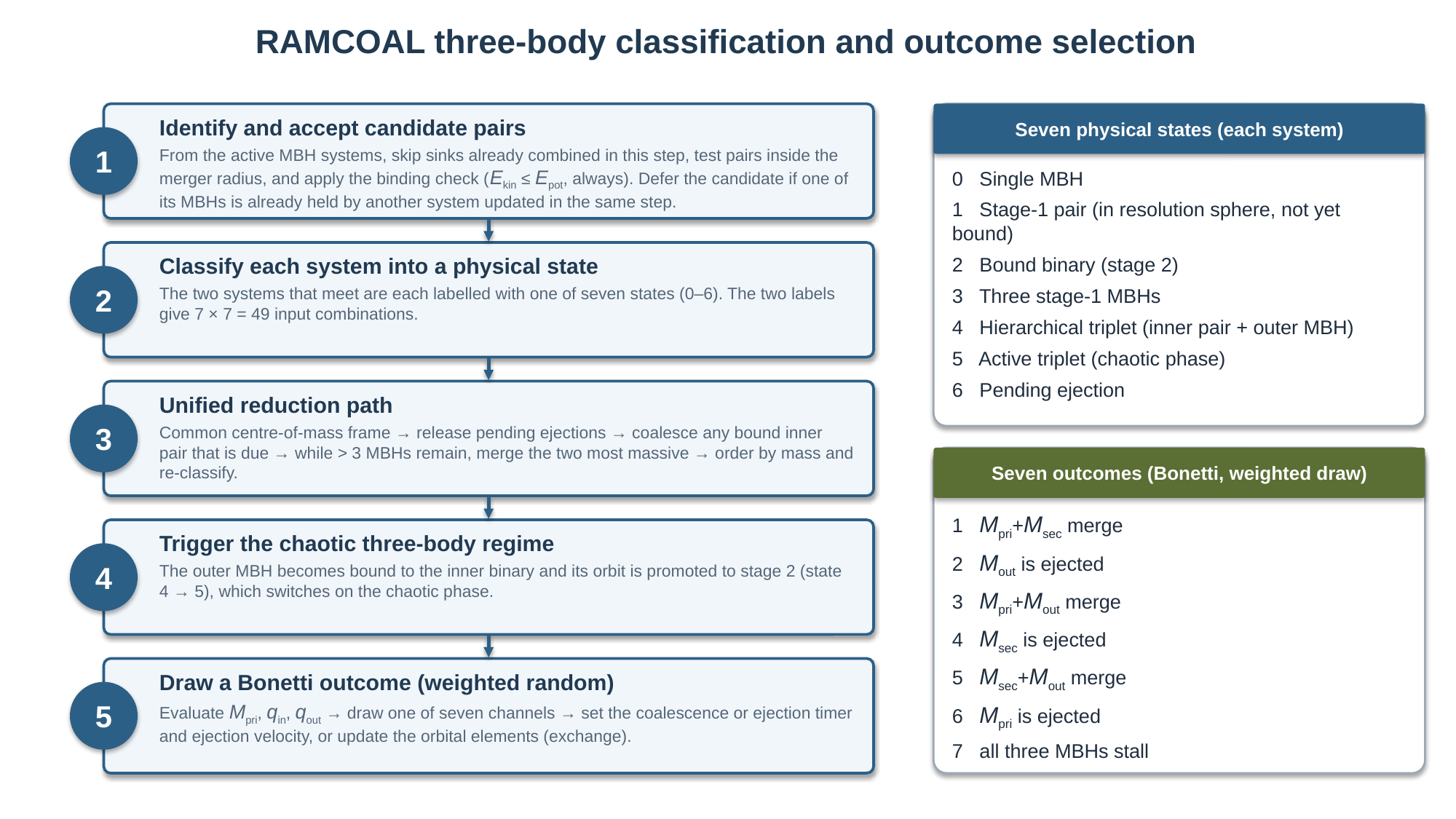}
\caption{Flowchart of the RAMCOAL classification of chaotic three-body systems and of the selection of their outcome, in five numbered steps (left), with two legends (right) for the seven physical states and the seven interaction outcomes. (1)~Candidate pairs are identified from the active MBH systems at that time step. MBHs that have already been combined into subgrid systems in the current step are skipped, pairs within the resolution-sphere radius are tested, and a relative-velocity binding check ($E_{\rm kin}\le E_{\rm pot}$) is applied. (2)~Each of the two systems that meet in an encounter is classified into one of seven physical states ($0$--$6$, see the upper legend and Table~\ref{tab:mergerfinderstates}). Together, the two labels give $7\times7=49$ input combinations. A pairing such as (single MBH, active triplet) therefore means an isolated MBH encountering an active triplet, not a single object in two states at once. (3)~In the reduction path, the systems are placed in a common centre-of-mass frame. Pending three-body ejections are released as resolved RAMSES sinks. Any bound inner pair that must coalesce first is merged. If more than three MBHs remain, the two most massive MBHs are merged, so that the system is reduced to at most three MBHs, which are then ordered by mass and re-classified (Section~\ref{sec:mergerclass}). (4)~The chaotic three-body regime is triggered. (5)~A single outcome is then drawn from the seven \citet{Bonetti2018} channels (lower legend) by a weighted random selection from the tabulated probabilities. This outcome sets the coalescence or ejection timer and the ejection velocity, or updates the orbital elements for an exchange.}
    \label{fig:flowchart3body}
\end{figure*}

Figure~\ref{fig:flowchart3body} shows the flowchart of the three-body treatment, including the 49 combinations of the merger classification. A triplet enters the chaotic three-body regime (state~5 of Table~\ref{tab:mergerfinderstates}) when its outer MBH is captured by the inner binary (Section~\ref{sec:chaotic}). When that happens RAMCOAL classifies the subgrid system as a chaotic three-body system, draws a triplet interaction outcome from the \Bonetti table, and updates the MBH configuration.

Figure~\ref{fig:flowchart3body} summarizes the procedure in five numbered steps. These are candidate identification and acceptance, physical-state classification, the unified reduction path, triggering of the chaotic three-body regime, and the weighted \Bonetti outcome draw. The two side legends list the seven physical states and the seven interaction outcomes. RAMCOAL first identifies a candidate triplet once three MBHs are within the same resolution sphere and are carried by a single sink particle. It then evaluates whether the triplet enters the chaotic three-body regime, and it decides the final outcome based on the triplet properties (Section~\ref{section:bonetti}). %

\subsubsection{Capture of the outer MBH}
\label{sec:chaotic}
The chaotic phase starts when the outer MBH is captured by the inner binary. While the inner pair is a bound binary and the outer MBH is still in stage~1 (state~4 of Table~\ref{tab:mergerfinderstates}), RAMCOAL tests the outer orbit with the analogues of the two criteria of equation~(\ref{equ:boundcrit}), applied to the whole triplet,
\begin{equation}
M_{\rm enc,out} \lesssim 2(M_{\rm pri}+M_{\rm sec}+M_{\rm out}),
\qquad {\rm or}\qquad
|E_{\rm K, out}| \lesssim |E_{\rm P, out}|.
\label{equ:outerbound}
\end{equation}
If either criterion is satisfied and the hardening rate that the outer orbit would have in stage~2 exceeds its dynamical-friction decay rate, the outer orbit is promoted to stage~2. The subgrid system then becomes an active triplet (state~5), and the chaotic phase is switched on.

A later galaxy merger can deliver a third MBH before an existing binary coalesces, with dynamical friction bringing the intruder into the nucleus. To get a feel for the physical scale of this trigger, consider a representative triplet with $(M_{\rm pri},M_{\rm sec},M_{\rm out})=(3,1,1)\times10^6\,{\rm M_\odot}$ and $a_{\rm in}=0.01$\,pc in a nucleus with a singular isothermal density profile and a velocity dispersion $\sigma=100\,{\rm km\,s^{-1}}$, the enclosed-mass capture criterion is satisfied within $r_{\rm inf}\simeq2$\,pc. Once the additional hardening condition for the outer MBH is met, RAMCOAL activates the three-body treatment. Subsequent outer-orbit decay can bring the intruder into close, chaotic encounters with the inner binary.

A typical pathway to a chaotic triplet starts with a galaxy merger that delivers an MBH binary. While the inner binary evolves, a later merger brings in a third MBH before the binary coalesces, as expected for massive galaxies at high redshift \citep{Kulkarni2012,Hoffman2023ASTRID}. The intruder sinks to the nucleus by dynamical friction until it is captured by the binary, and the triplet enters the chaotic phase.

\subsection{Bonetti-based triplet outcome}
\label{section:bonetti}

The chaotic three-body outcome is not computed by direct $N$-body integration during the simulation. Instead, RAMCOAL uses a discrete outcome table derived from the scattering survey of \citet{Bonetti2018}. The Bonetti experiments are grouped by primary MBH mass, inner-binary mass ratio, and outer mass ratio, and the relative frequencies of the merger and ejection channels are converted into a tabulated probability model.

We adopt six primary-mass bins,
\begin{equation}
\log_{10}\left(\frac{M_{\rm pri}}{M_\odot}\right)=5,6,7,8,9,10,
\end{equation}
and four values for each ratio,
\begin{equation}
q_{\rm in}, q_{\rm out}\in \{0.0316, 0.1, 0.3162, 1.0\}.
\end{equation}

For a given triplet, these three parameters are computed from the masses of the bound inner binary and of the outer MBH. $M_{\rm pri}$ is the more massive of the two bound members and $q_{\rm in}=M_{\rm sec}/M_{\rm pri}\le1$ is the binary mass ratio. The outer mass ratio is $q_{\rm out}=M_{\rm out}/(M_{\rm pri}+M_{\rm sec})$. The \Bonetti survey is constructed for triplets in which the inner binary is more massive than the incoming third MBH, $M_{\rm out}<M_{\rm pri}+M_{\rm sec}$ (so that $q_{\rm out}\le1$). The present model is therefore valid within this assumption, which is expected to be the most common configuration. It is, however, possible that the incoming MBH is more massive than the inner binary. If the third MBH arrives while the first pair is still in stage~1, the inner binary is the first pair that becomes bound. If it arrives when the first pair is already a bound binary, this binary is kept as the inner binary and the new MBH becomes the outer MBH, whatever its mass. In both cases the primary is the more massive member of the inner binary, so that $q_{\rm in}\le1$. The outer mass ratio exceeds unity only when the outer MBH is more massive than the inner binary. In that case the table is evaluated at the reciprocal, $q_{\rm out}\rightarrow(M_{\rm pri}+M_{\rm sec})/M_{\rm out}<1$. The channel labels (1)--(7) below still refer to the members as defined above. This fallback is an extrapolation of the \Bonetti results and not a calibrated prescription, because for an intruder heavier than the binary, exchanges are expected to dominate \citep{Sigurdsson1993}. In these exchanges, the intruder captures one of the inner members and the lightest MBH is ejected.

For each $(M_{\rm pri}, q_{\rm in}, q_{\rm out})$ cell, the procedure counts the number of realizations that end in each merger or ejection channel and stores the corresponding percentages in the triplet outcome table. The seven implemented outcomes are listed below.
\begin{enumerate}
    \item[(1)] $M_{\rm pri}+M_{\rm sec}$ merge,
    \item[(2)] $M_{\rm out}$ is ejected,
    \item[(3)] $M_{\rm pri}+M_{\rm out}$ merge,
    \item[(4)] $M_{\rm sec}$ is ejected,
    \item[(5)] $M_{\rm sec}+M_{\rm out}$ merge,
    \item[(6)] $M_{\rm pri}$ is ejected,
    \item[(7)] all three MBHs stall.
\end{enumerate}

This post-processing also computes the mean interaction completion time for the non-stalled channels. In practice, the \Bonetti survey is reduced to a direct census of outcomes in each mass-ratio bin. The explicit merger and ejection channels are counted, the residual fraction is assigned to the stalled channel, and both the branching ratios and characteristic completion times are retained for use in RAMCOAL. The resulting triplet-interaction outcome probabilities are shown as two-dimensional heatmaps over the primary mass and the two mass ratios in Appendix~\ref{app:heatmaps}.

The outcome is then chosen by a weighted random draw over the seven outcomes, using the tabulated \Bonetti probabilities for that cell. This stochastic selection is verified in Section~\ref{sec:selectortest}. Note that when the triplet state is activated and an outcome is assigned, a chaotic interaction timer, $t_{\rm chaos}$, is assigned to the triplet. This timer measures the interval between entry into the chaotic regime and execution of the outcome, which is either the coalescence of one pair or the ejection of one MBH. 

We adopt a fixed value, $t_{\rm chaos}=100$\,yr, in all three test cases (A, B and C, although only test~C reaches the chaotic phase). This is an ad hoc prescription that sets when the sampled outcome is applied, rather than a physically calibrated estimate of the interaction duration.

For an ejection outcome, the ejection is executed at the end of this timer, without an additional delay. The ejected MBH is re-inserted as a separate sink particle. The timer is not an estimate of the time required for the ejected MBH to leave the nucleus. The ejection velocity is not drawn from the scattering tables. Instead, it is set to the local escape speed, computed from the total mass enclosed within the resolution sphere (including the mass of MBHs). In this way, the ejected MBH is placed just outside the resolution sphere on an escaping trajectory. The ejected MBH is returned to the resolved sink population and the surviving pair is retained as the active binary or final remnant.

A chaotic triplet can undergo an exchange of partners, depending on the outcome drawn (outcomes 3--6). In such an exchange, the new inner binary retains the semi-major axis of the previous inner binary rather than adopting a value from the scattering tables. Its orbital frequency is recomputed for the new mass pair. The orbital parameters of the new outer MBH are then set by conserving the total energy and angular momentum of the triplet.

The exchange channels also redistribute angular momentum. The total angular momentum of the inner and outer orbits is conserved. The new inner angular momentum is scaled from the previous inner binary according to the new pair mass, and the residual is assigned to the outer orbit. The component positions and velocities are then reassigned so that the active inner binary is represented consistently. For the stalled channel, the triplet remains active and continues its standard stage-2 evolution.

\subsection{MBHB accretion in a circumbinary disc and feedback regulation}
\label{sec:accretion}

Before turning to the bound-binary accretion, we describe how the outer MBH accretes in RAMCOAL. In stage~1, the three MBHs are on orbits dominated by dynamical friction. Each of them, including the outer MBH, accretes individually following the BHL accretion model of paper~I. In stage~2, the bound inner pair accretes through the circumbinary disc and mini-disc channels described below. If the subgrid system is in state~4, the outer MBH is not bound to the binary and lies outside the inner binary's circumbinary region. It then accretes separately in stage~1 mode. Once the outer MBH becomes bound to the inner binary, the chaotic three-body phase begins, and accretion onto all three MBHs is suspended for its short duration. An MBH that is ejected from the system, whether by a three-body interaction or by a gravitational-wave recoil, no longer accretes, given its high relative velocity.

The stage-2 accretion model is also revised to account for MBHB accretion in the presence of a CBD. Hydrodynamic studies indicate that circumbinary inflow does not feed the two BHs as a smooth, axisymmetric Bondi-like flow. The cavity edge can host an overdense lump, gas may cross the gap in narrow streams, and only part of the inflow is captured by the two mini-discs \citep{Farris2014,Bowen2018,Duffell2020}. RAMCOAL represents this unresolved structure by splitting the inflow into two channels. One component of the inflow is delivered to the inner binary, and the residual is stored in the subgrid gap/CBD reservoir,
\begin{equation}
\label{equ:acc_inner}
\dot M_{12} = \dot M_{\rm inner}
\left[p_0 + p_1 \log_{10} q + p_2 \left(\log_{10} q\right)^2\right],
\end{equation}
where $q=M_2/M_1\le 1$, and the coefficients adopted in the present model are
\begin{equation}
p_0=0.8054,\qquad p_1=0.9840,\qquad p_2=0.3818.
\end{equation}

\begin{figure}
\centering
\includegraphics[width=\columnwidth]{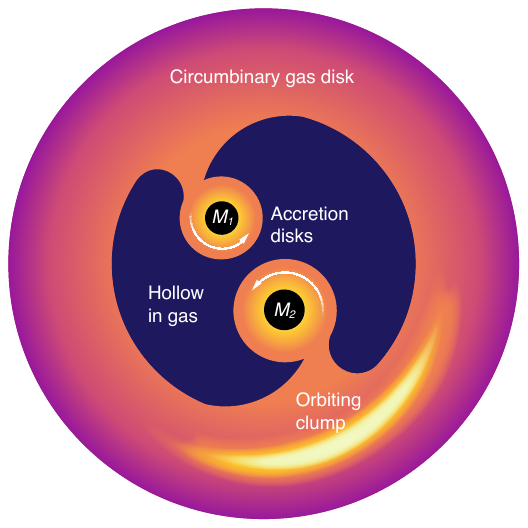}
\caption{Schematic representation of the stage-2 circumbinary-disc (CBD) accretion partition. The primary ($M_1$) and secondary ($M_2$) are each surrounded by a mini-disc and are embedded in a circumbinary gas disc with a central cavity, while an overdense lump orbits near the cavity edge. In the subgrid model, the subgrid inflow $\dot M_{\rm inner}$ is divided into a cavity-edge / CBD component, $\dot M_{\rm gap}$, and a mini-disc-fed component, $\dot M_{12}$. The mini-disc-fed component is then partitioned between the primary and secondary according to the preferential-accretion prescription adopted in the model. The figure illustrates this subgrid interpretation rather than a resolved hydrodynamical snapshot.}
\label{fig:cbdsplit}
\end{figure}

The residual inflow, which does not cross the cavity, is then
\begin{equation}
\dot M_{\rm gap} = \dot M_{\rm inner}-\dot M_{12}.
\end{equation}
Here $\dot M_{12}$ should be interpreted as the portion of the CBD supply that ultimately reaches the two mini-discs\citep{DOrazio2013, Gerosa2015}, while $\dot M_{\rm gap}$ represents the mass that remains in the subgrid cavity-edge / lump reservoir. The mass reaching the two mini-discs is further split using the preferential-accretion fit calibrated to the binary-accretion experiments summarized by \citet{Duffell2020},
\begin{equation}
\label{equ:pref}
\dot M_1 =
\dot M_{12}\,
\frac{0.1+0.9q}{1.1+0.9q},
\qquad
\dot M_2 = \dot M_{12}-\dot M_1.
\end{equation}
This parameterization captures the reported tendency of the secondary in unequal-mass binaries to accrete more efficiently than the primary while still allowing a finite residual supply to remain in the CBD/gap reservoir. The two-channel subgrid partition is illustrated schematically in Fig.~\ref{fig:cbdsplit}.

In stage~2, RAMCOAL applies a feedback-regulated accretion rule. For each MBH, the Eddington ratio is computed from
\begin{equation}
f_{{\rm Edd},i} = \frac{\dot M_i}{\dot M_{{\rm Edd},i}},
\qquad
\dot M_{{\rm Edd},i} =
\frac{4\pi G M_i m_{\rm p}}{\varepsilon_{{\rm r},i}\sigma_{\rm T} c},
\end{equation}
with $\varepsilon_{{\rm r},i}$ derived from the current spin-dependent innermost stable circular orbit (ISCO). If both MBHs satisfy $f_{{\rm Edd},i}<\eta_{\rm s2}$, both mini-discs are allowed to accrete. Here $\eta_{\rm s2}$ is a tunable parameter, which we set to $0.01$ in this work. The two mini-discs split the accretion rate that passes through the gap (see equations~(\ref{equ:acc_inner}) and (\ref{equ:pref})). If one MBH exceeds the threshold, its mini-disc accretion is suppressed for that substep and the corresponding supply is redirected into the CBD reservoir. If both exceed the threshold, the full stage-2 inflow is placed into the CBD and neither mini-disc accretes during that substep. The CBD reservoir evolves as
\begin{equation}
M_{\rm CBD}(t+\Delta t)=M_{\rm CBD}(t)+\dot M_{\rm CBD}\Delta t.
\end{equation}
The CBD reservoir therefore accumulates the gas that piles up when AGN feedback suppresses the direct feeding of the mini-discs.

In gas-rich binary models, accretion partition and spin evolution are often sensitive to disc geometry, mini-disc supply, and the relative orientation between the spin and orbital angular momenta \citep{Dotti2013,Perego2009,Gerosa2015,Farris2014}. 

Spin evolution is treated differently in the two RAMCOAL stages. The stage-1 treatment reproduces the standard RAMSES MBH-spin prescription and is summarized below only for completeness. The only RAMCOAL-specific modification to the spin model enters in stage~2, through a binary-forced warp radius that has no counterpart in the single-MBH RAMSES treatment. During stage~1, the spin evolution follows the RAMSES MBH-spin model summarized by \citet{Beckmann2025} in their sections 2.1--2.3 and based on the earlier models of \citet{Dubois2014b,Dubois2021}. In this formulation, each coarse accretion event is treated as the formation of a subgrid accretion disc whose angular-momentum direction is set by that of the gas measured on resolved scales,
\begin{equation}
\hat{\boldsymbol{j}}_{\rm d} = \hat{\boldsymbol{j}}_{\rm gas}.
\end{equation}
 Only the direction of the resolved gas angular momentum within the accretion kernel is used to orient the subgrid disc. The angular-momentum transfer and spin-evolution prescriptions are described in \citet{Dubois2014b, Beckmann2025}.

For each accretion substep, we derive the radiative efficiency from the current Kerr ISCO radius, compute the Eddington ratio $\chi$, and evaluate the~\cite{BardeenPetterson1975} radius using the spin-alignment scaling discussed by \citet{Barausse2012} and \citet{Gerosa2015}. We follow the standard Shakura--Sunyaev thin-disc solution \citep{ShakuraSunyaev1973} for the warped disc \citep{BardeenPetterson1975}.

This radius provides the disc angular momentum accreted in the given substep. The disc mass is capped at the self-gravitating disc mass~\citep{Dotti2013}, whenever the self-gravitating radius is smaller than the warp radius.

The mass within the~\cite{BardeenPetterson1975} radius is then the disc mass that takes part in the spin update. Counter-alignment of the MBH spin occurs when~\citep{King2005}
\begin{equation}
\cos\theta < -\frac{J_d}{2J_{\rm BH}},
\end{equation}
where $J_{\rm BH}=a_\ast M_{\rm BH}^{3/2}R_{\rm BH}$ ($R_{\rm BH}=2GM_{\rm BH}/c^2$ is the Schwarzschild radius and $a_\ast$ is the spin magnitude) is the MBH angular momentum and $\theta$ is the angle between the MBH spin and the accreted gas angular momentum ($J_d$).

Once the alignment state is known, the spin direction is rotated towards the total angular momentum of the MBH-plus-disc system,
\begin{equation}
\boldsymbol{J}_{\rm tot} = \boldsymbol{J}_{\rm BH} + \boldsymbol{J}_{\rm d},
\end{equation}
following the standard RAMSES spin prescription \citep{Dubois2014b,Beckmann2025}. The spin magnitude is then updated according to the local accretion mode. For efficiently accreting MBHs, the model adopts the thin-disc spin-up prescription, in which prograde accretion generally drives the spin towards the Thorne limit of $0.998$ \citep{Bardeen1970,Thorne1974}. For low-Eddington accretion, RAMCOAL switches to the thick-disc regime, in which jet extraction can spin the MBH down through the Blandford--Znajek mechanism. In that regime the spin evolution is taken from \citet{McKinney2012} for magnetically arrested discs, as also described by \citet{Dubois2021} and \citet{Beckmann2025}. As a result, BHs in stage~1 can be spun either up or down depending on the instantaneous Eddington ratio, the disc angular momentum, and the alignment history.

For stage~2, the spin evolution model keeps the same overall structure but replaces the single-MBH warp radius. The new warp radius is the smallest of the individual-MBH $r_{\rm BP, i}, i=1,2$, the self-gravitating radius, and the companion-MBH forced warp scale $r_{\rm warp,bin, i}$ \citep{MillerKrolik2013},
\begin{equation}
r_{\rm warp, i} = \min(r_{\rm BP,i},r_{\rm sg, i}, r_{\rm warp,bin, i}),
\end{equation}
with
\begin{equation}
r_{{\rm warp,bin, i}}
\propto
\left[\frac{8a_\ast M_{\rm i}}{3\beta(M_{\rm bin}-M_{\rm i})}\right]^{2/9}
a_{\rm bin}^{2/3}r_{\rm sg, i}^{1/3}, i=1,2
\end{equation}
where $a_{\rm bin}$ is the binary separation and $\beta$ is the cosine of the angle between the gas angular momentum and the binary orbital angular momentum. 

In a bound binary the two MBHs are evolved independently. The inflow is first divided between them by the circumbinary preferential-accretion split described above, so that each MBH has its own accretion rate. Its spin then evolves from the angular momentum of the gas it accretes, and the disc angular momentum it receives therefore scales with that accreted mass. The alignment test is applied to each member separately, with $a_{{\rm acc},i}=J_{{\rm d},i}/(2J_{{\rm BH},i})$ and counter-alignment when $\cos\theta_i<-a_{{\rm acc},i}$.

This companion-MBH forced warp radius is the main RAMCOAL-specific addition to the stage~2 spin model. The bound companion can reduce the effective warp radius and so change both the alignment timescale and the mass participating in each alignment event. The rest of the spin update is identical to the single-MBH RAMSES prescription.

In the current RAMCOAL model, the outer MBH accretes while it is still in stage~1. Once its outer orbit is promoted to stage~2, its accretion is suppressed rather than computed with a circumbinary-disc prescription adapted to triplets, because no hydrodynamical simulation of accreting MBH triplets has been performed yet. We leave this for future work. 

\subsection{Coalescence remnant spin and recoil kick}
\label{sec:recoil}

When two MBHs coalesce, RAMCOAL updates the remnant mass, spin, and recoil velocity. The calculation uses the binary mass ratio $q=M_{\rm sec}/M_{\rm pri}\leq 1$, the symmetric mass ratio
\begin{equation}
\eta = \frac{q}{(1+q)^2},
\end{equation}
the progenitor spin magnitudes $a_{\ast,1}$ and $a_{\ast,2}$, and the projections of the two spin vectors onto each other and onto the orbital-angular-momentum direction. The remnant-spin magnitude is then evaluated with the phenomenological fit of \citet{Rezzolla2008}, written as
\begin{equation}
\begin{split}
\ell ={}&
\frac{s_4}{(1+q^2)^2}
\left(a_{\ast,1}^2+a_{\ast,2}^2q^4+2a_{\ast,1}a_{\ast,2}q^2\cos\alpha\right) \\
&+
\frac{s_5\eta+t_0+2}{1+q^2}
\left(a_{\ast,1}\cos\beta_1+a_{\ast,2}q^2\cos\beta_2\right) \\
&+
2\sqrt{3}+t_2\eta+t_3\eta^2 ,
\end{split}
\end{equation}
where $s_4=-0.129$, $s_5=-0.384$, $t_0=-2.686$, $t_2=-3.454$, and $t_3=2.353$ \citep{Rezzolla2008}. The spin of the MBH remnant is
\begin{equation}
\boldsymbol{a}_{\rm rem} =
\frac{\boldsymbol{a}_{\ast,1} + q^2\boldsymbol{a}_{\ast,2} + q\,\ell\,\hat{\boldsymbol{L}}}
{(1+q)^2}.
\end{equation}
Here $\alpha$ is the angle between the two MBH spins, while $\beta_1$ ($\beta_2$) is the angle between the spin of MBH1 (MBH2) and the orbital-angular-momentum direction $\hat{\boldsymbol{L}}$.

The gravitational-wave recoil kick is implemented following the cosmological MBH-recoil treatment described by \citet{DongPaez2023}, using the decomposition and coefficients from~\cite{Lousto2013}. In the orthonormal basis $(\hat{\boldsymbol{e}}_1,\hat{\boldsymbol{e}}_2,\hat{\boldsymbol{L}})$ associated with the binary separation and orbital angular momentum, the kick is written as
\begin{equation}
\boldsymbol{v}_{\rm kick}
=
v_{\rm m} \hat{\boldsymbol{e}}_1
+ v_\perp\left(\cos\xi\,\hat{\boldsymbol{e}}_1+\sin\xi\,\hat{\boldsymbol{e}}_2\right)
+ v_\parallel \hat{\boldsymbol{L}},
\end{equation}
with
\begin{align}
&v_{\rm m} = A_{\rm m} \eta^2 \frac{1-q}{1+q}(1+B_{\rm m}\eta),\\
&v_\perp = H \eta^2 \frac{a_{\ast,1,\parallel}-q a_{\ast,2,\parallel}}{1+q},\\
&v_\parallel =
\frac{16\eta^2}{1+q}
\left(V_{11}+V_{\rm A} S_z+V_{\rm B} S_z^2+V_{\rm C} S_z^3\right)
\left|\boldsymbol{a}_{\ast,1,\perp}-q\boldsymbol{a}_{\ast,2,\perp}\right|
\cos\Delta\phi.
\end{align}
Here $a_{\ast,i,\parallel}$ and $\boldsymbol{a}_{\ast,i,\perp}$ are the components of the progenitor spins parallel and perpendicular to $\hat{\boldsymbol{L}}$, and $\Delta\phi$ is drawn uniformly in $[0,2\pi)$. The remaining quantities are fixed coefficients from the numerical-relativity fits of \citet{Lousto2013}, which are $A_{\rm m}=1.2\times10^{4}\,{\rm km\,s^{-1}}$, $B_{\rm m}=-0.93$, $H=6.9\times10^{3}\,{\rm km\,s^{-1}}$, $V_{11}=3677.76\,{\rm km\,s^{-1}}$, $V_{\rm A}=2481.21\,{\rm km\,s^{-1}}$, $V_{\rm B}=1792.45\,{\rm km\,s^{-1}}$, $V_{\rm C}=1506.52\,{\rm km\,s^{-1}}$, and $\xi=145^\circ$. Following the aligned-spin combination adopted in the model, we identify
\begin{equation}
S_z = \frac{2(a_{\ast,1,\parallel}+q^2a_{\ast,2,\parallel})}{(1+q)^2}.
\end{equation}
The kick is added to the remnant velocity. The random phase $\Delta\phi$ only enters the out-of-plane term $v_\parallel$, through $\cos\Delta\phi$, so its influence is set by the in-plane spin components. For non-spinning or aligned-spin progenitors, $\boldsymbol{a}_{\ast,i,\perp}=0$, the kick does not depend on $\Delta\phi$ at all. For large misaligned spins, in contrast, $v_\parallel$ dominates and $\Delta\phi$ controls the kick magnitude. For $q=1$ and in-plane, antiparallel spins with $a_{\ast,1}=a_{\ast,2}=0.9$, $v_{\rm m}=v_\perp=0$ and $|\boldsymbol{v}_{\rm kick}|\simeq3300\,|\cos\Delta\phi|\,{\rm km\,s^{-1}}$, so that a uniform $\Delta\phi$ gives a mean kick of $\simeq2100\,{\rm km\,s^{-1}}$ and exceeds $1000\,{\rm km\,s^{-1}}$ with a probability of 80\%. In such cases the retention or ejection of an individual remnant is effectively stochastic, and only the distribution over many mergers is meaningful, as in the cosmological recoil treatment of \citet{DongPaez2023}. For the small spins of our test cases ($a_\ast\lesssim0.1$, and $a_\ast\sim10^{-5}$ in test case~C), the $\Delta\phi$-dependent term is at most $\sim100\,{\rm km\,s^{-1}}$. The kick is then set mainly by the mass asymmetry, so the outcomes shown here are insensitive to $\Delta\phi$.

The velocity kick is then compared with a local escape velocity, estimated from the mass of the MBHs, gas, stars, and dark matter enclosed within the resolution sphere of the merger remnant. If the kick exceeds this escape velocity, the remnant is released back into the simulation as an additional sink particle. The new sink particle is placed at a spatial offset from the parent system, and the recoil velocity is added to it. We apply the kick in this way, and not through a timer like the chaotic-ejection timer, because the two processes differ physically. A GW recoil is imparted during the final orbits and plunge, that is, within minutes to days for MBHs of $10^5$--$10^8\,{\rm M_\odot}$. This is instantaneous compared with any simulation time step. A three-body ejection, in contrast, results from a resonant interaction of finite duration (Section~\ref{section:bonetti}). The spatial offset is purely numerical. The kicked remnant is re-inserted as a resolved sink, displaced from the parent sink by twice the numerical merger radius along each Cartesian axis (that is, by $\simeq3.5$ merger radii), so that it is not immediately re-captured by the numerical merger criterion in the same step. A remnant escaping at $v_{\rm kick}\gtrsim v_{\rm esc}$ moves at a few hundred ${\rm km\,s^{-1}}$ or more. At the resolution of our tests, it would cover this distance within a few Myr. The offset therefore amounts to a small shift in time of its later, resolved trajectory.

\subsection{Treatment of quadruplets and higher-order encounters}
\label{sec:highorder}

The present version of RAMCOAL is designed primarily as a subgrid model for triplets, but it also includes a simplified treatment for the case in which four or more MBHs enter a subgrid system. The central assumption is deliberately restrictive. The model does not attempt to follow a generic long-lived resonant four-body interaction. Instead, the updated merger classification reduces the encounter to an effective triplet hierarchy that can be handled by the three-body prescription.

As illustrated in the flowchart of Fig.~\ref{fig:flowchart3body}, this reduction is handled by the same procedure described above for the 49 combinations. After the two incoming systems are placed in a common centre-of-mass frame, any already-bound inner pair is allowed to merge first. MBHs with a pending three-body ejection are released, and, because the three-body prescription handles at most three MBHs, any excess is removed by merging the most massive pairs first until no more than three active MBHs remain. Three survivors, if present, define the next triplet state, two survivors are returned as a stage~1 pair, and a single survivor is returned to the standard single-sink population. The last case can happen when the merging systems have a pending ejection as well as an escaping recoiled MBH.

The same approach is extended to higher-order subgrid encounters that arise when multiple close systems overlap within one simulation step. RAMCOAL aims to accommodate rare $N\geq 4$ events without losing MBH identities, but it does so through hierarchical reduction rather than through a dedicated few-body integrator. This approximation should be kept in mind when interpreting the highest-order multiple systems in the model. The model provides a reduction prescription for cosmological simulations, not a calibrated theory of subgrid four-body scattering.

Treating encounters with $N>3$ MBHs is nevertheless necessary for cosmological applications, both for numerical robustness and because such encounters are not rare. Every combination of two subgrid systems must be mapped onto a valid state. Otherwise, MBHs would be lost or duplicated, or the simulation would have to stop. The classification of the 49 combinations guarantees a valid state for any input. In addition, at the kiloparsec-scale resolution of cosmological volumes, the resolution sphere of a massive high-redshift nucleus can contain more than three MBHs. In the ASTRID simulation at $z\sim2$, MBH quadruplets are 14--30\% as numerous as triplets (24 versus 174 systems with maximum separations below 30\,kpc, and 174 versus 588 below 200\,kpc, \citealp{Hoffman2023ASTRID}). Multiple-MBH nuclei are also expected to be common in massive haloes at $2\lesssim z\lesssim6$ \citep{Kulkarni2012}. The frequency of such encounters in cosmological volumes, and the fraction of MBHs affected by the forced mergers of step~(iii) of Section~\ref{sec:mergerclass}, will be quantified with the COSCOAL simulations (Section~\ref{sec:cosmo}). This fraction is expected to decrease at higher resolution, and the forced mergers are logged so that they can be treated separately in merger catalogues.

\section{Triplet Test Cases in an Isolated Galaxy}
\label{sec:testcases}

We test the triplet extension in an isolated galaxy set up to match the controlled RAMCOAL binary tests of paper~I. These tests aim to isolate the behaviour of the three-body model in a self-consistently evolving galaxy. They do not aim to sample the full cosmological diversity of MBH triplets, which we will cover in future work. The host galaxy consists of dark matter, stars, gas, and MBHs. The dark-matter halo follows a spherical~\cite{NFW1997} profile with virial mass $M_{\rm vir}=1.1\times10^{11}\,{\rm M_\odot}$, concentration $c=10$, and virial radius $R_{\rm vir}=95\,{\rm kpc}$ at redshift zero. The baryonic component consists of a \cite{Hernquist1990} stellar bulge with mass $3.3\times10^9\,{\rm M_\odot}$ and scale radius $293\,{\rm pc}$, plus a gas and stellar disc with total mass $7.78\times10^9\,{\rm M_\odot}$. The disc has an exponential radial profile with a scale length of $2.93\,{\rm kpc}$ and a vertical scale height of $293\,{\rm pc}$, and it is truncated at five scale lengths. The gas mass is $1.1\times10^9\,{\rm M_\odot}$, corresponding to 14.3\% of the disc mass, while the stellar disc mass is $6.67\times10^9\,{\rm M_\odot}$.

We place the galaxy at the centre of a $100\,{\rm kpc}$ box with a $128^3$ coarse grid. We refine the grid with a pseudo-Lagrangian mass criterion and a Jeans-length criterion, down to a finest cell size of $\Delta x\simeq98\,{\rm pc}$. The corresponding resolution sphere is $R_{\rm sp}=4\Delta x\simeq0.39\,{\rm kpc}$, and the gas mass resolution is $\sim10^{4}\,{\rm M_\odot}$. The galaxy is first relaxed for $100\,{\rm Myr}$ without MBHs, after which the MBHs are inserted. Table~\ref{tab:isolatedgalaxyic} summarizes the host-galaxy initial setup used in this work.

\begin{table}
\centering
\caption{Shared isolated-galaxy initial condition for the triplet tests, following the setup of paper~I.}
\label{tab:isolatedgalaxyic}
\begin{tabular}{ll}
\toprule
Quantity & Value \\
\midrule
Halo profile & NFW \\
$M_{\rm vir}$ & $1.1\times10^{11}\,{\rm M_\odot}$ \\
$R_{\rm vir}$ & $95\,{\rm kpc}$ \\
Halo concentration & 10 \\
Bulge mass & $3.3\times10^9\,{\rm M_\odot}$ \\
Bulge scale radius & $293\,{\rm pc}$ \\
Disc mass & $7.78\times10^9\,{\rm M_\odot}$ \\
Gas mass & $1.1\times10^9\,{\rm M_\odot}$ \\
Stellar disc mass & $6.67\times10^9\,{\rm M_\odot}$ \\
Disc scale length & $2.93\,{\rm kpc}$ \\
Disc scale height & $293\,{\rm pc}$ \\
Box size & $100\,{\rm kpc}$ \\
Coarse grid & $128^3$ \\
Pre-MBH relaxation & $100\,{\rm Myr}$ \\
\bottomrule
\end{tabular}
\end{table}

\begin{table*}
\centering
\caption{Initial configuration of the three isolated-galaxy triplet tests (cases~A, B, and~C). In cases~A and~B the three MBHs have masses $M_1=3.33\times 10^6\,\rm M_\odot$, $M_2=1.67\times 10^6\,\rm M_\odot$, and $M_3=10^6\,\rm M_\odot$, with the primary MBH1 initialized at the galaxy centre. For each test we list the initial separation $d$ of MBH2 and MBH3 from MBH1. We also list the radial ($v_r$) and tangential ($v_t$) components of their initial velocity relative to MBH1, and the corresponding initial orbital eccentricity $e$. A negative $v_r$ denotes motion towards MBH1, and $e$ is estimated from the orbital energy and angular momentum in the host-galaxy potential. MBH2 is launched on an almost tangential, near-circular orbit ($v_r\simeq0$, $e\simeq0.1$), whereas MBH3 carries a substantial inward radial component and is set on a highly eccentric, plunging orbit ($e\simeq0.8$). Case~C is a more compact configuration in which the three MBHs instead have equal masses, $M_1=M_2=M_3\simeq3.0\times10^6\,{\rm M_\odot}$. Their initial spins differ in orientation, with MBH1 aligned with the disc angular momentum, $\hat{a_\star}=(0,0,1)$, and MBH2 and MBH3 randomly misaligned, $\hat{a_\star}=(0.50,0.30,0.81)$ and $(-0.31,0.72,0.62)$ respectively.}

\label{tab:tripletic}
\begin{tabular}{llcccccccc}
\toprule
 &  & \multicolumn{4}{c}{MBH2 -- MBH1} & \multicolumn{4}{c}{MBH3 -- MBH1} \\
\cmidrule(lr){3-6}\cmidrule(lr){7-10}
Case & Configuration & $d$ (kpc) & $v_r$ (km\,s$^{-1}$) & $v_t$ (km\,s$^{-1}$) & $e$ & $d$ (kpc) & $v_r$ (km\,s$^{-1}$) & $v_t$ (km\,s$^{-1}$) & $e$ \\
\midrule
A & prograde in-plane ($i=0$) & 1.00 & $0$ & 100 & 0.08 & 1.82 & $-150$ & 100 & 0.8 \\
B & prograde inclined ($i=\pi/2$) & 1.00 & $0$ & 100 & 0.08 & 1.73 & $-140$ & 100 & 0.8 \\
C & prograde in-plane compact triplet ($i=0$) & 0.22 & 0 & 7 & 0.05 & 0.24 & 1 & 16 & 0.02 \\
\bottomrule
\end{tabular}
\end{table*}

\subsection{Validation of the triplet interaction routine}
\label{sec:selectortest}

We verify that the chaotic three-body routine itself reproduces the intended outcome statistics. For this test the triplet is initialized in the isolated galaxy with a single, fixed configuration. In this way, the only quantity under examination is the mapping from a hierarchical triplet onto the seven-channel \Bonetti outcome distribution. We fix a single representative configuration with the primary mass $M_1=10^{7}\,{\rm M_\odot}$, inner mass ratio $q_{\rm in}=0.80$, and outer mass ratio $q_{\rm out}=0.556$. The inner and outer orbits are coplanar ($\cos i=1$), with eccentricities $e_{\rm in}=0.35$ and $e_{\rm out}=0.55$ and semi-major axes $a_{\rm in}=5\,{\rm mpc}$ and $a_{\rm out}=25\,{\rm mpc}$. The MBHs are non-spinning ($a_{\rm _\ast,1}=a_{\rm _\ast,2}=a_{\rm _\ast,3}=0$). We repeat the \Bonetti outcome draw with $N_{\rm seed}=100$ independent random seeds, each used for one run, and all runs were completed.

\begin{figure}
\centering
\includegraphics[width=\columnwidth]{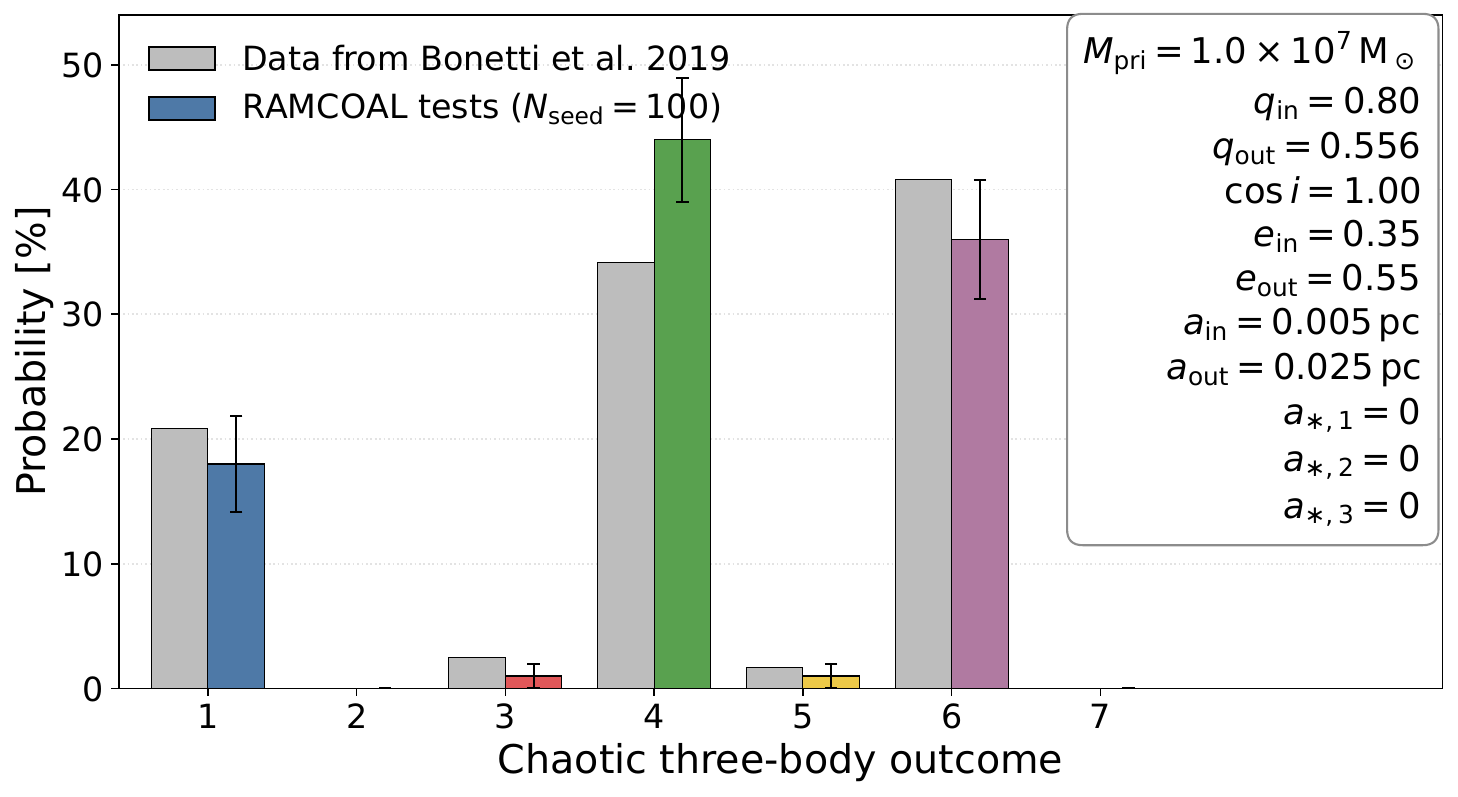}
\caption{Validation of the selection of chaotic three-body outcomes in RAMCOAL. The parameters of the fixed hierarchical triplet are listed in the panel ($M_{\rm pri}=10^{7}\,{\rm M_\odot}$, $q_{\rm in}=0.80$, $q_{\rm out}=0.556$, $\cos i=1$, $e_{\rm in}=0.35$, $e_{\rm out}=0.55$, $a_{\rm in}=0.005\,{\rm pc}$, $a_{\rm out}=0.025\,{\rm pc}$, $a_{\ast, i}=0$). The coloured bars, with Poisson errors, show the outcome frequencies returned by $N_{\rm seed}=100$ independent RAMCOAL draws for the seven chaotic three-body channels. These frequencies are compared with the tabulated probabilities of \citet{Bonetti2019} (grey bars). The channels are the seven implemented outcomes shown in Fig.~\ref{fig:flowchart3body}.}
\label{fig:bonettiselector}
\end{figure}

Figure~\ref{fig:bonettiselector} compares the outcome frequencies recovered from these draws (coloured bars, with Poisson error bars) against the tabulated probabilities of \citet{Bonetti2019} for the corresponding $(M_{\rm pri},q_{\rm in},q_{\rm out})$ cell (grey bars). For this configuration the drawn outcomes are dominated by two ejection channels. The ejection of MBH2 leaves the MBH1--MBH3 binary (outcome~4, $44/100$), and the ejection of MBH1 leaves the MBH2--MBH3 binary (outcome~6, $36/100$). These are followed by the coalescence of the original inner binary (outcome~1, $18/100$). The exchange merger MBH1--MBH3 (outcome~3) and the merger MBH2--MBH3 (outcome~5) each occur once. The ejection of the incoming MBH3 (outcome~2) and the stalled channel (outcome~7) do not occur, because for this $(M_{\rm pri},q_{\rm in},q_{\rm out})$ cell the \citet{Bonetti2019} table assigns both of these channels zero probability. The recovered frequencies agree with the \Bonetti probabilities within the Poisson scatter expected for $100$ realizations, which confirms that the routine draws the seven channels with the correct weights.

This test does not probe an influence of the environment on the outcome, and it is not meant to. For a given $(M_{\rm pri},q_{\rm in},q_{\rm out})$ cell, the outcome probabilities are independent of the host galaxy by construction. The host galaxy only affects the time at which a triplet reaches the chaotic regime and, thus,  the values of $(M_{\rm pri},q_{\rm in},q_{\rm out})$ at this particular time. 

\subsection{Triplet test case~A (in-plane initial orbital configuration)}

The first triplet experiment places three MBHs in the relaxed isolated galaxy with a prograde in-plane initial orbital configuration. The inner two MBHs are initialized as two separate sink particles, close enough to form the first subgrid pair. The third MBH is introduced on a wider orbit. The initial conditions of the MBHs are listed in Table~\ref{tab:tripletic}. 

\begin{figure*}
\centering
\includegraphics[width=495.99bp]{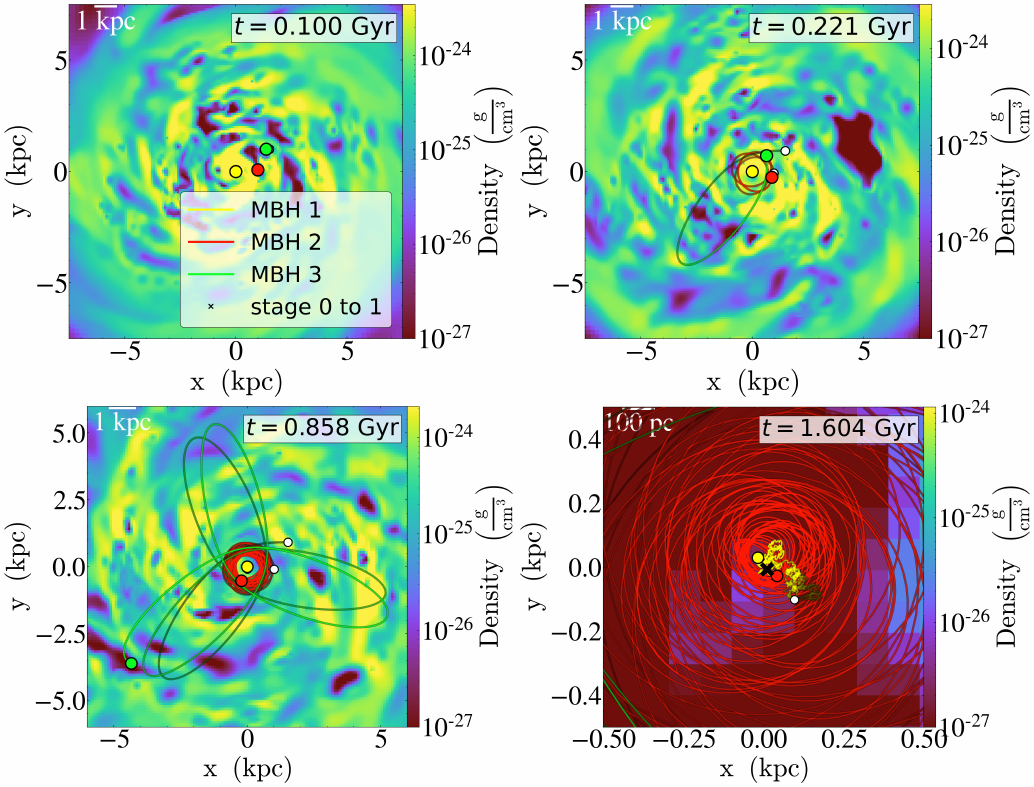}
\caption{Gas-density slices through the mid-plane of the galaxy disc with MBH trajectories for triplet test case~A (in-plane initial orbital configuration). The four panels each capture a qualitatively distinct phase of the evolution rather than a continuous time sequence, at $t=0.1$, $0.221$, $0.858$, and $1.604\,{\rm Gyr}$ (the last zoomed to the central $1\,{\rm kpc}$). The coloured curves are the projected sink-particle trajectories (MBH1 yellow, MBH2 red, MBH3 green). Each MBH is shown in the colour of its legend entry. Its trajectory runs from the time at which the MBHs are inserted ($t=0.1\,{\rm Gyr}$) to the time of the snapshot. It is faint at the start and becomes stronger towards the time of the snapshot. The older parts of a trajectory are partly transparent over the density map and have a thin dark outline, so they look darker than the marker. White circles mark the starting point of each trajectory, i.e. the position at which the MBH was inserted, and filled circles mark the position of each MBH at the time of the snapshot. In the first panel the trajectories consist of a single point, so only the filled circles are visible. The top left panel shows the initial configuration, with the three MBHs as separate resolved sink particles (stage~0). The top right and bottom left panels show the resolved dynamical-friction inspiral, as MBH1 and MBH2 spiral together while MBH3 moves inwards on a wider, more eccentric orbit (still stage~0). In the bottom right panel, once MBH1 and MBH2 fall within the resolution sphere, RAMCOAL combines them into a single sink (black cross). This sink internally carries the MBH1--2 pair through the subgrid evolution (stages~1 and~2). The subgrid evolution is not plotted here, because stage~1 is short and would be hard to see. The black cross marks the centre of mass of the pair once it enters the subgrid regime, while MBH3 stays a separate sink on its kiloparsec orbit.}

\label{fig:tripletcaseA_snapshots}
\end{figure*}

\begin{figure*}
\centering
\includegraphics[width=0.95\textwidth]{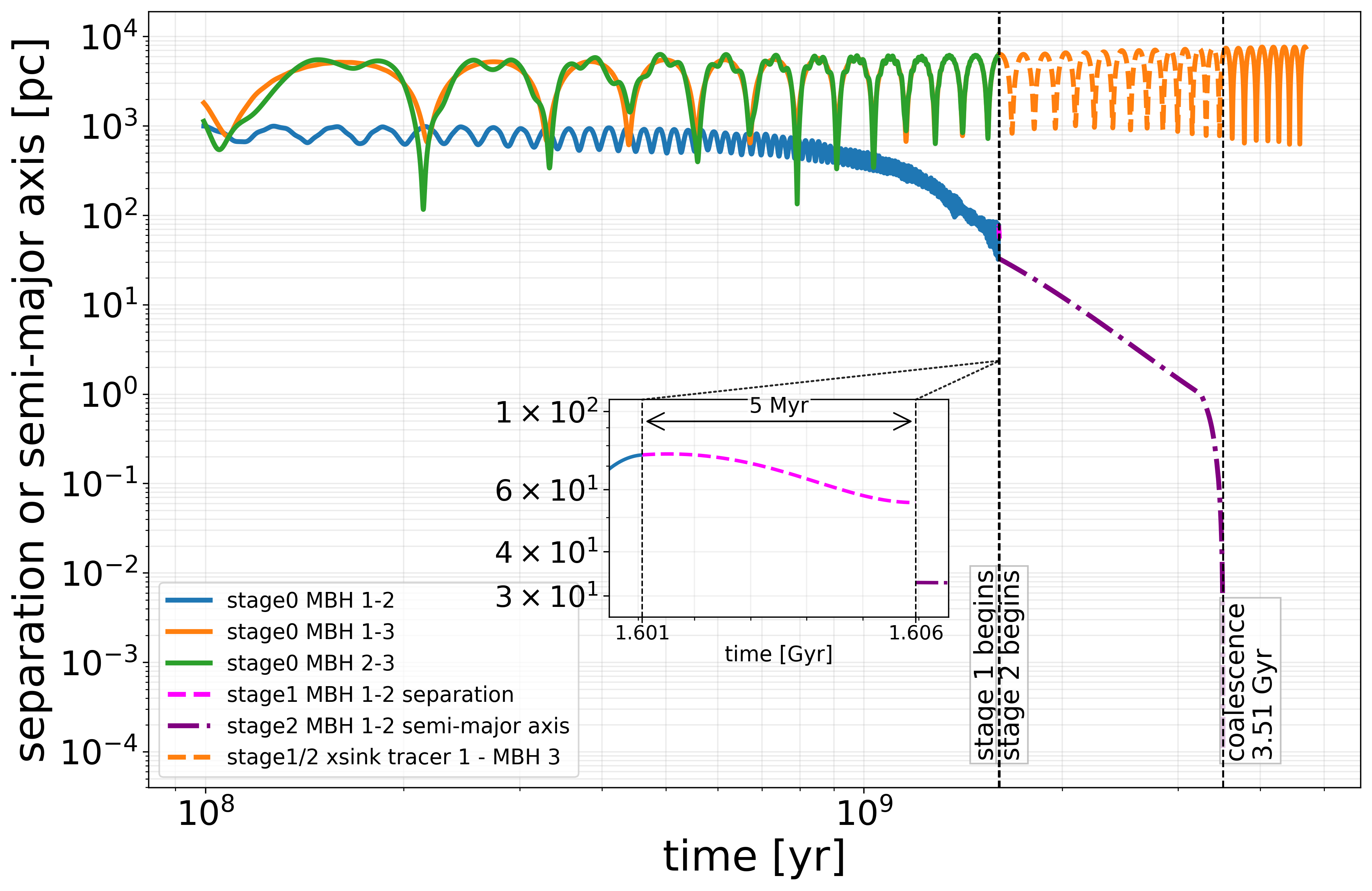}
\caption{Pairwise-separation evolution for triplet test case~A with the in-plane initial orbital configuration. The figure presents the resolved stage~0 pairwise separations of the three MBHs (MBH\,1--2, MBH\,1--3, and MBH\,2--3). It also shows the subsequent RAMCOAL stage~1 and stage~2 evolution of the selected inner pair and the labelled coalescence time. Because MBH1 and MBH2 remain within $\sim1\,{\rm kpc}$ of each other, the MBH\,2--3 curve follows the MBH\,1--3 curve except near the pericentre passages of MBH3. At these passages it shows which of the two inner MBHs MBH3 approaches. At six pericentre passages MBH3 comes within $\simeq120$--$400\,{\rm pc}$ of MBH2, that is, inside or at the edge of the resolution sphere (e.g. 117 and $135\,{\rm pc}$ at $t\simeq0.21$ and $0.79\,{\rm Gyr}$). In contrast, it never comes closer than $\simeq620\,{\rm pc}$ to MBH1. These passages occur at relative velocities of 110--$220\,{\rm km\,s^{-1}}$, more than ten times the mutual escape speed of MBH2 and MBH3. These unbound fly-bys do not lead to a subgrid pair, and the pair that enters RAMCOAL is instead the slowly decaying MBH1--MBH2 pair. The curve labelled ``xsink tracer~1 -- MBH3'' is the separation between MBH3 and the centre of mass of the inner MBH1--MBH2 pair. Once that pair is stored as a single subgrid sink, its centre-of-mass position is followed by the combined sink particle (the ``xsink tracer''). This curve therefore shows the distance between the third MBH and the hardening inner binary through stages~1 and~2.}
\label{fig:tripletcaseA_separation}
\end{figure*}

Figure~\ref{fig:tripletcaseA_snapshots} summarizes the evolution in four panels, each chosen to show a qualitatively distinct phase rather than a continuous time sequence. The first panel ($t=0.1\,{\rm Gyr}$) shows the initial configuration, with the three MBHs released near the galaxy centre and tracked as three separate resolved sink particles. The second and third panels ($0.221, 0.858\,{\rm Gyr}$) show the resolved dynamical-friction inspiral, with MBH1 and MBH2 spiralling together towards the nucleus while MBH3 moves inwards on a wider, more eccentric orbit. All three are still resolved sinks (stage~0). The fourth panel ($1.604\,{\rm Gyr}$) marks the qualitative change introduced by RAMCOAL. MBH1 and MBH2 have entered the subgrid regime and are now represented by a single combined sink indicated by the black cross at the nucleus, while MBH3 remains on its kiloparsec orbit. In test case~A, stage~1 is very short ($5$\,Myr, Fig.~\ref{fig:tripletcaseA_separation}), so we do not show the stage~1 orbits in the snapshots. Test case~B has a longer stage~1, and its subgrid MBH trajectories are shown in Fig.~\ref{fig:tripletcaseB_snapshots_subgrid}. 

Here we clarify what the plotted trajectories represent. While the three MBHs remain outside each other's resolution spheres, they are three independent RAMSES sink particles, and each curve is the trajectory of one resolved sink. When MBH1 and MBH2 approach within the resolution sphere, RAMCOAL does not merge them numerically. Instead, the pair is stored inside a single sink particle whose position is the binary centre of mass. From that point on, the tightly wound central curve is therefore the trajectory of this one combined sink. It is not the trajectory of two separately resolved MBHs, nor that of two distinct sink particles orbiting each other. The MBH1--2 relative orbit is thereafter evolved as a subgrid binary (Fig.~\ref{fig:tripletcaseA_separation}) and is not spatially resolved in the snapshots, while MBH3 remains a distinct sink on its wider orbit.

Over the sequence the MBH1--2 pair therefore passes through RAMCOAL stages~1 and~2. Stage~1 (subgrid dynamical friction, before the pair becomes bound) is a very short period of the evolution in this in-plane case. It falls between the top and bottom rows of the figure, so the case~A trajectories do not illustrate it. These trajectories step directly from three resolved sinks (stage~0) to the tightly wound stage~2 track. Case~B has a substantially longer stage~1, and its snapshot sequence (Fig.~\ref{fig:tripletcaseB_snapshots}) gives a clearer demonstration of the RAMCOAL tracks during stage~1.

The separation history in Fig.~\ref{fig:tripletcaseA_separation} connects the sink trajectories shown in the snapshots to the subgrid orbital evolution. During stage~0, the three pairwise separations oscillate as the MBHs orbit through the live galaxy potential. The inner pair enters the subgrid RAMCOAL treatment, with the stage~1 and stage~2 transitions marked by vertical dashed lines. In this case, stage~1 is short and lasts only about $5\,{\rm Myr}$, as shown in the inset. Stage~2, by contrast, lasts much longer, about $2\,{\rm Gyr}$ from its onset near $1.4\,{\rm Gyr}$ to the coalescence. The inner semi-major axis then decreases from roughly $30\,{\rm pc}$ to about $1\,{\rm pc}$ over more than a Gyr of gradual hardening before plunging to the ISCO radius $a_{\rm coal}\simeq1.4\times10^{-6}\,{\rm pc}$ at $3.51\,{\rm Gyr}$.
 
Throughout this evolution, the wider orbit of MBH3 stays at kiloparsec scales. In this configuration the outer MBH never becomes part of an active triplet, and the coalescence is that of the original MBH1--MBH2 pair, with the third MBH retained on a wide orbit. The third MBH remains dynamically relevant, at a separation of $\sim 1 $\,kpc from the MBH1--MBH2 remnant. In this test case, MBH3 will eventually merge with the remnant, but only on a much longer timescale ($ >$\,Gyr).

Note that in Fig.~\ref{fig:tripletcaseA_separation} the plotted quantity changes across the transition. In stages~0 and~1 the curve is the \emph{instantaneous} pair separation, whereas in stage~2 it is the binary \emph{semi-major axis} $a$ (see the legend). The apparent jump at the stage~1--stage~2 boundary is therefore largely a change of variable, since a bound eccentric binary is summarized by a single semi-major axis rather than by its oscillating separation.

The stage~1 evolution is relatively short in this scenario because the promotion to subgrid stage~1 depends on a binding criterion in addition to the separation criterion. RAMCOAL requires two conditions to be met at the same time. The pair must lie within the resolution sphere, and it must be gravitationally bound ($E_{\rm kin}\le E_{\rm pot}$). In this case the separation condition is satisfied well before the transition. The two MBHs are already far inside the resolution sphere, so in this test case it is the binding condition that sets the moment of transition to stage~1.

Since stage~1 entry already requires the pair to be bound, the binding criterion holds from the start of stage~1. The only remaining condition for stage~2 is the enclosed-mass criterion (enclosed galaxy mass less than twice the binary mass), which is why stage~2 follows so quickly after stage~1 here. We note that in this test the binary is relatively massive compared with the surrounding galactic nucleus. As a result, the enclosed-mass criterion is met at a comparatively large separation, and stage~2 here begins near $30\,{\rm pc}$. In a Milky-Way-like MBH system, the binary is far less massive relative to its host nucleus. The same criterion is therefore typically satisfied only at $\sim{\rm pc}$ scales, and stage~2 would begin at correspondingly smaller separations. The stage~2 evolution time would also be much shorter, as found in our previous work \citep{RAMCOAL2024}.

This sequence follows the standard ordering of MBHB evolution, from large-scale orbital decay to the formation of a bound binary, its hardening by environmental interactions, and the final GW-driven inspiral \citep{Begelman1980,Quinlan1996,Peters1964,SesanaKhan2015}. The difference is that, here, the stages are embedded in a hydrodynamical simulation rather than imposed entirely in post-processing. The interval between entry into the subgrid RAMCOAL regime (start of stage~1) and the labelled coalescence (end of stage~2) illustrates why this evolution matters for merger catalogues built from numerical mergers \citep{Kelley2017, Volonteri2020, Li2022TNG50, DongPaez2023, Li2023TNG50DualAGN, Sayeb2024Triples}. In a conventional sink-merger catalogue, a grid-scale close pair would have been removed at the numerical merger stage. RAMCOAL instead preserves an evolving binary and keeps updating its separation, eccentricity, accretion state, and spin state. This provides information closer to that required by semi-analytic delay models \citep{IzquierdoVillalba2026}, but with the environmental history taken from the simulated galaxy rather than from a prescribed host parameter alone.

\begin{figure*}
\centering
\includegraphics[width=0.82\textwidth]{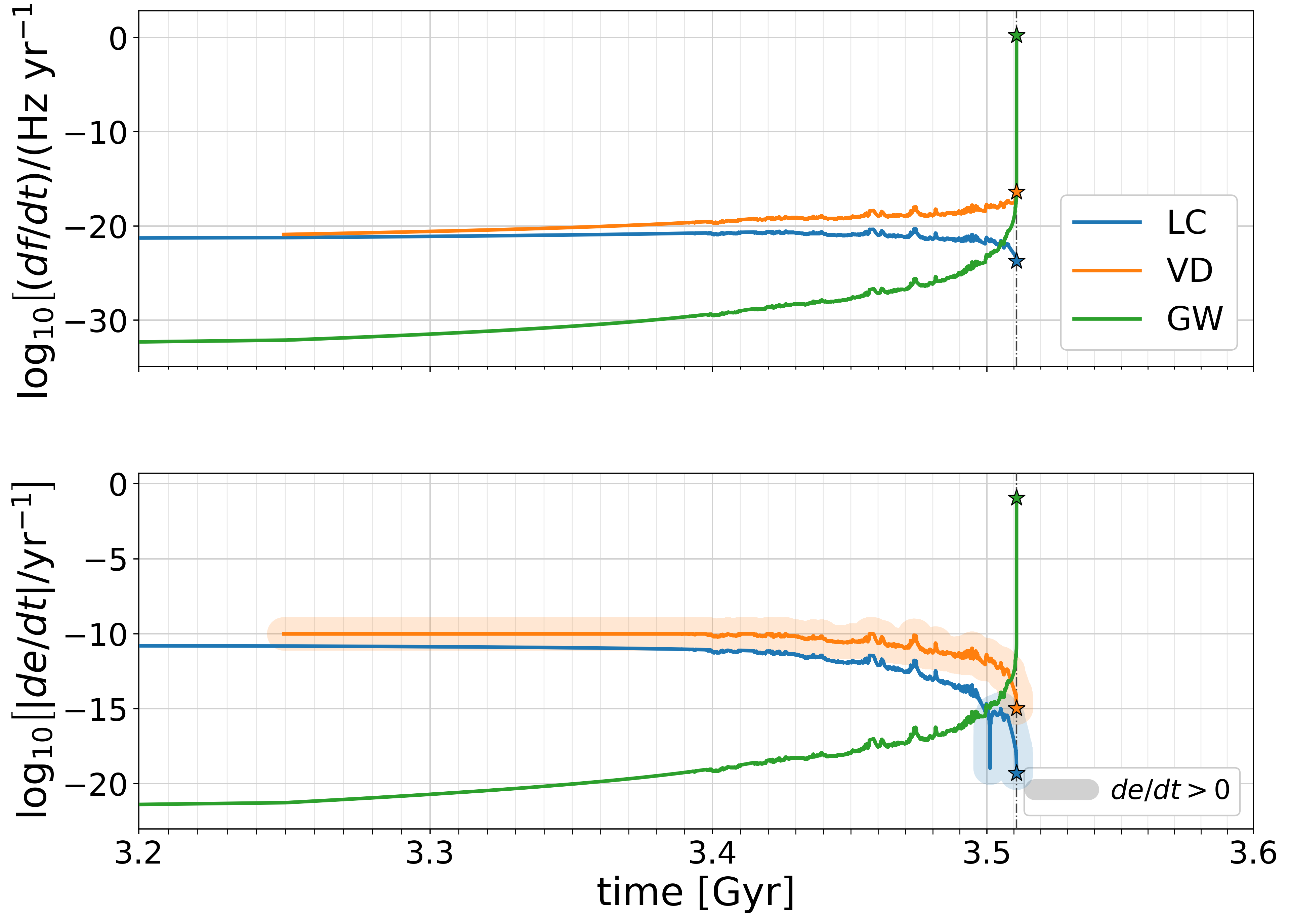}
\caption{Evolution of the hardening terms for triplet test case~A. The panels show the loss-cone (LC), viscous-disc (VD), and gravitational-wave (GW) contributions to the semi-major-axis and eccentricity evolution.}
\label{fig:tripletcaseA_df}
\end{figure*}

\begin{figure}
\centering
\includegraphics[width=\columnwidth]{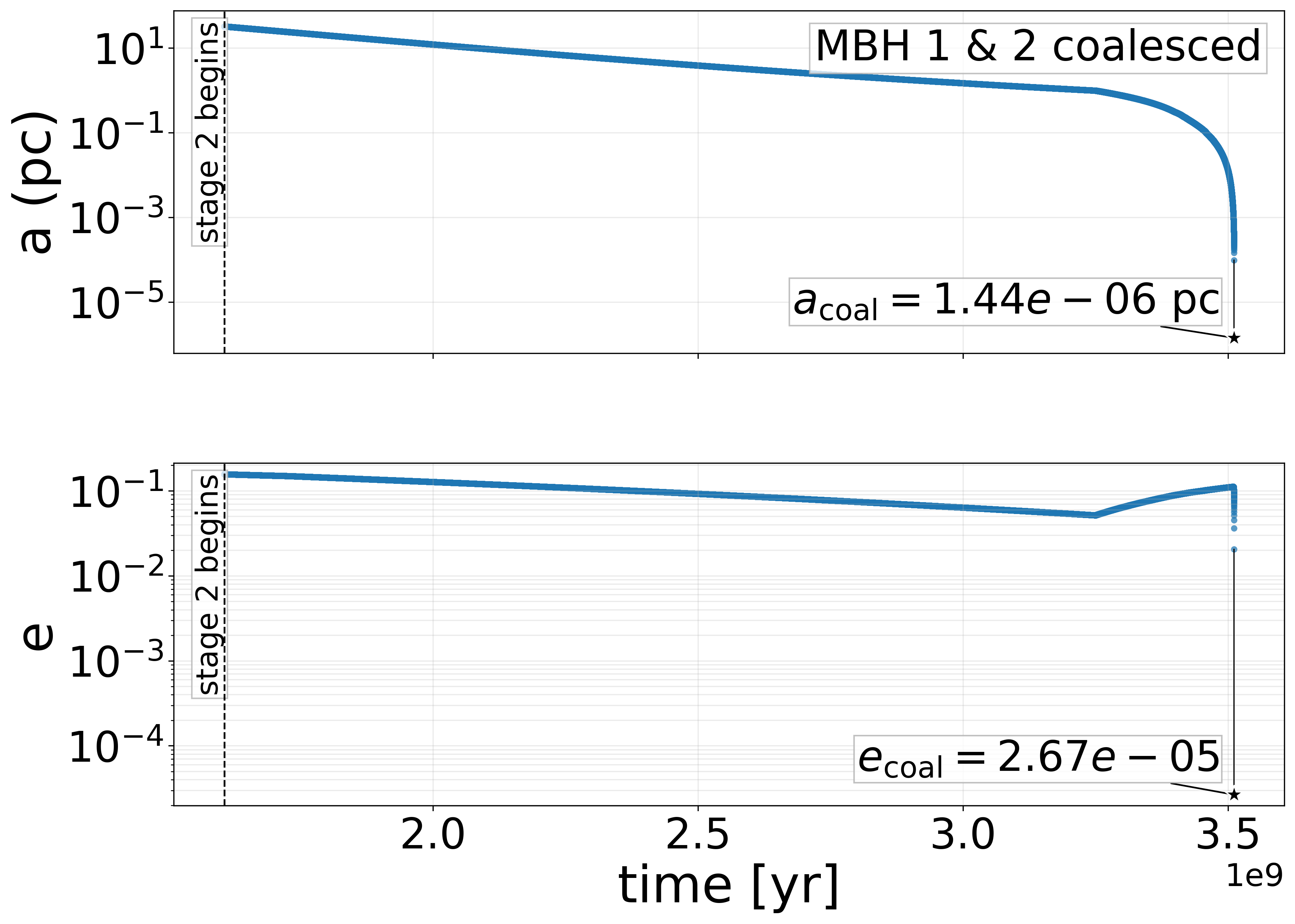}
\caption{Subgrid orbital elements for the inner binary in triplet test case~A. The figure presents the stage~2 semi-major axis and eccentricity evolution and marks their values at coalescence.}
\label{fig:tripletcaseA_orbital}
\end{figure}

\begin{figure}
\centering
\includegraphics[width=\columnwidth]{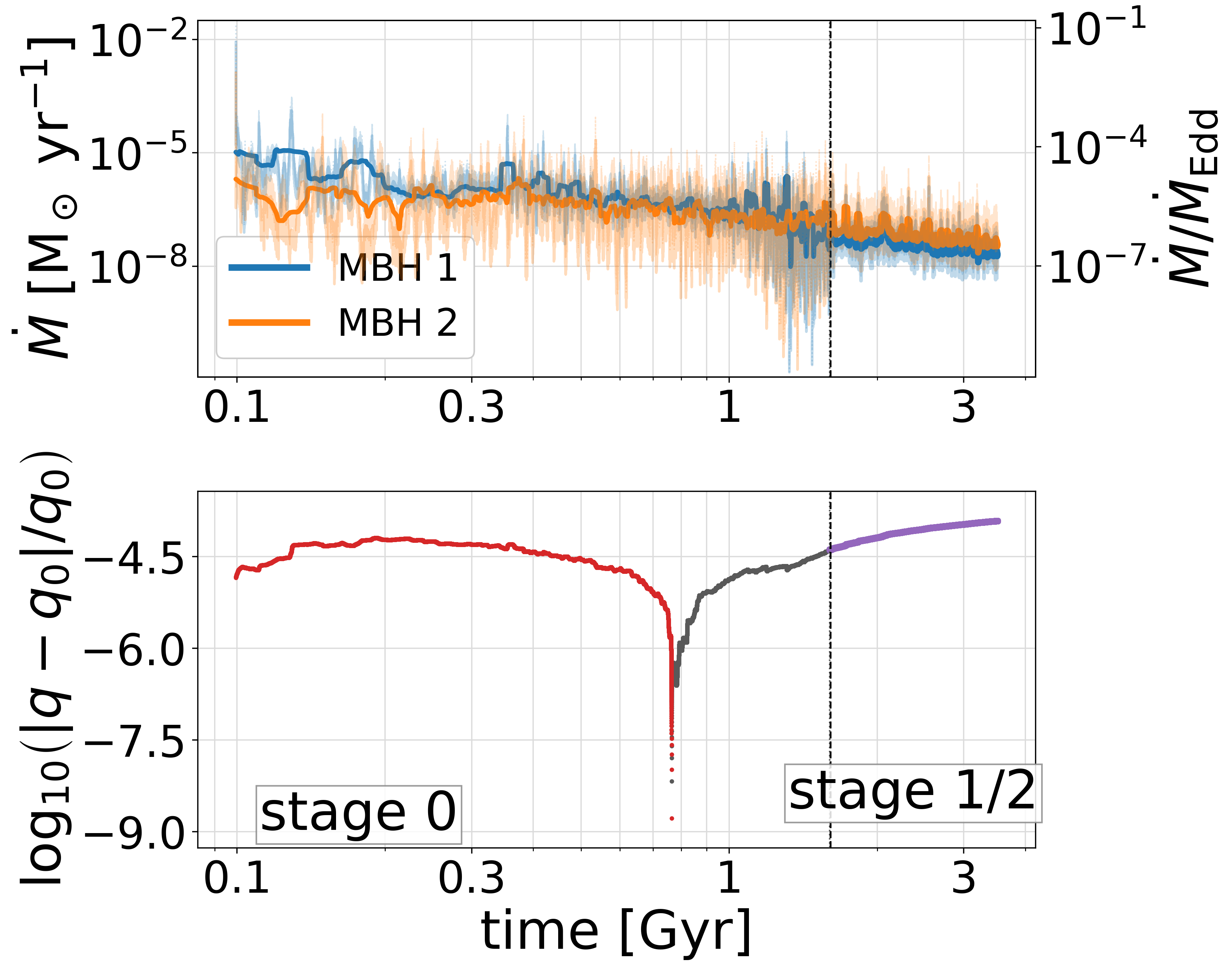}
\caption{Accretion and mass-ratio evolution for triplet test case~A. The upper panel shows the component accretion rates of MBH1 (blue) and MBH2 (orange). The light shaded tracks show the accretion rate at each time step. The solid, smoothed tracks plotted on top show the centred running mean over a $20\,{\rm Myr}$ window. Each point therefore averages the rate within $\pm0.01\,{\rm Gyr}$ of that time. This is a local running mean rather than an average over a whole stage. The left axis gives the absolute rate $\dot{M}$ in ${\rm M_\odot\,yr^{-1}}$ and the right axis the Eddington ratio $\dot{M}/\dot{M}_{\rm Edd}$, showing that accretion remains strongly sub-Eddington throughout. The lower panel tracks the fractional deviation of the inner-binary mass ratio from its initial value, $\log_{10}(|q-q_0|/q_0)$, where $q$ is the instantaneous mass ratio and $q_0$ its initial value. Red markers indicate a decreasing mass ratio (negative $(q-q_0)/q_0$). The mass ratio first decreases because MBH1 is in the centre of the galaxy, where the gas reservoir is richer. Later the mass ratio starts to increase because of preferential accretion onto MBH2 and the self-regulated accretion of MBH1. In both panels the vertical dashed line marks the transition from the resolved stage~0 to the subgrid stage~1/2 evolution.}
\label{fig:tripletcaseA_accretion}
\end{figure}

\begin{figure}
\centering
\includegraphics[width=\columnwidth]{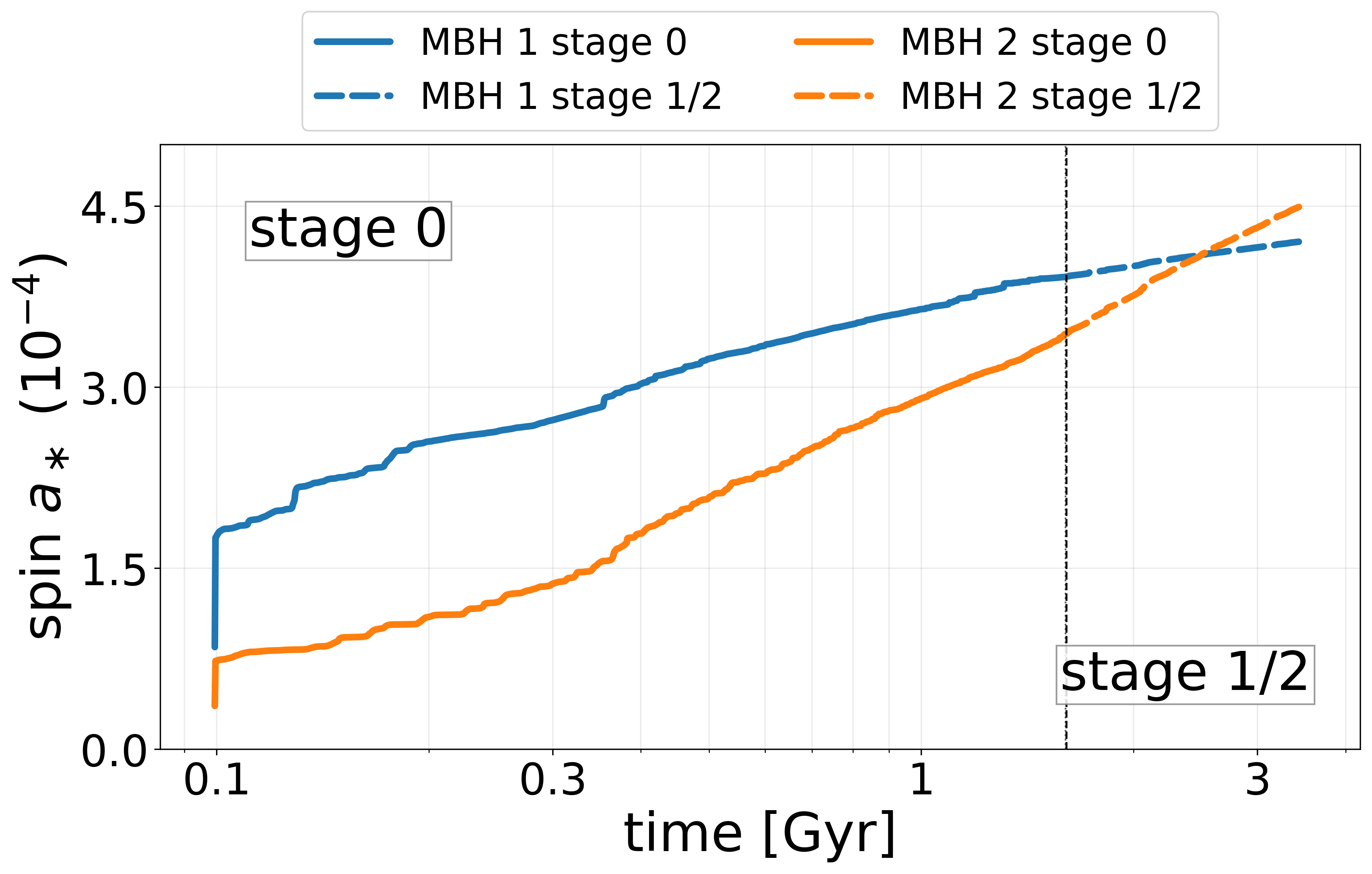}
\caption{Spin-magnitude evolution for triplet test case~A. The vertical dashed line marks the transition from the resolved stage~0 to the subgrid stage~1/2 evolution. The spin of the secondary MBH becomes larger than that of the primary, because preferential accretion makes it grow faster.}
\label{fig:tripletcaseA_spin}
\end{figure}

Figs~\ref{fig:tripletcaseA_df}--\ref{fig:tripletcaseA_spin} show the orbital, accretion, and spin evolution of the inner binary in stage~2.
Figure~\ref{fig:tripletcaseA_df} presents the hardening terms for this gas-rich coplanar configuration. It indicates that the viscous-disc term is comparable to or somewhat larger than the loss-cone term during most of stage~2. The gravitational-wave term overtakes both only in the final approach to merger. This ordering matches the usual expectation that stellar and gaseous processes govern the wider hard-binary phase, whereas the GW term dominates only at the smallest separations \citep{Peters1964,Quinlan1996,SesanaKhan2015}. The bottom panel of Fig.~\ref{fig:tripletcaseA_df} shows the contributions of LC, VD, and GW to the eccentricity evolution in the hardening phase. The shaded regions indicate when a mechanism increases the eccentricity instead of circularizing the orbit. The circumbinary disc always increases the eccentricity in this case, because the eccentricity is lower than the attractor eccentricity found in several CBD simulations \citep{Zrake2021,DOrazio2021,Siwek2023b}. Loss-cone scattering, in contrast, decreases the eccentricity at the early hardening stage. Towards the end (at $3.5$\,Gyr), it starts to make the orbit more eccentric. This is when the blue line enters the shaded region in the bottom panel of Fig.~\ref{fig:tripletcaseA_df}. During the hardening phase, the effect of stellar loss-cone scattering on the binary eccentricity naturally changes direction. When the binary is still relatively wide, stellar encounters remove energy and angular momentum in a way that slightly circularizes the orbit. As the binary becomes harder and more tightly bound, close three-body encounters with stars become more efficient at extracting angular momentum than orbital energy, which makes the orbit more eccentric. Overall, however, the eccentricity still decreases, because GW emission dominates at the end and circularizes the orbit.

Figure~\ref{fig:tripletcaseA_orbital} traces the stage~2 decay of the semi-major axis and the evolution of the eccentricity towards the coalescence values labelled in the figure. It shows that the MBH1--MBH2 binary remains only mildly eccentric, with $e$ at the level of $0.05$--$0.15$ during most of the hard-binary phase. At coalescence the eccentricity is reduced to $e_{\rm coal}\simeq3\times10^{-5}$, as expected once gravitational-wave emission circularizes the orbit \citep{Peters1964}.

Figure~\ref{fig:tripletcaseA_accretion} shows the accretion rates and the mass-ratio evolution from stage~0 to stage~2. Preferential accretion onto the secondary MBH is clearly visible from about 0.2\,Gyr, which is also when the mass ratio starts to increase.  

Figure~\ref{fig:tripletcaseA_spin} illustrates the gradual growth of the spin magnitudes across stage~0 and stage~1/2. The spin magnitude of the secondary overtakes that of the primary, because preferential accretion (Fig.~\ref{fig:tripletcaseA_accretion}) boosts its accretion and therefore its spin growth. During stage~0 the MBHs accrete in the magnetically arrested disc (MAD) state. Naively, in the MAD regime Blandford--Znajek angular-momentum extraction tends to spin the MBH down, so one might expect the spins to decrease. The spins nonetheless grow here because both MBHs start with spins very close to zero, where the extraction torque is weak. In this regime, the equilibrium spin at which $da_\ast/{\rm d}t=0$ is small but positive. The MBHs sit below this threshold, so the net torque remains slightly positive and the spins increase rather than decrease.

Simulations have found that CBD structure, preferential accretion, and mini-disc variability can alter the respective growth history of the two BHs before merger \citep{Farris2014,Duffell2020,Siwek2023a,Siwek2023b}. Even though our simulations do not resolve the CBD, RAMCOAL follows the accretion, mass growth, spin evolution, and environmental hardening of the subgrid binary. These are among the quantities needed to turn a delayed coalescence time into a predicted GW source with component masses, spins, eccentricity, and possible electromagnetic signatures.

\subsection{Triplet test case~B (inclined initial orbital configuration)}

The second triplet experiment uses the same relaxed isolated galaxy but adopts an inclined initial orbital configuration. MBH1 and MBH2 are still orbiting in the galactic plane, while the orbit of MBH3 is now inclined by $i=\pi/2$ with respect to the galactic plane (its initial velocity has a non-zero vertical $z$-component). The initial conditions of the MBHs are listed in Table~\ref{tab:tripletic}. This case probes a plunging encounter following an inclined orbit, in which the incoming outer MBH can perturb, or exchange with, the initially bound pair before the system reaches the final GW-dominated regime.

\begin{figure*}
\centering
\includegraphics[width=494.98bp]{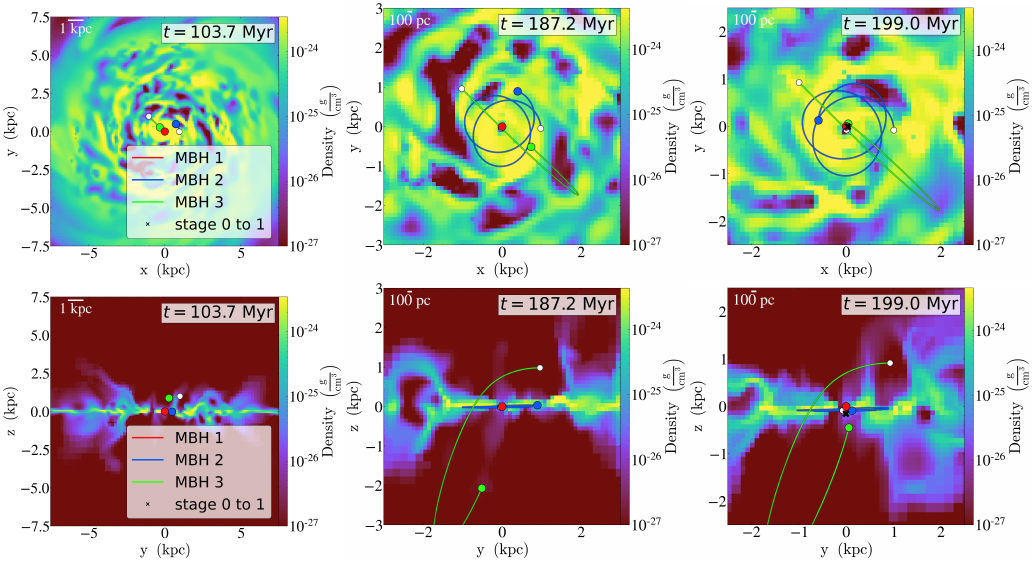}
\caption{Gas-density slice snapshots with MBH tracks for triplet test case~B, using the inclined initial orbital configuration. The top row presents the $z$-direction slices for the three selected snapshots, and the bottom row displays the corresponding $x$-direction slices. MBH1, MBH2, and MBH3 are shown in red, blue, and green, respectively, the colours of their legend entries. Each trajectory runs from the time at which the MBHs are inserted ($t=100\,{\rm Myr}$) to the time of the snapshot. It is faint at the start and becomes stronger towards the time of the snapshot. The older parts of a trajectory are partly transparent over the density map and have a thin dark outline, so they look darker than the marker. White circles mark the starting point of each trajectory, i.e. the position at which the MBH was inserted, and filled circles mark the position of each MBH at the time of the snapshot. The black cross in the last panel at $t=199$\,Myr marks the starting position of the stage~1 RAMCOAL subgrid evolution. The trajectories of MBH1 and MBH3 during RAMCOAL stage~1 are shown in Fig.~\ref{fig:tripletcaseB_snapshots_subgrid} on $y$-direction slices for clarity.}
\label{fig:tripletcaseB_snapshots}
\end{figure*}

\begin{figure*}
\centering
\includegraphics[width=455.50bp]{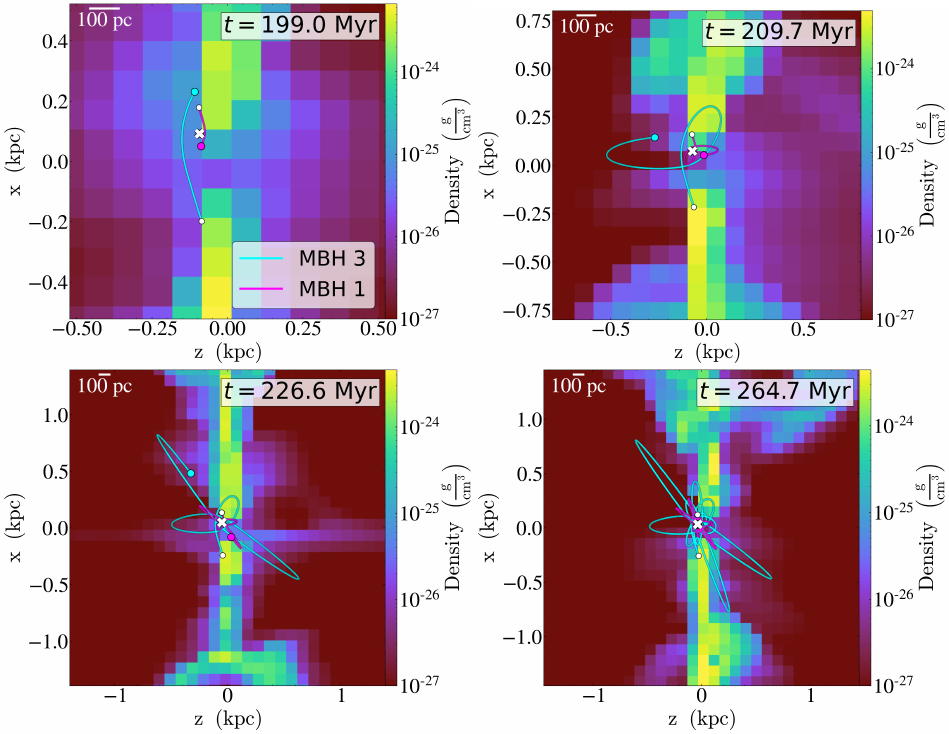}
\caption{Gas-density slice snapshots with subgrid MBH tracks for test case~B. The cyan track shows the orbit of MBH3 around the centre of mass of the pair evolving in stage~1 of RAMCOAL. The magenta track shows the corresponding orbit of MBH1. These panels show the subgrid evolution of the pair after it enters the RAMCOAL regime, following on from the last panel of Fig.~\ref{fig:tripletcaseB_snapshots}. They are viewed along the $y$-direction (unlike Fig.~\ref{fig:tripletcaseB_snapshots}), which shows the orbit more clearly. The last panel, at $t=264.7$\,Myr, shows MBH1 and MBH3 just before they become gravitationally bound and enter RAMCOAL stage~2. In stage~2 only the orbital parameters are available.}
\label{fig:tripletcaseB_snapshots_subgrid}
\end{figure*}

Figure~\ref{fig:tripletcaseB_snapshots} shows snapshots with the MBH trajectories seen from two directions, because the initial configuration of the MBHs is no longer in-plane. The $z$- and $x$-direction slices reveal that the incoming MBH3 approaches the nuclear region along a plunging trajectory perpendicular to the galactic disc and to the orbital plane of the inner binary. MBH2, in contrast, follows a less-eccentric in-plane orbit around MBH1. In the last panel of Fig.~\ref{fig:tripletcaseB_snapshots}, at $t=199$\,Myr, the black cross marks the centre of mass of MBH1 and MBH3 as they enter RAMCOAL stage~1. Their stage~1 evolution around this centre of mass, as modelled by RAMCOAL, is shown in Fig.~\ref{fig:tripletcaseB_snapshots_subgrid}. These panels show the subgrid evolution of the pair after it enters the RAMCOAL regime, following on from the last panel of Fig.~\ref{fig:tripletcaseB_snapshots}. They are viewed along the $y$-direction (unlike Fig.~\ref{fig:tripletcaseB_snapshots}), which shows the orbit more clearly. The orbital plane of MBH1 and MBH3 does not lie in the plane of the galactic disc, because the orbit of MBH3 is inclined. The last panel, at $t=264.7$\,Myr, shows MBH1 and MBH3 just before they become gravitationally bound and enter RAMCOAL stage~2. In stage~2 only the orbital parameters are available.

Changing the orbital configuration can change the close-passage sequence, the identity of the surviving or merging binary, and the outcome of a possible chaotic interaction \citep{Bonetti2018}. It also alters the galaxy environment in which the MBHs evolve and changes the evolution time and the MBH orbital parameters. This can result in very different GW signals.

\begin{figure*}
\centering
\includegraphics[width=0.95\textwidth]{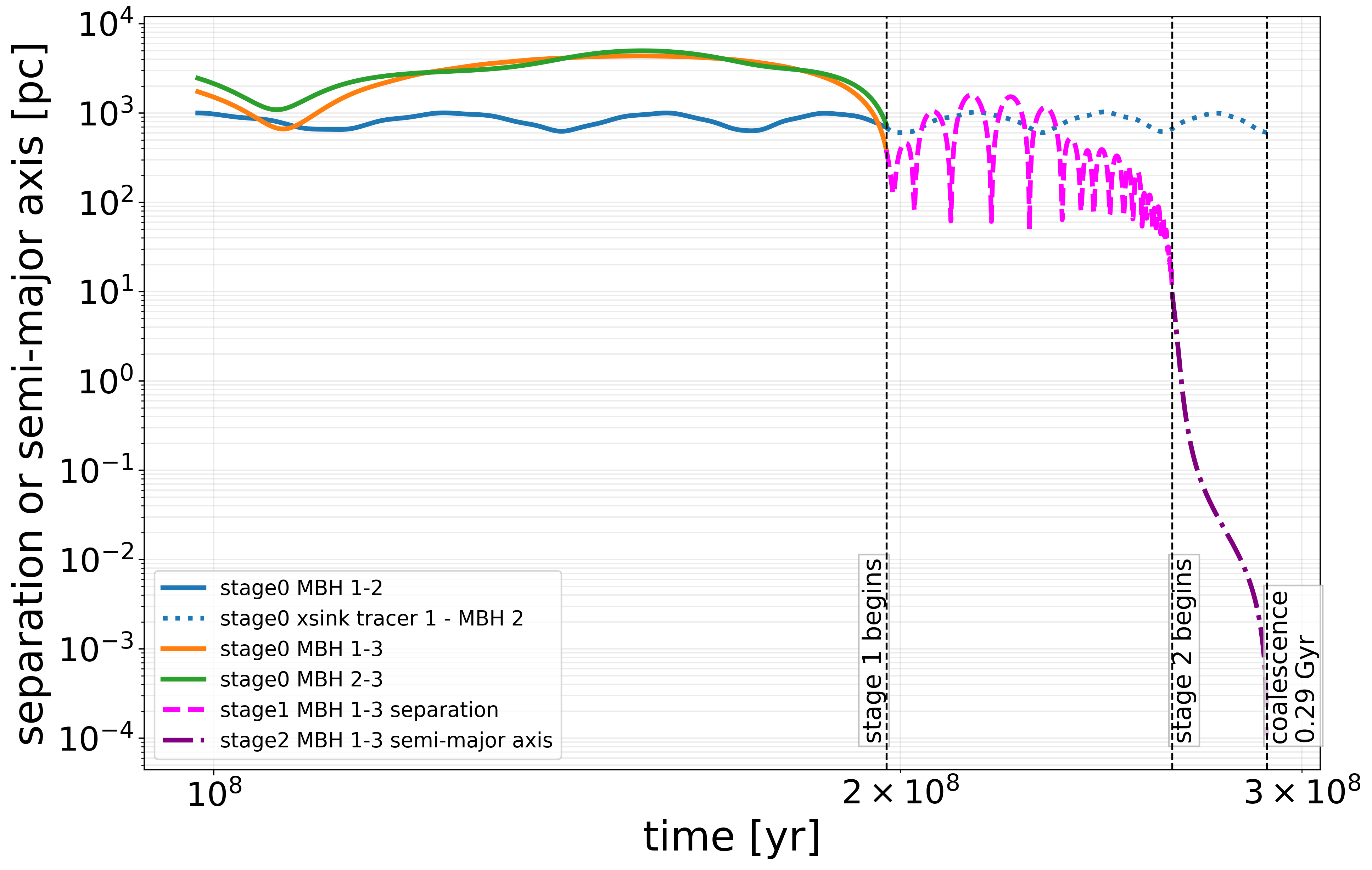}
\caption{Pairwise-separation evolution for triplet test case~B with the inclined initial orbital configuration. The figure presents the resolved stage~0 pairwise separations of the three MBHs (MBH\,1--2, MBH\,1--3, and MBH\,2--3). It also shows the stage~1 and stage~2 evolution of the MBH1--MBH3 inner binary, the separation between this binary and the surviving MBH2 on its wider orbit, and the labelled coalescence time. It should be read together with the snapshots in Fig.~\ref{fig:tripletcaseB_snapshots}. The blue dotted line shows the separation evolution between MBH2 and the centre of mass of MBH1 and MBH3, which already form a subgrid pair. The position of this centre of mass is traced by the sink particle carrying the pair (xsink tracer 1). The MBH\,2--3 curve shows which pair forms. When MBH3 plunges through the nucleus at $t\simeq0.197\,{\rm Gyr}$, it enters the resolution sphere of MBH1 (at a separation of $\simeq390\,{\rm pc}$). At the same time it remains $\simeq750\,{\rm pc}$ from MBH2, which orbits MBH1 in the disc plane at $\simeq0.6$--$1\,{\rm kpc}$. The pair that enters RAMCOAL is therefore MBH1--MBH3 and not the original inner pair MBH1--MBH2, and during stage~0 MBH3 never comes closer than $\simeq750\,{\rm pc}$ to MBH2.}
\label{fig:tripletcaseB_separation}
\end{figure*}

The separation evolution in Fig.~\ref{fig:tripletcaseB_separation} shows that the inclined case enters the subgrid RAMCOAL stages much earlier than the in-plane case and completes the subgrid evolution far more rapidly. Stage~1 begins at about $200 \,{\rm Myr}$, stage~2 begins at about $263 \,{\rm Myr}$, and the coalescence happens at $290\,{\rm Myr}$, roughly an order of magnitude earlier than in the in-plane case. Note that in this case it is MBH1 and MBH3 that become bound and eventually coalesce, whereas in test case~A it is MBH1 and MBH2.

The incoming MBH3 is captured by the primary, so the pair that finally merges is not the pair initialized as the inner binary. The displaced MBH2 is left on a wider orbit, which is traced separately in Fig.~\ref{fig:tripletcaseB_separation}. The inner separation undergoes pronounced oscillations during stage~1, indicating a more eccentric and dynamically active encounter. This change of coalescing partner, together with the much shorter delay, is the principal qualitative difference between tests~A and~B.

Within these two controlled experiments, the different outcome is associated with the imposed initial orbital configuration. The host galaxy setup is the same, while the inclined orbit changes the sequence of close passages and the pair selected for the final subgrid evolution. The result should therefore be read as a demonstration of sensitivity to the orbital configuration, not as a statistical statement that inclined encounters generally coalesce more rapidly. As found in paper~I, the outcome can also depend on the random seed, so a single realization of each geometry does not capture the full scatter. It is likewise not a substitute for a direct few-body integration of the resonant phase. Instead, it illustrates how a probabilistic approach can be embedded in a hydrodynamical simulation while still determining which MBHs merge, the merger time, and the pre-coalescence binary state. 

In both cases, however, the outer MBH never enters the subgrid RAMCOAL regime. The comparison of test cases~A and~B shows that the evolution time and the resulting remnant can be completely different, even without a chaotic triplet interaction and even without forming an active triplet. We demonstrate the evolution of an active triplet and the modelling of the chaotic interaction in test case~C below.

\begin{figure*}
\centering
\includegraphics[width=0.82\textwidth]{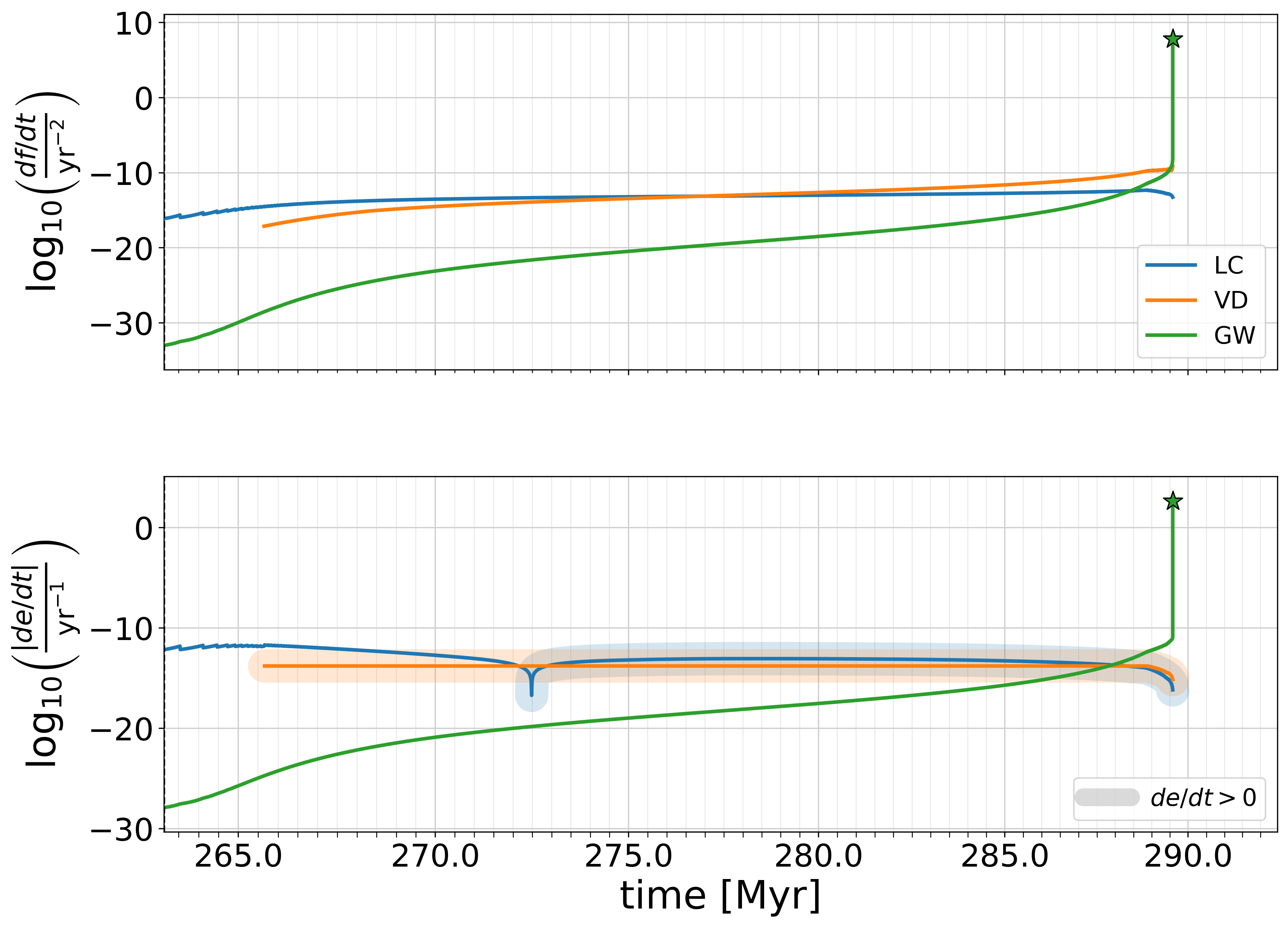}
\caption{Evolution of the hardening terms for triplet test case~B. The panels show the loss-cone (LC), viscous-disc (VD), and gravitational-wave (GW) contributions to the semi-major-axis and eccentricity evolution for the inclined case.}
\label{fig:tripletcaseB_df}
\end{figure*}

\begin{figure}
\centering
\includegraphics[width=\columnwidth]{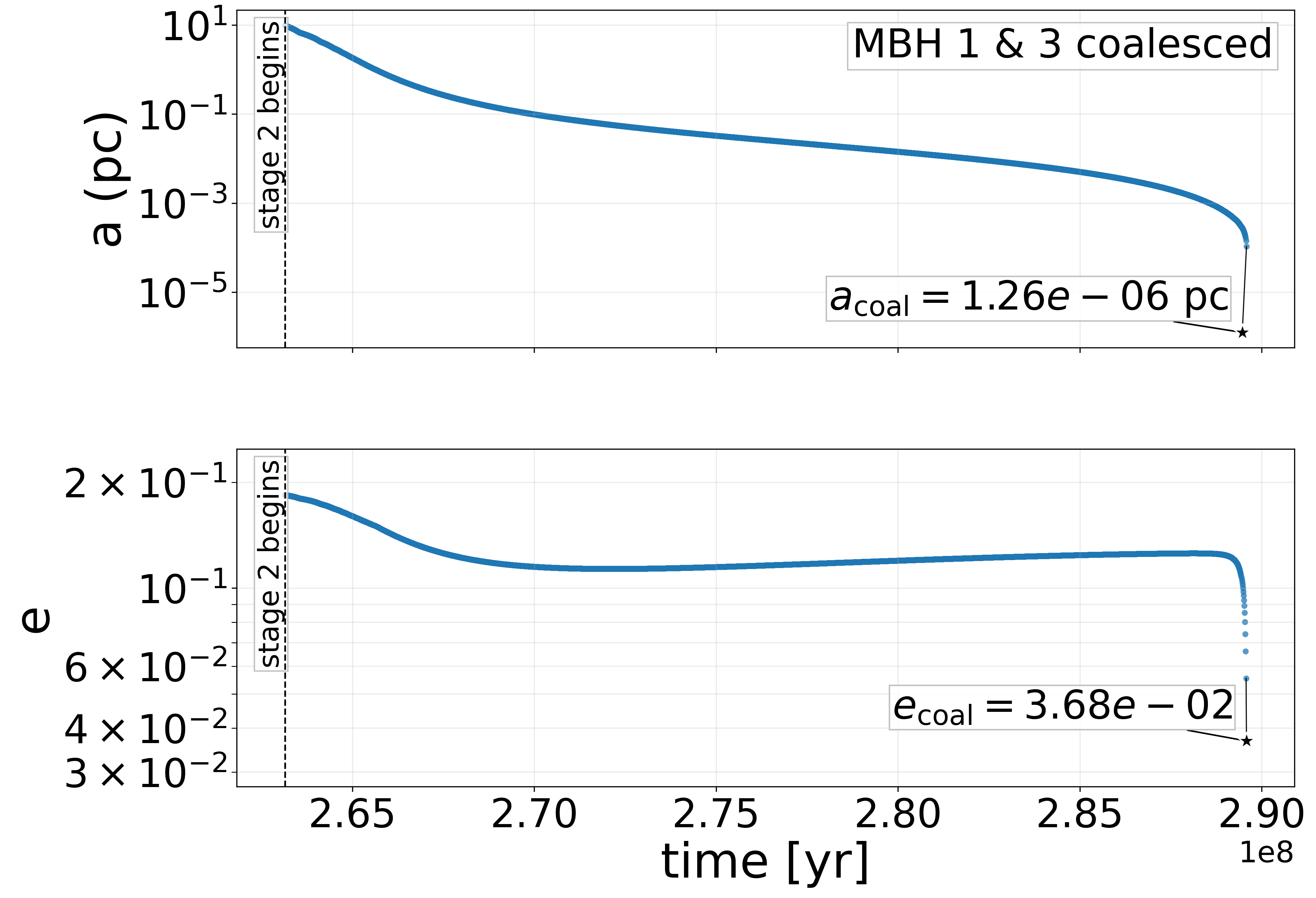}
\caption{Subgrid orbital elements for the final binary in triplet test case~B. The figure traces the stage~2 semi-major axis and eccentricity evolution and labels the MBH1--MBH3 coalescence.}
\label{fig:tripletcaseB_orbital}
\end{figure}

\begin{figure}
\centering
\includegraphics[width=\columnwidth]{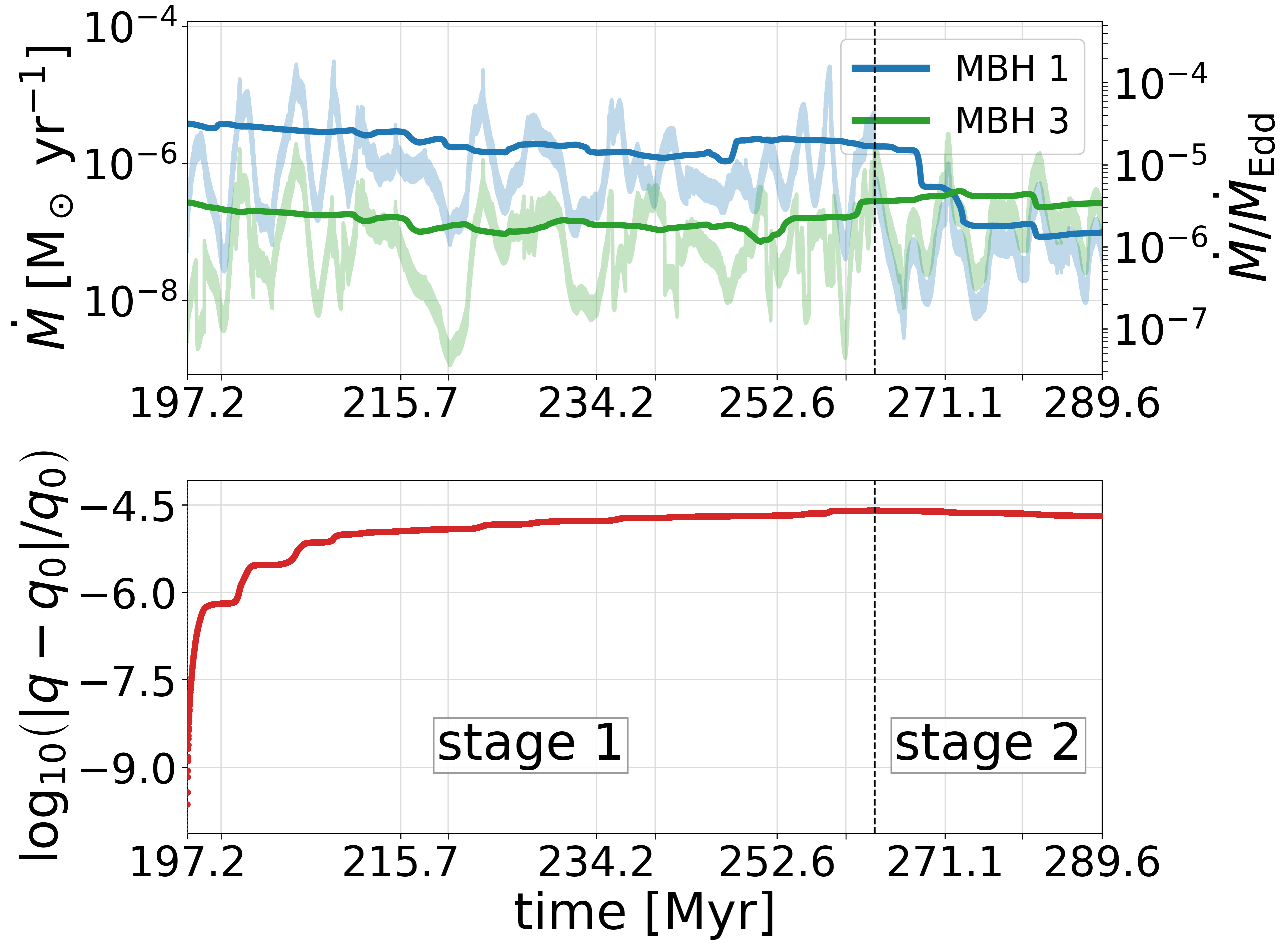}
\caption{Accretion and mass-ratio evolution for triplet test case~B. The upper panel displays the component accretion histories of MBH1 and MBH3, while the lower panel tracks the corresponding mass ratio.}
\label{fig:tripletcaseB_accretion}
\end{figure}

\begin{figure}
\centering
\includegraphics[width=\columnwidth]{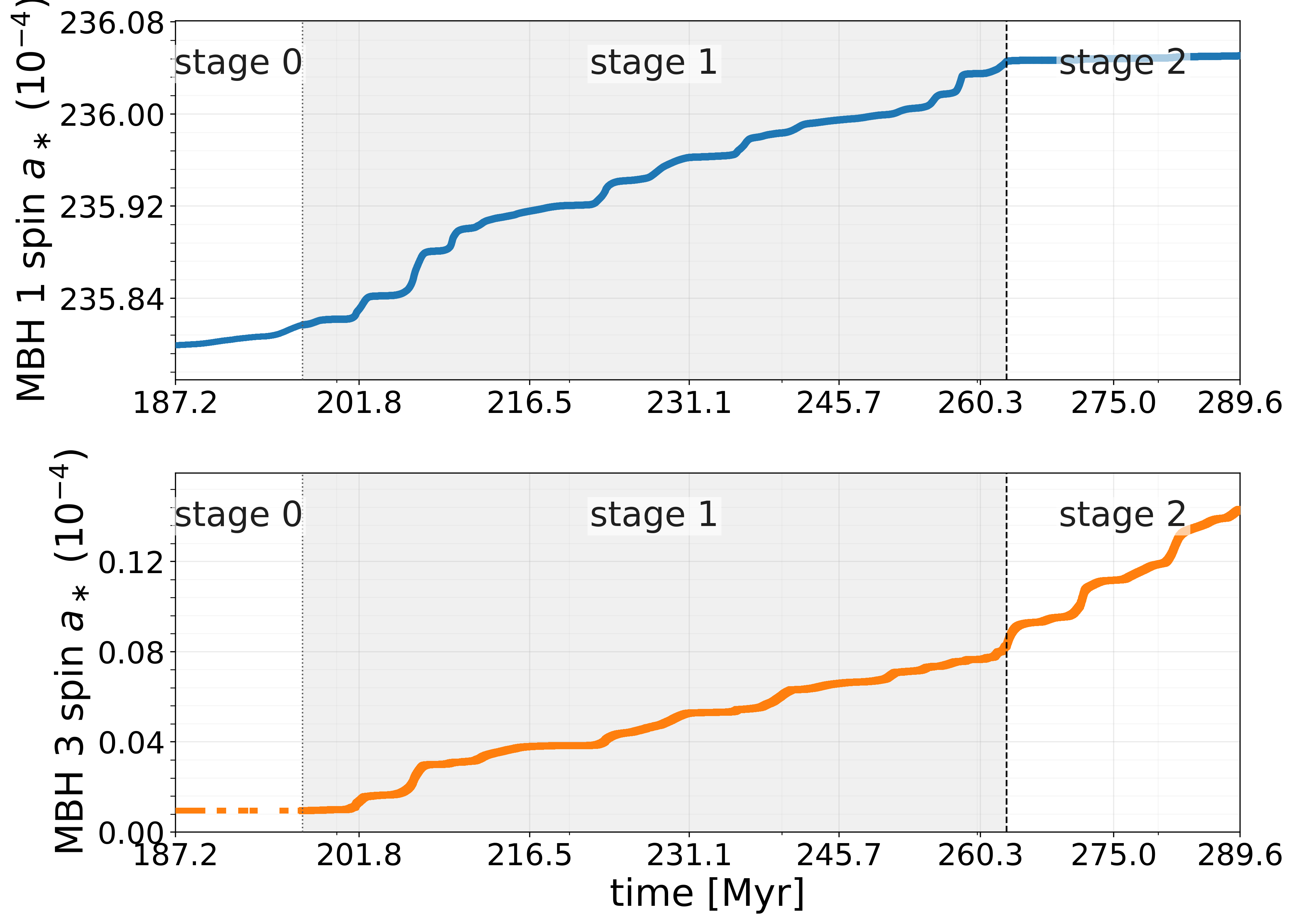}
\caption{Spin-magnitude evolution for triplet test case~B. The figure separates the resolved stage~0 spin history from the later RAMCOAL stage~1/2 evolution and marks the onset of stage~2.}
\label{fig:tripletcaseB_spin}
\end{figure}

\begin{figure}
\centering
\includegraphics[width=\columnwidth]{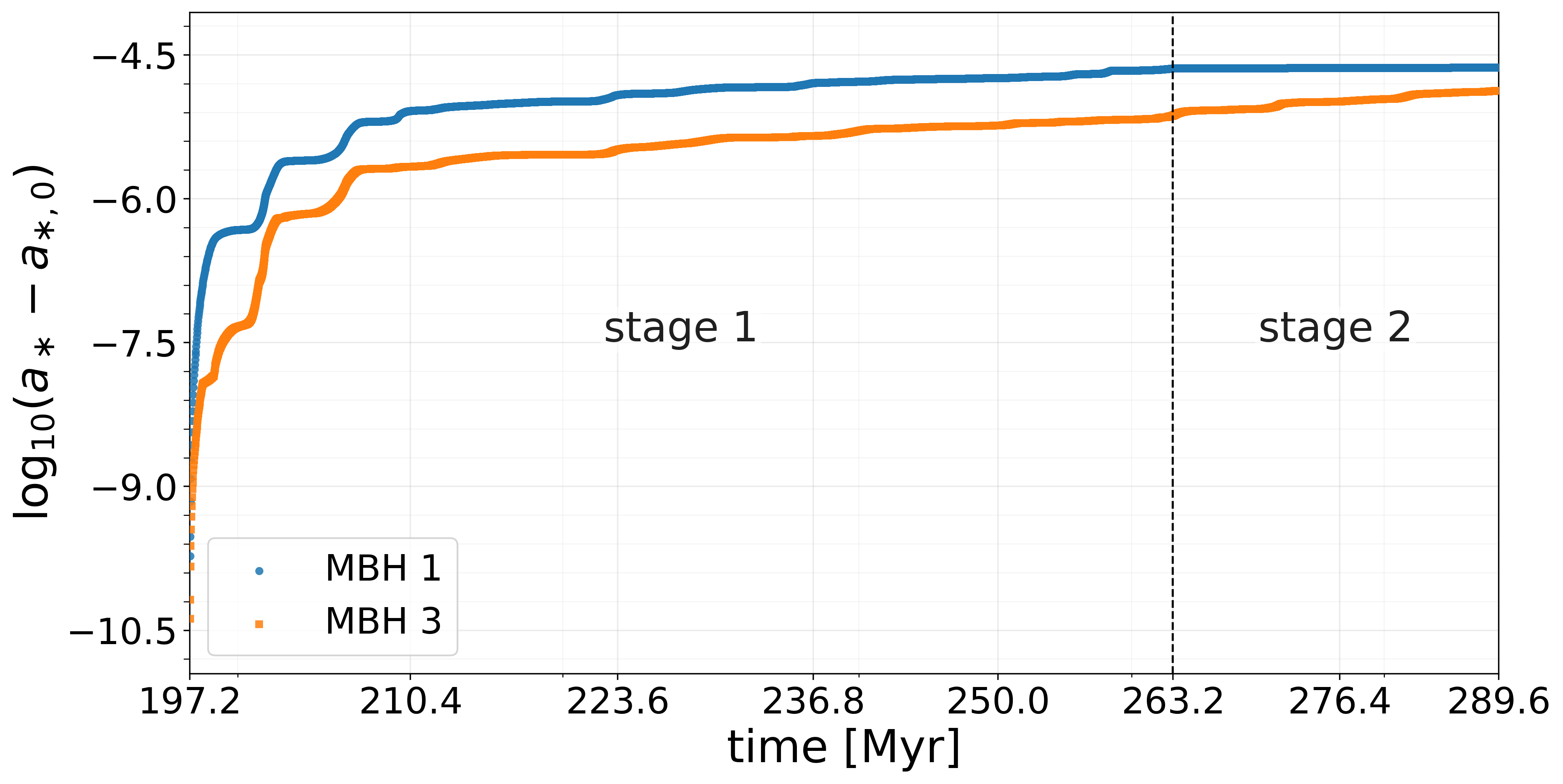}
\caption{
Evolution of the subgrid spin growth for MBH1 and MBH3 after the pair enters stage~1.
The plotted quantity is $\log_{10}(a_\ast-a_{\ast,0})$, where $a_{\ast,0}$ is the spin magnitude of each MBH at the onset of stage~1.
The vertical dashed line marks the transition from stage~1 to stage~2.
}
\label{fig:tripletcaseB_spin_growth}
\end{figure}

\begin{figure}
\centering
\includegraphics[width=\columnwidth]{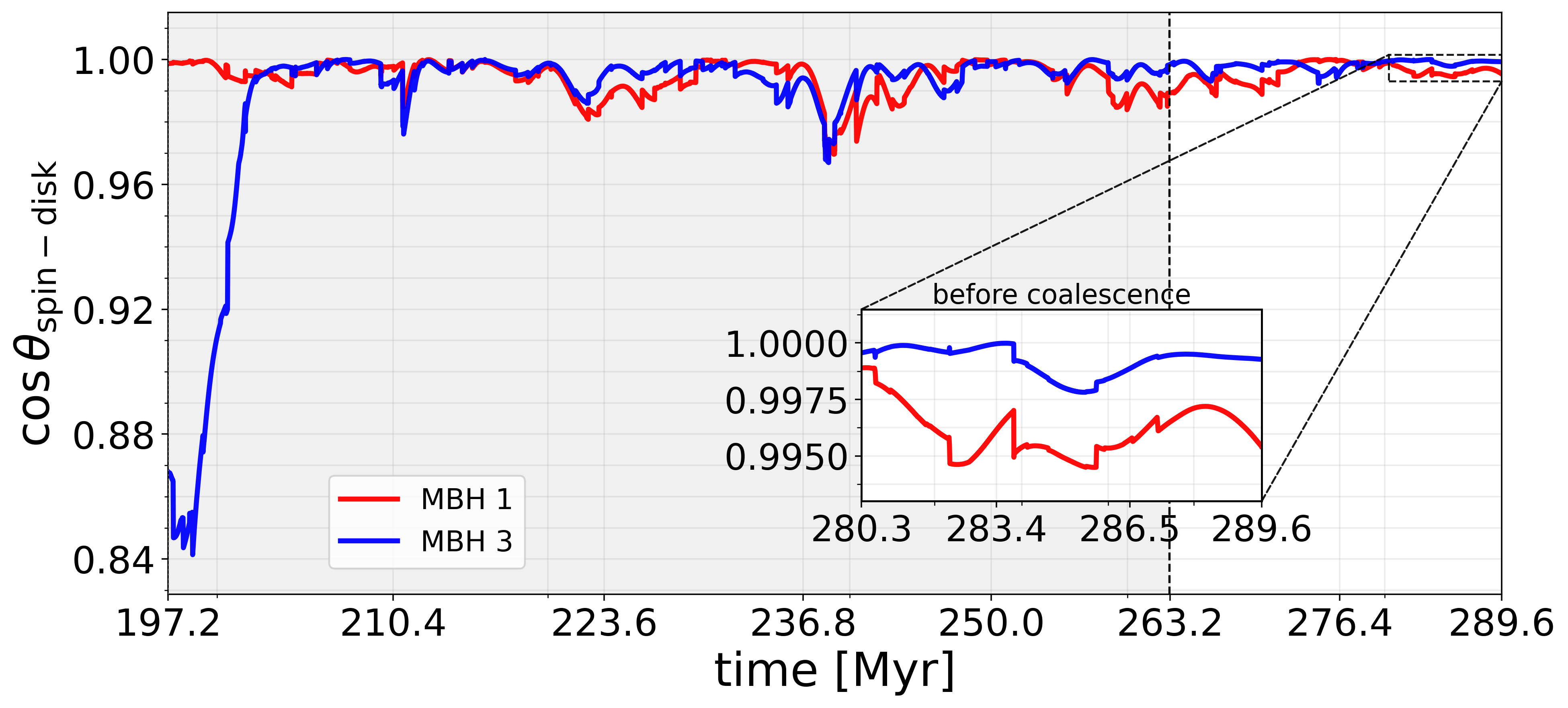}
\caption{Spin--gas-disc alignment for triplet test case~B. The panel follows $\cos\theta_{\rm spin-disk}$, the cosine of the angle between each MBH spin and the local gas-disc angular-momentum direction. It is shown for MBH1 (red) and MBH3 (blue) as a function of time, with the stage~1 and stage~2 transitions marked. The inset zooms in on the final approach to coalescence.}
\label{fig:tripletcaseB_spin_alignment}
\end{figure}

Figures~\ref{fig:tripletcaseB_df}--\ref{fig:tripletcaseB_spin_alignment} present the same orbital, accretion, and spin evolution tracked by RAMCOAL as for case~A, together with the spin alignment shown in Fig.~\ref{fig:tripletcaseB_spin_alignment}. 

As in the in-plane case, Fig.~\ref{fig:tripletcaseB_df} shows the loss-cone, viscous-disc, and GW terms during the stage~2 evolution. In test case~B, however, stage~2 is much shorter. Towards the end of the hardening phase, the viscous-disc drag still dominates over loss-cone scattering, even though MBH3 started on a highly inclined orbit. This is because we always assume that a bound MBHB is embedded in a CBD, which may be optimistic. We leave the development of an environment-dependent criterion for CBD formation in RAMCOAL to future work. 

The circumbinary disc always increases the eccentricity in this case, because the eccentricity is lower than the attractor eccentricity when the binary enters the hardening phase. Loss-cone scattering, by contrast, first reduces the eccentricity during the early hardening stage and then drives the orbit more eccentric towards the end. Once the binary becomes hard enough, close three-body encounters with stars remove angular momentum more efficiently than orbital energy, which increases the eccentricity. In this system, that transition occurs when the binary separation is $0.05$\,pc. After that, loss-cone scattering modestly increases the eccentricity until gravitational-wave emission takes over and circularizes the binary before coalescence.

The orbital-elements plot traces the semi-major axis decreasing from parsec scales to the coalescence scale, while the eccentricity remains modest and then drops near the end of the plotted evolution. In more detail, the MBH1--MBH3 binary, which started from a more inclined orbit, retains a higher residual eccentricity than the binary of the in-plane case. The eccentricity remains near $0.1$--$0.2$ during the short hard-binary phase. It decreases only to $e_{\rm coal}\simeq4\times10^{-2}$ at coalescence (Fig.~\ref{fig:tripletcaseB_orbital}), compared with $e_{\rm coal}\sim10^{-5}$ in the coplanar case~A. Figure~\ref{fig:tripletcaseB_df} shows the eccentricity evolution rate due to each physical mechanism in the hardening phase. The rates indicate that the loss-cone term becomes positive earlier in the inclined orbit case and stays positive for longer, which results in a higher eccentricity at coalescence. 

As shown in Fig.~\ref{fig:tripletcaseB_accretion}, MBH3 accretes at a higher rate once it is captured by MBH1 and the two evolve together in the disc as a subgrid pair. In stage~0, MBH1 at the centre of the galaxy potential accretes at a higher rate than MBH3. This is because the inclined orbit keeps MBH3 outside the galaxy disc most of the time, where there is no gas reservoir for it to accrete from. The situation changes when MBH3 is captured by MBH1 and starts to orbit in the disc plane. The mass ratio then starts to grow because of preferential accretion in the hardening phase.

Both spins remain low throughout (magnitudes $<0.1$, Fig.~\ref{fig:tripletcaseB_spin}). MBH3 has the lower spin of the two because it barely accretes in stage~0 and only begins to spin up after it is captured near the galactic centre. Figure~\ref{fig:tripletcaseB_spin_growth} shows that the spin of MBH1 grows faster, because the richer gas environment at the centre of the galaxy gives MBH1 a higher accretion rate. The difference between the spin growth of MBH1 and MBH3 becomes smaller towards the end of stage~2, because of preferential accretion onto MBH3.

The spin-alignment figure of test case~B provides information that the separations alone do not give. It shows $\cos\theta_{\rm spin-disk}$, the cosine of the angle between each MBH spin and the local gas-disc angular-momentum direction. The panel shows that MBH3 is misaligned during the early stage~1 phase (down to $\cos\theta_{\rm spin-disk}\approx0.84$). After $t\sim200\,{\rm Myr}$, both MBH1 and MBH3 settle into near-alignment with the gas disc and remain so through stage~2.

\subsection{Triplet test case~C (active triplet configuration)}
\label{sec:testC}

In test cases~A and~B, the system settles into a bound inner binary with a distant third MBH, so the chaotic three-body channel is never activated in these simulations. To demonstrate the chaotic three-body treatment of RAMCOAL, we set up a third, more compact configuration (test case~C of Table~\ref{tab:tripletic}). Its three MBHs are placed within $\sim240\,{\rm pc}$ of one another, so that they enter the resolution sphere together and form a subgrid hierarchical triplet. For this more compact test the resolution sphere is set to $R_{\rm sp}=2\Delta x\simeq200\,{\rm pc}$, rather than the fiducial $4\Delta x$ used in tests~A and~B. The three MBHs have equal mass, $M_1\simeq M_2= M_3=3.0\times10^6\,{\rm M_\odot}$. The MBHs start on near-coplanar orbits with small relative velocities ($\lesssim20\,{\rm km\,s^{-1}}$). Their initial spins are small in magnitude but differ in orientation, with MBH1 aligned with the disc angular momentum and MBH2 and MBH3 initially misaligned in random directions.

\begin{figure}
\centering
\includegraphics[width=\columnwidth]{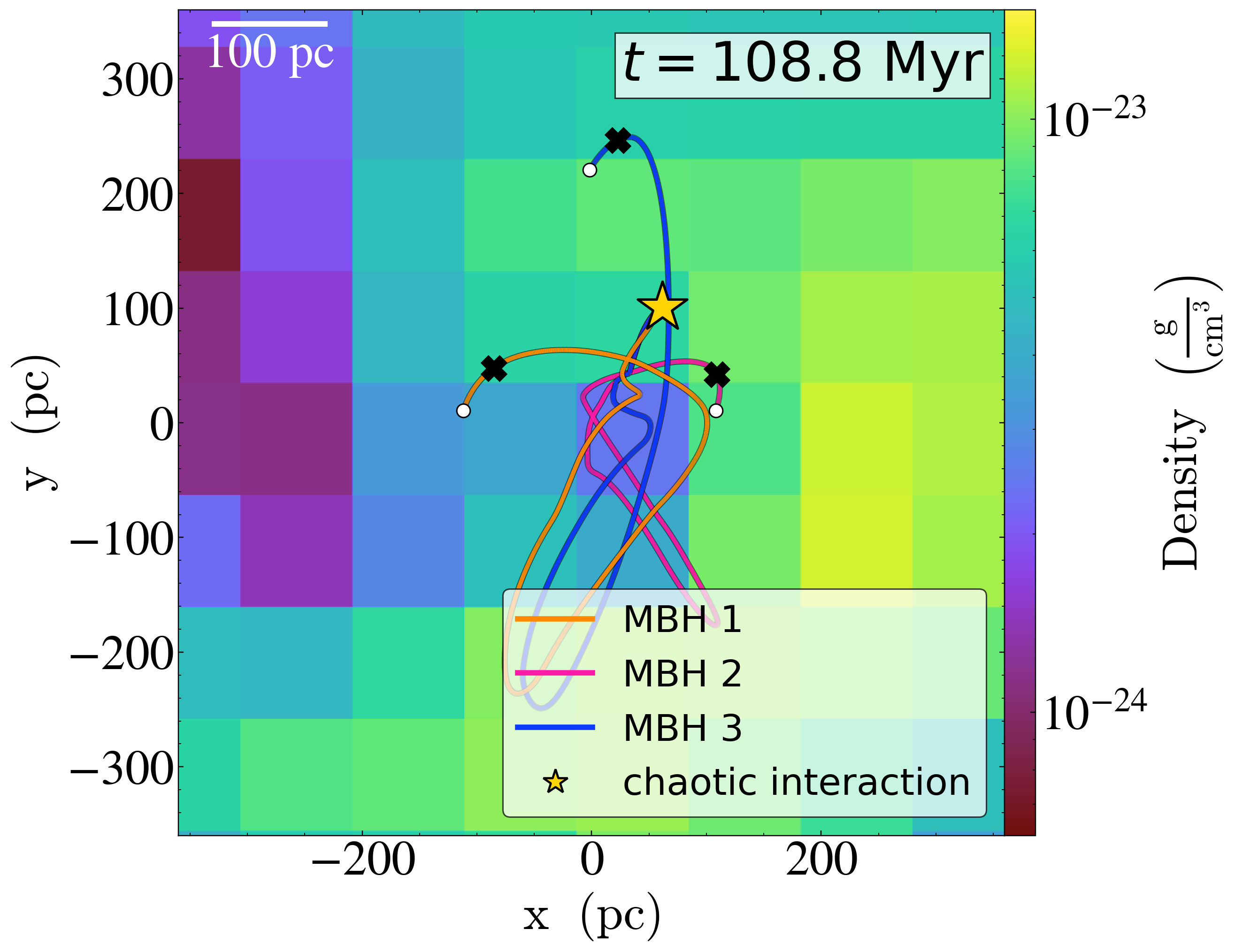}
\caption{Gas-density slice at $t=108.8\,{\rm Myr}$ for triplet test case~C, at the moment of the chaotic triplet interaction (marked by the yellow star). The projected trajectories of the three MBHs (MBH1 orange, MBH2 magenta, MBH3 blue) evolving under dynamical friction in subgrid stage~1 converge into an active triplet (yellow star), where the chaotic three-body interaction happens. }
\label{fig:tripletcaseC_snapshot}
\end{figure}

\begin{figure*}
\centering
\includegraphics[width=\textwidth]{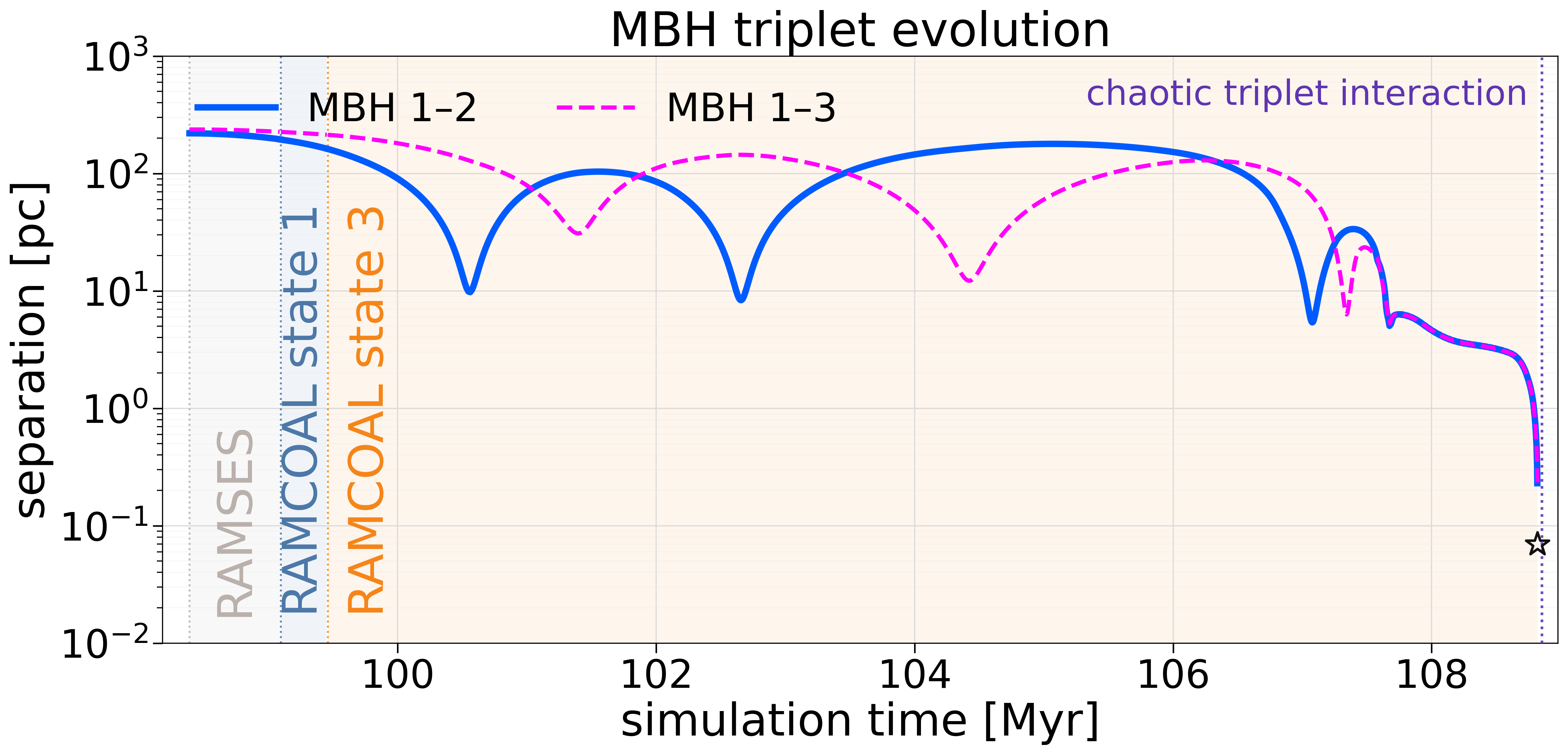}
\caption{Separation evolution of the three MBHs in triplet test case~C, shown as the separations of MBH2 and MBH3 from MBH1 (labelled MBH\,1--2 and MBH\,1--3, as in Figs~\ref{fig:tripletcaseA_separation} and~\ref{fig:tripletcaseB_separation}). The MBH2--MBH3 separation is omitted, and the semi-major axis of the inner MBH2--MBH3 binary at the onset of the chaotic phase is given in Table~\ref{tab:tripletCparams}. The system passes from resolved RAMSES evolution (grey shaded) to a subgrid stage-1 pair of MBH1 and MBH2, while MBH3 is still resolved (blue shaded, RAMCOAL state~1, from $\simeq99.1\,{\rm Myr}$). It then passes to three subgrid stage-1 MBHs (orange shaded, RAMCOAL state~3, from $\simeq99.5\,{\rm Myr}$). The curves end when MBH2 and MBH3 become bound at $\simeq108.82\,{\rm Myr}$ (open star, plotted at the semi-major axis $a_{\rm in}\simeq0.069\,{\rm pc}$ of the new inner binary), shortly before the chaotic triplet interaction at $\simeq108.85\,{\rm Myr}$ (dotted line). The outcome drawn at the chaotic interaction is an exchange merger, in which MBH2 leaves its original partner MBH3 and coalesces with the originally outer MBH1.}
\label{fig:tripletcaseC_separation}
\end{figure*}

\begin{figure}
\centering
\includegraphics[width=\columnwidth]{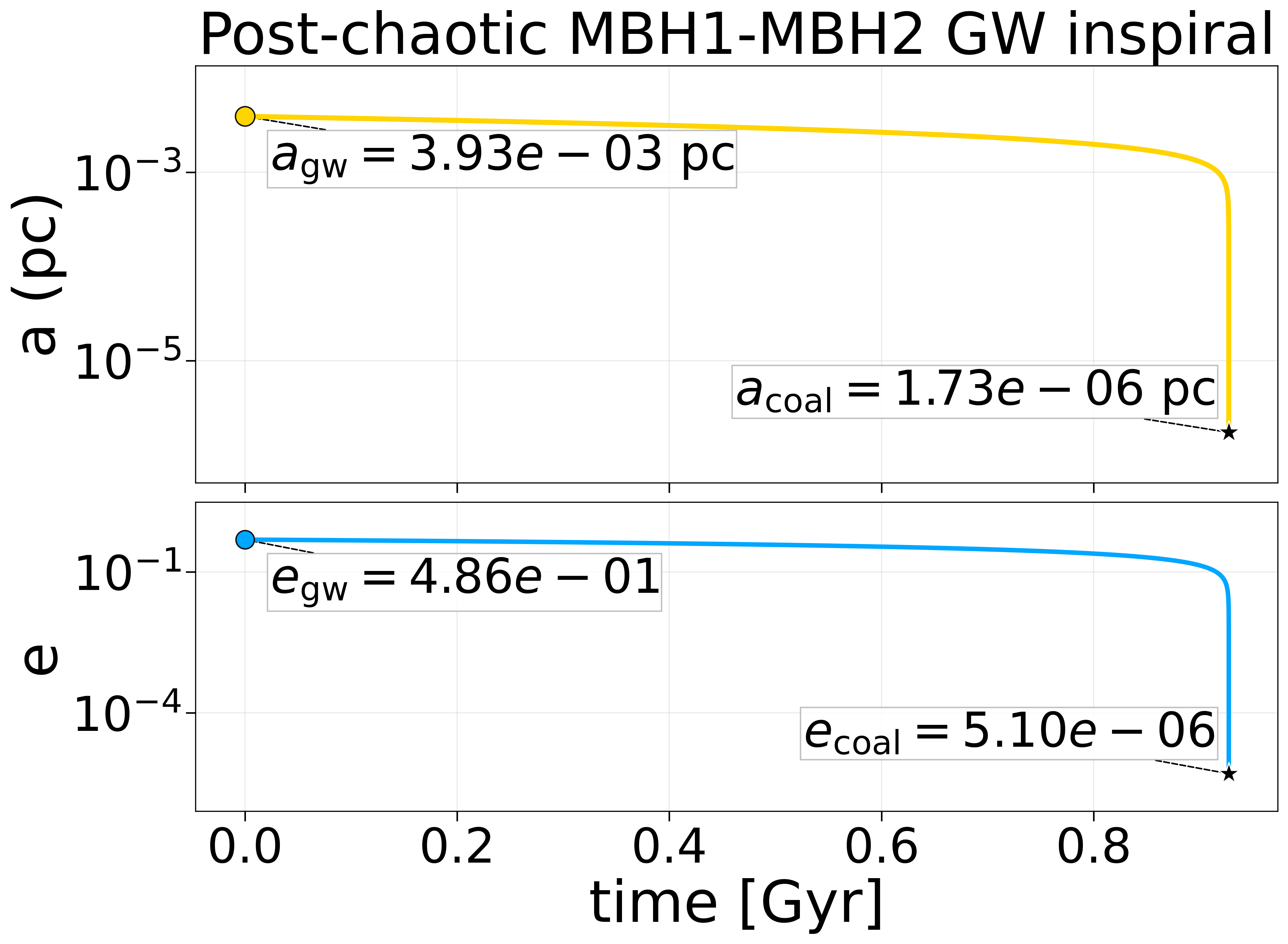}
\caption{Post-processed gravitational-wave inspiral of the MBH1--MBH2 remnant binary in test case~C, reconstructed from the recorded triplet parameters. In the simulation the coalescence is assumed to occur when the chaotic phase starts and the outcome is drawn. Here we follow the GW-driven inspiral in detail, starting the binary at the onset of the GW-dominated phase, $a_{\rm gw}=3.9\times10^{-3}\,{\rm pc}$ (top), and evolving it under GW emission alone. Over $\sim0.9\,{\rm Gyr}$ the orbit hardens to the ISCO scale $a_{\rm coal}=1.7\times10^{-6}\,{\rm pc}$ while the eccentricity (bottom) decreases from $e_{\rm gw}=0.49$ to $e_{\rm coal}\simeq5\times10^{-6}$.}
\label{fig:tripletcaseC_gw}
\end{figure}

Figures~\ref{fig:tripletcaseC_snapshot} and~\ref{fig:tripletcaseC_separation} follow the evolution through the RAMCOAL phases and states. The system begins in resolved RAMSES evolution (stage~0) at separations of $\sim200\,{\rm pc}$. MBH1 and MBH2 enter the subgrid treatment first, as a stage-1 pair (state~1), at $\simeq99.1\,{\rm Myr}$. At $\simeq99.5\,{\rm Myr}$ all three MBHs are within the resolution sphere and are treated as three unbound stage-1 objects (RAMCOAL state~3). Continued dynamical-friction inspiral then binds the MBH2--MBH3 pair into an inner binary, at $\simeq108.82\,{\rm Myr}$ and with $a_{\rm in}\simeq0.069\,{\rm pc}$. As a result, the system becomes a hierarchical triplet, made of the bound MBH2--MBH3 inner binary and the outer MBH1, which is still in stage~1 (state~4). Shortly afterwards, at $\simeq108.85\,{\rm Myr}$, the triplet is promoted to an active, chaotic triplet (state~5) (dotted line in Fig.~\ref{fig:tripletcaseC_separation}). The promotion is triggered by the capture of MBH1 described in Section~\ref{sec:chaotic}. At this step MBH1 satisfies equation~(\ref{equ:outerbound}) with respect to the inner binary, and the stage-2 hardening rate of its orbit exceeds its dynamical-friction decay rate. Its orbit is therefore promoted to stage~2, and the chaotic phase is switched on. The outer-orbit elements of Table~\ref{tab:tripletCparams} are the turning-point estimates of Section~\ref{sec:chaotic} and do not describe the instantaneous position of MBH1. At the trigger, MBH1 is $\simeq9$\,pc from the centre of mass of the inner binary, beyond $a_{\rm out}(1+e_{\rm out})\simeq3.9$\,pc. At this point RAMCOAL computes the three parameters of the outcome table (Table~\ref{tab:tripletCparams}) and draws a single \Bonetti outcome. The drawn channel is an exchange merger (outcome~3, an $M_{\rm pri}+M_{\rm out}$ coalescence). The incoming, originally outer MBH1 merges with the inner-binary primary MBH2, and MBH3, originally a member of the inner pair, becomes the new outer MBH. The coalescing pair (MBH1--MBH2) is therefore not the original inner binary (MBH2--MBH3), because the third body has changed which pair ultimately merges. This is the full stage~0~$\rightarrow$~state~1~$\rightarrow$~state~3~$\rightarrow$~state~4~$\rightarrow$~chaotic-interaction pathway. It shows that every step of the triplet module of RAMCOAL works within a hydrodynamical galaxy simulation.

\begin{table}
\centering
\caption{Parameters of the active MBH triplet in test case~C at the moment the chaotic triplet phase is activated (RAMCOAL state~5, $t\simeq108.8\,{\rm Myr}$). These are the quantities passed to the Bonetti lookup table to draw the outcome. The inner binary is MBH2--MBH3 and the outer MBH is MBH1. The outer-orbit elements $a_{\rm out}$ and $e_{\rm out}$ are the turning-point estimates used by RAMCOAL at the capture of MBH1 (Section~\ref{sec:chaotic}).}
\label{tab:tripletCparams}
\setlength{\tabcolsep}{4pt}
\begin{tabular}{ll}
\toprule
Quantity & Value \\
\midrule
Component masses $M_1,M_2,M_3$ & $\simeq3.0\times10^{6}\,{\rm M_\odot}$ each \\
Inner binary & MBH2--MBH3 \\
Inner mass ratio $q_{\rm in}$ & $1.00$ \\
Outer mass ratio $q_{\rm out}$ & $0.50$ \\
Inner semi-major axis $a_{\rm in}$ & $0.069\,{\rm pc}$ \\
Outer semi-major axis $a_{\rm out}$ & $2.59\,{\rm pc}$ \\
Inner eccentricity $e_{\rm in}$ & $0.19$ \\
Outer eccentricity $e_{\rm out}$ & $0.49$ \\
Mutual inclination $\cos i$ & $1.00$ \\
Spins $a_{\ast,1},a_{\ast,2},a_{\ast,3}$ & $1.4\times10^{-5},\,3.2\times10^{-5},\,1.2\times10^{-5}$ \\
Drawn \Bonetti outcome & exchange merger (MBH1$+$MBH2) \\
\bottomrule
\end{tabular}
\end{table}

Table~\ref{tab:tripletCparams} lists these MBH parameters at the onset of the chaotic three-body interaction. The three MBHs are still of comparable mass ($\simeq3\times10^6\,{\rm M_\odot}$), giving an equal-mass inner binary ($q_{\rm in}=1$) and an outer mass ratio $q_{\rm out}=0.5$. At activation, the inner binary has hardened to $a_{\rm in}\simeq0.07\,{\rm pc}$ and has a modest eccentricity $e_{\rm in}\simeq0.19$, while the outer MBH is on a wider ($a_{\rm out}\simeq2.6\,{\rm pc}$), more eccentric ($e_{\rm out}\simeq0.49$), coplanar ($\cos i=1$) orbit. The spins remain small ($a_\ast\sim10^{-5}$) because accretion has stayed sub-Eddington throughout. From the corresponding \Bonetti cell the model draws the exchange-merger channel and coalesces MBH1 with MBH2. This example shows the complete sequence of the three-body treatment, from the computation of the table parameters and the weighted outcome draw to the reconfiguration of the surviving system, all executed self-consistently within the galaxy simulation.

A chaotic MBH triplet has been shown to speed up the coalescence of an MBH binary relative to GW emission alone, by driving the binary to high eccentricity and much smaller separations, where GW emission is faster. Studies have shown that Kozai--Lidov excitation shortens the GW merger time by more than an order of magnitude in over half of near-equal-mass systems \citep{Blaes2002}. Direct chaotic-triple simulations report coalescence fractions as high as $\sim85\%$ \citep{Hoffman2007}. The \Bonetti post-Newtonian suite, on which the RAMCOAL outcome probabilities are based, finds that $\sim20$--$30\%$ of otherwise stalled binaries coalesce within a Hubble time \citep{Bonetti2018}. Because RAMCOAL does not resolve the chaotic three-body interaction itself, we assume that the selected coalescence occurs as soon as the chaotic phase starts. At that moment RAMCOAL records the relevant parameters of the triplet configuration (Table~\ref{tab:tripletCparams}). This allows the detailed three-body interaction and the subsequent GW-driven inspiral to be reconstructed in post-processing whenever a detailed coalescence track is required.

The \Bonetti outcomes~1, 3, and~5 do not involve an ejection. In these outcomes, two of the MBHs coalesce, and the merger remnant is kept together with the third MBH. The system then becomes an ordinary RAMCOAL stage-2 binary, with the remnant as one component and the third MBH as the other. This binary evolves under the stage~2 hardening mechanisms. In the stalled outcome (outcome~7), both the inner and the outer binary follow the stage~2 evolution. For the case~C outcome shown here (outcome~3), the MBH1--MBH2 merger produces a remnant $M_{12}$, while MBH3 remains bound to it and continues to evolve as a new RAMCOAL binary. This differs from the ejection channels, in which the third MBH is removed from the subgrid system. In the coalescence channel the post-triplet evolution continues through the standard stage~2 binary hardening phase.

Figure~\ref{fig:tripletcaseC_gw} illustrates such a post-processed reconstruction for the MBH1--MBH2 remnant of test case~C. Starting from the recorded configuration, the remnant binary is placed at the onset of the GW-dominated phase, $a_{\rm gw}=3.9\times10^{-3}\,{\rm pc}$, and is then evolved under GW emission alone. Over $\sim0.9\,{\rm Gyr}$ the orbit hardens to the ISCO scale $a_{\rm coal}=1.7\times10^{-6}\,{\rm pc}$ while the eccentricity decreases from $e_{\rm gw}=0.49$ to $e_{\rm coal}\sim5\times10^{-6}$. In the simulation, the coalescence is taken to be instantaneous once the chaotic three-body interaction is activated, but the full GW-driven inspiral can be reconstructed from the stored parameters whenever needed.

\begin{figure}
\centering
\includegraphics[width=\columnwidth]{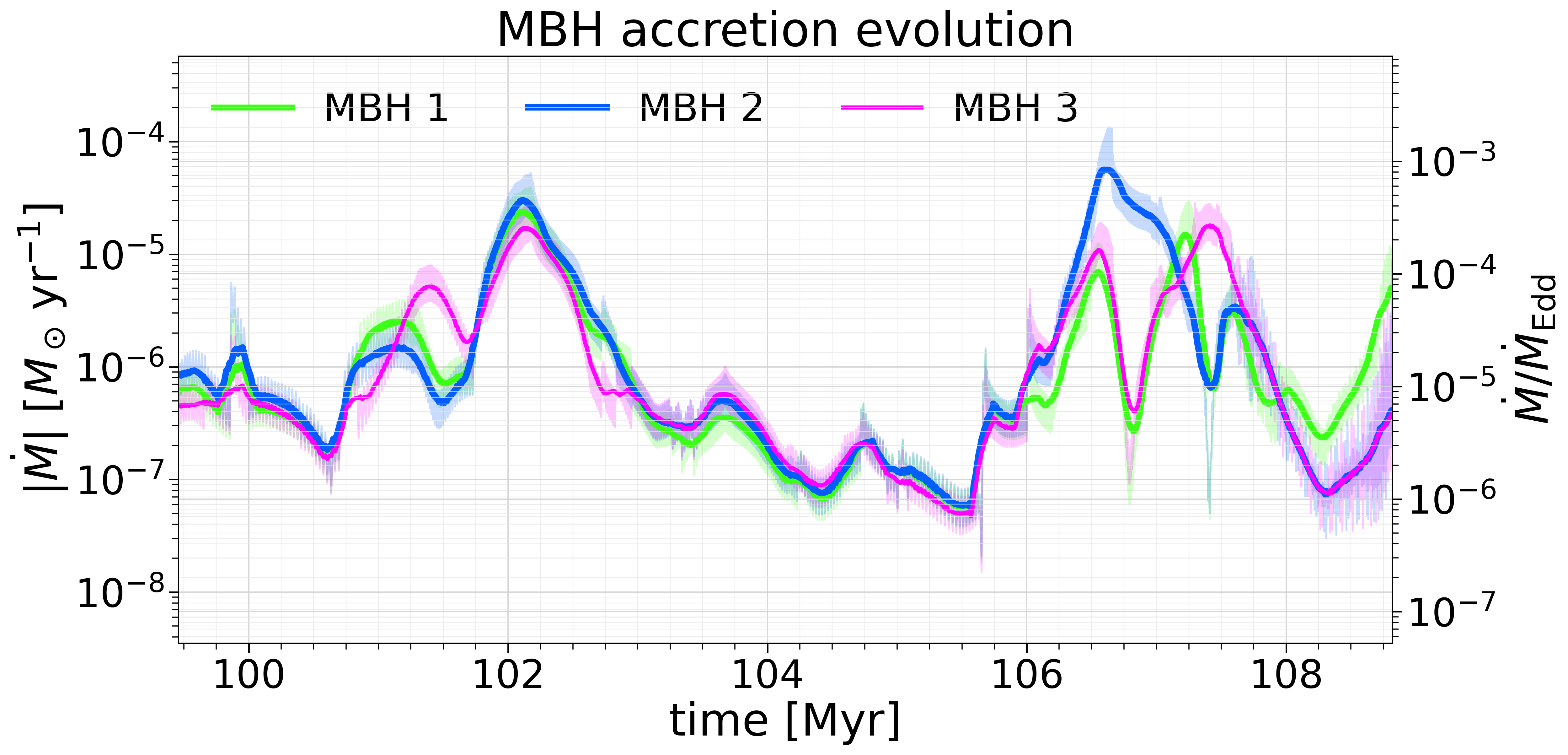}
\caption{Component accretion-rate evolution for triplet test case~C (MBH1 green, MBH2 blue, MBH3 magenta). The left axis gives the absolute rate $\dot M$ and the right axis the Eddington ratio $\dot M/\dot M_{\rm Edd}$. Raw histories are shown in light shading, and running means over a $0.18$\,Myr window are shown as solid lines. Accretion remains strongly sub-Eddington throughout.}
\label{fig:tripletcaseC_accretion}
\end{figure}

\begin{figure}
\centering
\includegraphics[width=\columnwidth]{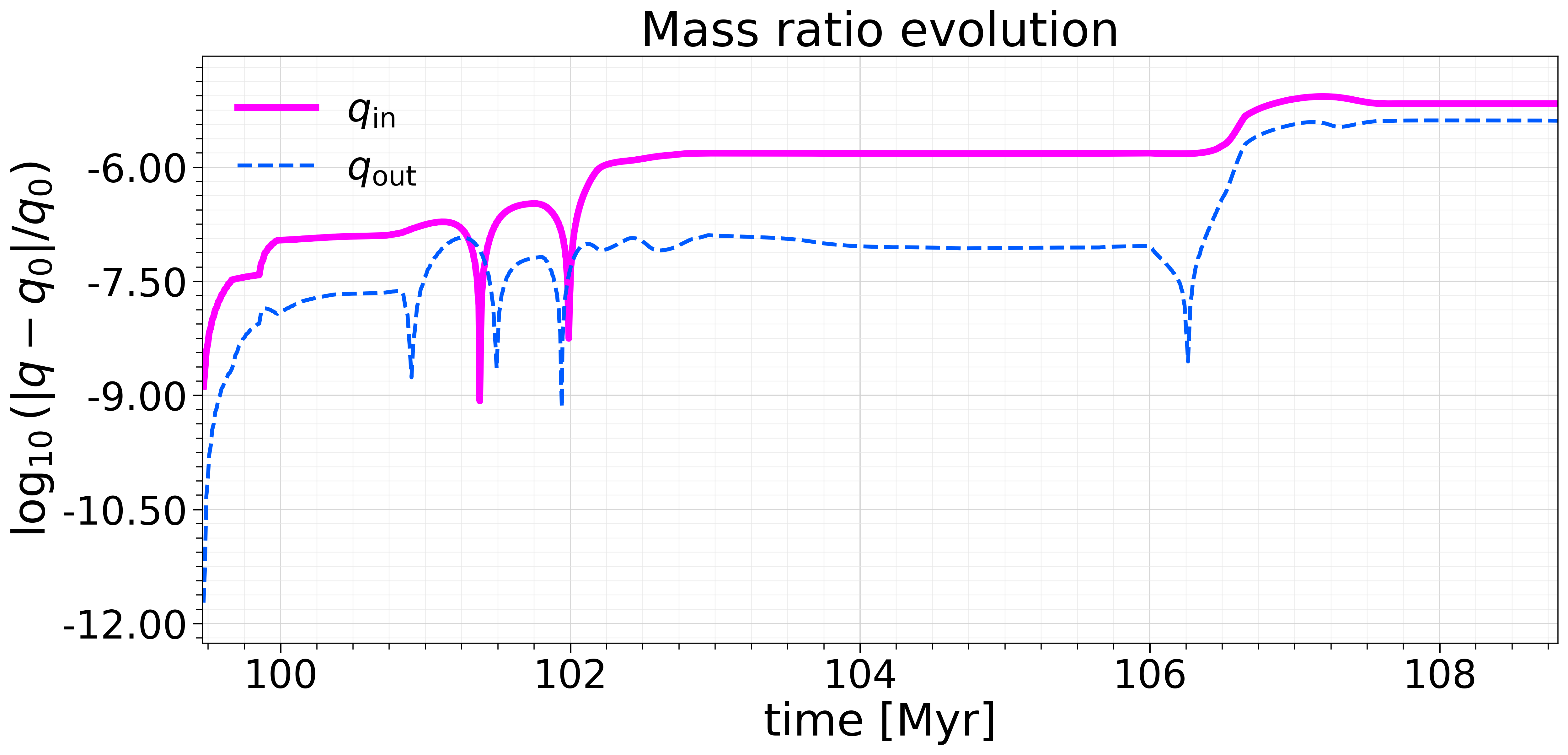}
\caption{Mass-ratio evolution for triplet test case~C, shown as fractional deviations from the initial values of the inner-pair ratio $q_{\rm in}=\min(M_2,M_3)/\max(M_2,M_3)$ (initial $q_0=1$) and the outer ratio $q_{\rm out}=M_1/(M_2+M_3)$ (initial $q_0=0.5$).}
\label{fig:tripletcaseC_massratio}
\end{figure}

\begin{figure}
\centering
\includegraphics[width=\columnwidth]{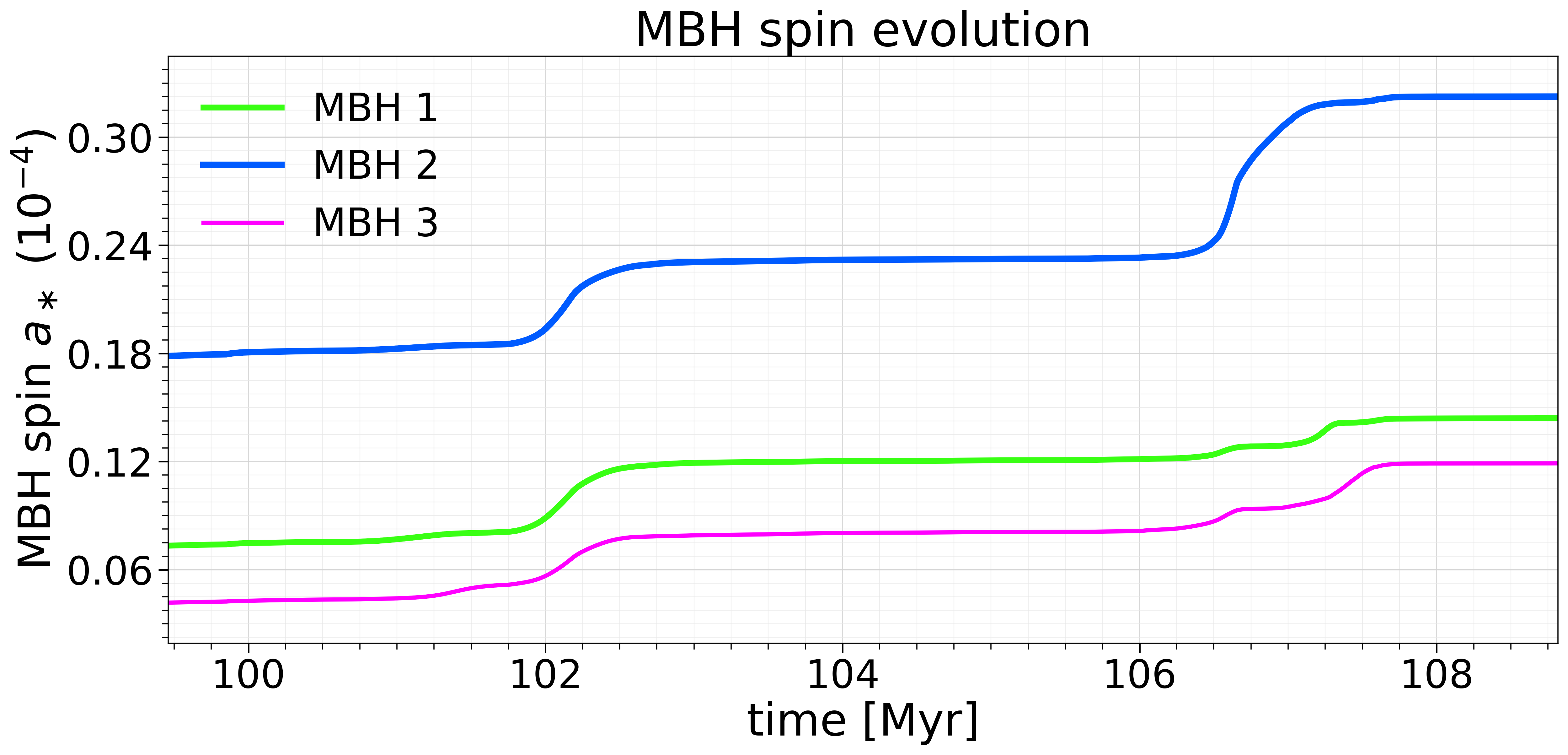}
\caption{Spin-magnitude evolution for triplet test case~C (MBH1 green, MBH2 blue, MBH3 magenta). The spins grow slowly, in steps tied to the accretion episodes, and remain small ($a_\ast\sim10^{-4}$).}
\label{fig:tripletcaseC_spin}
\end{figure}

The accompanying accretion, mass-ratio, and spin evolution of the unbound triplet phase (RAMCOAL state~3) are shown in Figs~\ref{fig:tripletcaseC_accretion}--\ref{fig:tripletcaseC_spin}. All three MBHs accrete at sub-Eddington rates. The fluctuations in these rates reflect the orbital motion of the MBHs through relatively gas-rich and gas-poor regions of the galaxy. The most prominent accretion episodes occur near $102$ and $106$--$107\,{\rm Myr}$. The mass ratios evolve only mildly from their initial values. The spin magnitudes grow slowly and in steps tied to the accretion episodes, and they remain small ($a_\ast\sim10^{-4}$) throughout. This is consistent with the low-accretion, MAD spin regime discussed for test case~A.

\subsection{Comparison with previous work}

A distinctive feature of tests~A and~B is that an inclined incoming MBH can change which pair ultimately coalesces. Test~C shows the most extreme case of this effect, a chaotic three-body interaction whose drawn outcome is an exchange merger. Idealized scattering and post-Newtonian triplet simulations have long found that a bound MBH binary perturbed by a third MBH can undergo exchanges, prompt or Kozai--Lidov-driven mergers, and ejections \citep{Hoffman2007,Bonetti2018,Bonetti2019}. The coalescing pair therefore need not be the original binary. \citet{Hoffman2007} found that, in most of their three-body integrations of triplet MBH systems in massive galaxies, the original binary coalesces. In a minority of runs, a new and typically more eccentric binary forms, and the lightest MBH is frequently displaced to a wide orbit or ejected. The outcomes of all three test cases fall within this qualitative range. In test case~B, MBH3 is captured by the primary, and in test case~A, MBH3 remains on a wide orbit. In test case~C, an exchange merger is triggered, in which the outer MBH displaces an original inner member and coalesces with the primary.

Cosmological simulations that resolve or model MBH triplets report similar behaviour. Using the regularized KETJU code in a cosmological zoom-in, \citet{Mannerkoski2021} followed an MBH triplet that formed in a multiple-galaxy merger. They identified three-body interactions and Kozai--Lidov oscillations that displaced one MBH onto a kiloparsec-scale orbit. One pair coalesced a few gigayears after the galaxy merger, while the remaining binary persisted at parsec separations. RAMCOAL differs from KETJU in that it does not resolve the few-body dynamics directly. It uses calibrated subgrid prescriptions instead, which makes it less accurate during the resonant phase but applicable to larger samples. The cost of this approach is quantified in Section~\ref{sec:cosmo}. In a cosmological simulation of the COSCOAL suite, the subgrid integration takes $\simeq1\%$ of the total run time, at a mass resolution where a regularized treatment would not be applicable.

\citet{Sayeb2024Triples} similarly found, in the Illustris simulations, that a non-negligible fraction of binaries encounter a third MBH before coalescence, and that such encounters can either drive otherwise stalled binaries to coalescence or leave wandering MBHs. The three RAMCOAL tests reproduce several of these behaviours in a controlled isolated-galaxy setting, including gigayear-scale delays and an evolution path and merger time that depend on the orbital configuration.

These comparisons should be read as qualitative consistency checks rather than quantitative validation. The three experiments are single realizations in one host galaxy. A statistical comparison with the merger fractions, ejection fractions, and eccentricity distributions reported by \citet{Hoffman2007}, \citet{Bonetti2019}, \citet{Mannerkoski2021}, and \citet{Sayeb2024Triples} would require a population of RAMCOAL triplets and is deferred to future work. 

\section{Computational cost}
\label{sec:cosmo}

The main motivation for a subgrid treatment of MBH pairs and multiplets is that it can be applied to cosmological volumes, in which many MBHs and multiplets must be followed at the same time, at a mass resolution that does not allow a direct integration of their few-body dynamics. We have produced a suite of COSCOAL cosmological simulations with the full RAMCOAL model (Li et al., in prep., which detailed analysis is beyond the scope of this paper), that we use, here in this paper, only to quantify the computational cost of RAMCOAL and of its triplet module. We summarize its performance in a $15\,h^{-1}\,{\rm Mpc}$ box size simulation of the suite (hereafter COSCOAL-15), which is described in Appendix~\ref{app:cosmo}. 

At every fine time step, each subgrid system is advanced through ten integration sub-cycles. Single-core measurements for pairs and triplet systems in a range of environments give a cost of 40--70\,$\mu$s per call, with little dependence on the configuration. A triplet system costs approximately $10\%$ more than a pair. Ten calls therefore require approximately 0.4--0.7\,ms of single-core time, independently of the particle resolution of the simulation. At $z\simeq2.4$, a fine step in COSCOAL-15 takes approximately 13\,s, so this integration contributes approximately 0.003--0.005\% of the step time per subgrid system.

The additional triplet operations comprise the classification and reduction described in Section~\ref{sec:mergerclass} and the selection of a \Bonetti outcome. These are performed only when subgrid systems meet or a triplet becomes active, and their cost is negligible compared with the orbital integration. Bound binaries in stage~2 require a sub-step $\min(P_{\rm peri},\epsilon\,\Delta t)$, where $P_{\rm peri}$ is the orbital period at pericentre, $\Delta t$ is the hydrodynamical time step, and $\epsilon$ is a free parameter. Their cost per fine step therefore depends on $\epsilon$. The workload in the COSCOAL-15 interval examined here is dominated by stage-1 systems. 

COSCOAL-15 accumulated 66.1\,h of wall-clock time on 960 MPI tasks between $z=50$ and $z=2.31$. Approximately 230 subgrid systems formed during this interval, most of which remained present at its end. Three-MBH systems accounted for $34\%$ of the subgrid updates. All subgrid systems were advanced by a single MPI task, so their integration time contributed directly to the wall-clock time. We estimate this contribution from the logged number of integration calls and the measured single-core cost per call.

\begin{figure}
\centering
\includegraphics[width=\columnwidth]{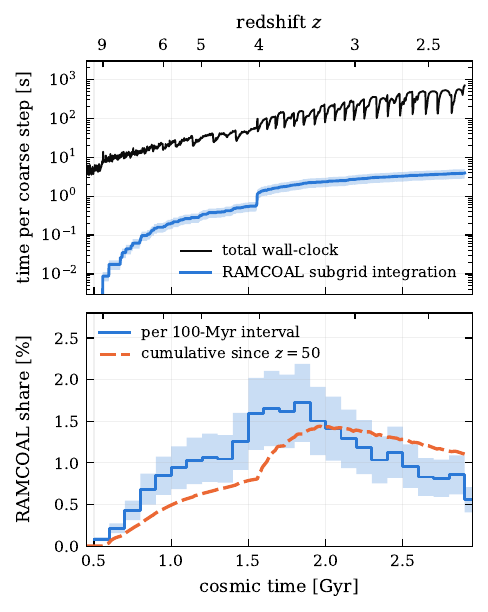}
\caption{Computational cost of RAMCOAL in COSCOAL-15. The upper panel shows the wall-clock time per coarse step of the full simulation (black) and the estimated subgrid integration time (blue). The shaded band corresponds to the measured cost per integration call, 40--70\,$\mu$s. Periodic spikes reflect output and load-balancing steps. The jump at $z\simeq4$ marks the release of the finest refinement level, which doubles the number of fine steps per coarse step. The lower panel shows the estimated fraction of wall-clock time spent in the subgrid integration over 100-Myr intervals (blue) and cumulatively since $z=50$ (dashed orange).}
\label{fig:cosmo_cost}
\end{figure}

Table~\ref{tab:cosmocost} summarizes the computational
performance of RAMCOAL in COSCOAL-15, while
Figure~\ref{fig:cosmo_cost} shows the estimated subgrid integration
time alongside the timing of the full simulation. Over the run,
the subgrid integration accounts for $1.1\%$ of the total
wall-clock time, with a range of 0.8--1.4\% corresponding to the
measured call costs. Averaging the fine-step fractions within
each coarse step gives a median fraction of $1.0\%$ (0.7--1.3\%)
and a maximum of $3.2\%$ (2.4--4.1\%). At $z\simeq2.4$, there
are approximately 200 subgrid systems, which require about
0.1\,s within a fine step lasting approximately 13\,s. The
relative overhead peaks near $z\simeq3.5$ and subsequently
declines as the cost of hydrodynamics, gravity, and
stellar-particle evolution grows faster than the number
of subgrid systems.

\begin{table}
\centering
\caption{Computational performance of RAMCOAL in the COSCOAL-15 simulation between $z=50$ and $z=2.31$. The fraction of wall-clock time of RAMCOAL are estimated from the number of integrator calls recorded in the run and the measured single-core cost of one call (range in brackets). The fractions of a fine time step are averaged over the fine steps of each coarse step.}
\label{tab:cosmocost}
\setlength{\tabcolsep}{4pt}
\begin{tabular}{lc}
\toprule
Quantity & Value \\
\midrule
Box size & $14.7\,h^{-1}$\,Mpc ($21.1$\,Mpc) \\
Dark-matter particles & $256^3$, $1.8\times10^7\,{\rm M_\odot}$ \\
$R_{\rm sp}$ (physical) & $\simeq1.3$\,kpc \\
Redshift range & $50\rightarrow2.31$ \\
MPI tasks and wall-clock time & 960 and 66.1\,h \\
Subgrid systems at $z\simeq2.4$ & $\simeq200$ \\
Fraction of updates for three-MBH systems & 34\% \\
Fraction of total wall-clock time & 1.1\% (0.8--1.4\%) \\
Median fraction of a fine time step & 1.0\% (0.7--1.3\%) \\
Maximum fraction of a fine time step & 3.2\% (2.4--4.1\%) \\
\bottomrule
\end{tabular}
\end{table}

\section{Discussion}

The preceding sections return to two questions raised in the Introduction. The first is how subgrid MBHB and triplet evolution can be followed without identifying a numerical sink merger with a physical coalescence. The second is how such a model can retain the environmental information needed to connect MBH orbital decay with accretion, feedback, spin evolution, recoil, and the GW source population. The RAMCOAL triplet extension addresses these questions with a 3-stage binary and triplet model within a RAMSES simulation. In this way, it follows the galactic environment explicitly and treats the final orbital evolution below the grid scale. It is complementary to analytic theory, semi-analytic population modelling, hydrodynamical simulations, and direct $N$-body simulations, each of which is better suited to a different part of the problem.

\subsection{Relation to existing modelling approaches}

Existing treatments of MBHB evolution differ mainly in where they place the boundary between resolved dynamics and subgrid orbital modelling. Classical analytic and semi-analytic theory provides the physical foundation. In this framework, binary evolution is described through dynamical friction, stellar hardening, viscous drag in a circumbinary disc, and GW emission, with transitions between regimes determined by idealized or locally estimated environmental quantities \citep{Begelman1980,Quinlan1996,Peters1964,SesanaKhan2015}. RAMCOAL follows the same physical sequence, but takes the environmental inputs from the hydrodynamical simulation rather than imposing them solely as external parameters. This allows the hardening model to respond to the simulated host galaxy, although the sub-resolution orbital terms necessarily remain prescription based.

Semi-analytic and semi-empirical population models have different priorities. By combining MBH populations, merger trees, and parametrized hardening or delay prescriptions, they can examine how assumptions about seeding, growth, galaxy assembly, and environmental delay propagate into PTA and LISA predictions \citep{Barausse2012,IzquierdoVillalba2026}. Their main advantage is that broad parameter spaces can be explored at modest computational cost, with the detailed environmental history of an individual binary typically reduced to an effective timescale or phenomenological prescription. RAMCOAL operates at the opposite end of this trade-off, following each subgrid binary together with the evolving gas, stars, feedback, and remnant dynamics of its host.

Hydrodynamical galaxy simulations follow the evolving environment that semi-analytic models generally approximate. They can follow gas inflow, star formation, AGN feedback, galaxy mergers, and MBH growth self-consistently on resolved scales \citep{Chapon2013,Pfister2019,Massonneau2023,Beckmann2025}. Their difficulty is the remaining scale separation, since even refined simulations do not resolve the regime in which hard binaries interact with stellar loss cones, circumbinary discs, and GW emission. A standard sink-particle merger is therefore a numerical event, not necessarily the physical coalescence of two MBHs. RAMCOAL changes this interpretation by treating the numerical close-pair event as the starting point of a subgrid evolution. In this way, the hydrodynamical context is retained, while the time at which the grid-scale merger criterion is met is no longer treated as the physical merger time.

Direct $N$-body simulations lie at the other extreme. These methods follow close MBH and stellar dynamics with high force accuracy and can describe stellar scattering, eccentricity evolution, and post-Newtonian binary evolution in selected systems \citep{Sesana2011,Varisco2021,Gualandris2022,Mannerkoski2023}. KETJU is particularly relevant because it embeds a regularized treatment of small-scale MBH dynamics within a galaxy-simulation framework \citep{Mannerkoski2023}. RAMCOAL adopts a more approximate description of close few-body dynamics and is aimed instead at applications in which many galaxies or cosmological environments are evolved together with gas physics, accretion, feedback, and delayed coalescence. As quantified in Section~\ref{sec:cosmo}, the subgrid integration costs $\simeq0.5$\,ms of single-core time per subgrid system and fine time step, independently of the particle resolution. In a cosmological simulation, it takes $\simeq1\%$ of the total wall-clock time. The two methods are also designed for different problems. RAMCOAL includes, through subgrid prescriptions, the gas-driven orbital evolution, the accretion onto each MBH of a binary, the spin evolution, and the feedback below the resolution limit. A gravitational few-body integration does not include these processes, even when the stellar dynamics is fully resolved. The two approaches are complementary. Direct-dynamics simulations can help calibrate subgrid prescriptions, while RAMCOAL can carry calibrated physics into larger samples of simulated galaxies.

Three-body scattering simulations quantify merger, exchange, stalled, and ejection outcomes \citep{Bonetti2018}, while numerical-relativity recoil formulae relate the remnant mass and spin configuration to possible kicks \citep{Lousto2013}. RAMCOAL uses these results because large-scale simulations cannot directly integrate every chaotic three-body interaction or numerical-relativity merger. The trade-off is that each library result is mapped onto the instantaneous subgrid state, while environmental evolution during the resonant interaction is represented through the surrounding subgrid model.

This comparison shows both what RAMCOAL shares with previous approaches and how it differs from them. RAMCOAL adopts the physical sequence that underlies classical MBHB theory, from resolved orbital decay and subgrid dynamical friction to hard-binary evolution, environmental hardening, and GW inspiral. The main difference lies in how the model is coupled to the simulation. Rather than assigning the full delay after a numerical merger catalogue has been constructed, RAMCOAL tracks the evolution of each binary or triplet on the fly in the parent hydrodynamical simulation. The local gas density, stellar properties, accretion state, feedback state, and remnant motion can therefore continue to influence the subgrid system. RAMCOAL is thus closer to hydrodynamical galaxy simulations than to purely analytic delay models, while retaining the efficiency and interpretability of semi-analytic prescriptions.

\subsection{Connection to current observations}

Several observational probes connect directly to the quantities tracked in RAMCOAL. These quantities include the evolution of the eccentricity, mass ratio, spin, and orbital angular momentum during the late inspiral and at coalescence. Evolution of this kind sets the amplitude and spectral shape of the nanohertz gravitational-wave background now being probed by pulsar timing arrays \citep{Agazie2023,EPTA2023,Reardon2023,Xu2023,Agazie2023SMBHB}. MBHs that remain on wide orbits can become offset or wandering MBHs. If such an MBH continues to accrete, it could appear as a dual or offset AGN of the kind targeted by current surveys \citep{DeRosa2020,Li2022TNG50,Li2023TNG50DualAGN,Saeedzadeh2024,Perna2025}. Although no such ejection occurs in our tests, MBHs ejected in a strong triplet interaction are, like recoiling MBHs, another source of offset or wandering MBHs. RAMCOAL retains the MBHs ejected in triplet interactions and tracks their mass, accretion state, and offset, instead of spuriously merging them at the numerical merger step.

In each of these regimes, the model is useful because it provides the pre-coalescence parameters that the predictions require. These are the residence times and eccentricities for the nanohertz band, the masses, spins, eccentricity, and time to merger for LISA and TianQin, and the gas, accretion, and recoil state relevant to electromagnetic counterparts. All of them are tied to a physical coalescence rather than to a numerical sink merger, and the triplet extension can change which pair merges and when \citep{Bonetti2018}.

\subsection{Limitations}
\label{sec:limitations}

Several limitations should be stated explicitly. RAMCOAL remains a sub-resolution model. It uses local properties measured in the simulation, but it does not resolve the stellar loss cone, the full circumbinary-disc structure, or the resonant few-body dynamics of the chaotic triplet phase. Its predictions therefore depend on the adopted prescriptions for dynamical friction, stellar hardening, gas torques, GW evolution, recoil, and ejection.

The triplet outcome model also inherits the finite resolution of the \Bonetti-based probability grid. Several interaction times are set to representative values rather than to the mean interaction times of the \Bonetti bins, and the triplet-ejection velocity is set to an approximate value scaled to the escape speed rather than drawn from the $N$-body simulation data. The \Bonetti scattering survey also assumes that the inner binary is more massive than the incoming third MBH, $M_3<M_1+M_2$ (equivalently $q_{\rm out}\le1$). The present model is therefore strictly valid in this regime, which is expected to be the most common configuration. The opposite case, in which the intruder outweighs the inner binary ($M_3>M_1+M_2$), lies outside the tabulated grid. In the current implementation such triplets are handled by the reciprocal-$q_{\rm out}$ fallback of Section~\ref{section:bonetti}, which is an extrapolation of the tabulated outcomes and not a calibrated prescription. 
Likewise, the fixed chaotic timer, $t_{\rm chaos}=100$\,yr, is an ad hoc prescription for scheduling the sampled outcome, not a physically calibrated interaction or nuclear escape timescale (Section~\ref{section:bonetti}).

The model further inherits the assumptions of the parent hydrodynamical simulation. The measured gas density, stellar density, velocity dispersion, accretion rate, and feedback state depend on numerical resolution, refinement strategy, sink-particle treatment, and the underlying galaxy-formation model. RAMCOAL reduces the need to equate numerical sink mergers with physical coalescences of MBHs, but it cannot remove uncertainties associated with MBH seeding, low-mass galaxy completeness, or the resolved structure of the nuclear environment. The spin magnitudes remain sensitive to the adopted prescriptions for alignment and the accretion history. We also note that the direction of the gas angular momentum that sets the spin alignment is measured on the resolution scale (Section~\ref{sec:accretion}).

Two further approximations matter mainly for cosmological applications. First, encounters that bring more than three MBHs into one resolution sphere are reduced to at most three by merging the most massive MBHs (Sections~\ref{sec:mergerclass} and~\ref{sec:highorder}). These forced mergers are flagged in the merger catalogue. Their frequency depends on the resolution and on the seeding model, and it will be quantified with the COSCOAL simulations. Second, at the kiloparsec resolution of cosmological volumes, stage~1 replaces the resolved orbit over a wide range of separations. The pairing times of light MBHs therefore depend on the accuracy of the subgrid dynamical-friction prescription and of the kernel-averaged environment on which it relies.

The three test cases should be read as controlled single realizations in one host galaxy rather than a representative sample. Tests~A and~B demonstrate the three-stage evolution of binaries and the dependence of the outcome on geometry. Test~C follows a triplet through its complete RAMCOAL evolution, from three resolved MBHs, through the stage-1 and hierarchical phases, to a chaotic three-body interaction whose \Bonetti-drawn outcome is executed. This tests every step of the triplet model within a galaxy simulation. Because these are individual runs, neither the absolute delay times nor the residual eccentricities should be taken as representative. The accretion rates also remain sub-Eddington with nearly constant mass ratios, so the preferential-accretion and feedback-regulation channels are only weakly probed here. A statistical characterization including merger, exchange, and ejection fractions and the resulting eccentricity and spin distributions requires a population of RAMCOAL triplets.  Full cosmological simulations (Section~\ref{sec:cosmo}) using the model presented here will be described by Li et al. (in preparation).

\section{Conclusions}

We have presented an extension of RAMCOAL that follows subgrid MBH triplets in hydrodynamical galaxy simulations. The extension includes MBH triplets in the 3-stage subgrid model previously developed for binaries. It keeps a numerical close-pair event distinct from a physical coalescence and adds a criterion for when a third MBH alters the binary. It also represents hierarchical triples, chaotic three-body outcomes, exchanges, and ejections through a classification into seven states. The orbital evolution is coupled to the quantities that shape observable populations. Circumbinary inflow is partitioned between the cavity reservoir and the two mini-discs, spin alignment and remnant spin are followed, and recoiling or ejected MBHs are returned to the resolved sink population. The model therefore tracks both whether MBHs coalesce and the masses, spins, kicks, and offsets of the remnants, which is important for low-frequency GW surveys.

Our main points are as follows.
\begin{itemize}

\item \textit{Three-stage subgrid evolution.} MBHs evolve as RAMSES sink particles, then through a stage-1 dynamical-friction phase, and finally as stage-2 bound binaries. These binaries are hardened by stellar scattering, gas torques, circumbinary-disc coupling, and gravitational-wave emission. In this way, a numerical sink merger is not equated with a physical coalescence.
\item \textit{Triplet prescription.} When a hierarchical triplet turns chaotic, the model maps it onto the scattering outcomes of \citet{Bonetti2018} and updates the surviving configuration, distinguishing inner-binary mergers, exchange-driven mergers, stalled triplets, and three-body ejections. The seven-state classification ensures that any combination of two subgrid systems is mapped onto a valid state.
\item \textit{Coupled physics.} The orbital evolution is coupled to the circumbinary mass partition, feedback-regulated mini-disc accretion, spin alignment, remnant spin, and the recoil or ejection of MBHs back into the sink population. The model therefore tracks the orbital evolution, masses, spins, and recoils of the MBHs on the fly, all the way to coalescence. This information is needed to generate accurate and physically motivated waveforms. 

\item \textit{Test cases.} Three isolated-galaxy tests demonstrate the model. Two contrasting geometries (in-plane and inclined) show that the encounter geometry alone can change which pair coalesces and how long this takes. A third, compact configuration follows a triplet through its complete RAMCOAL evolution, from three resolved MBHs to a chaotic three-body interaction and its \Bonetti exchange-merger outcome. For the first time, we have followed the dynamical evolution of an MBH triplet self-consistently within a hydrodynamical galaxy simulation, from three resolved MBHs all the way to coalescence. During this evolution, the triplet passes through the subgrid inspiral and the chaotic three-body interaction. This allows RAMCOAL to include triplet-driven coalescences in gravitational-wave merger catalogues for PTA, LISA, and TianQin, with physically motivated delays and with the remnant masses, spins, eccentricities, and recoils they produce.

\item \textit{Outlook.} The model is designed for cosmological simulations, to produce merger catalogues with physical delays, pre-coalescence parameters, and host-galaxy properties for PTA, LISA, TianQin, and multimessenger studies. 
\end{itemize}

\section*{Acknowledgements}

The Flatiron Institute is a division of the Simons Foundation. RSB acknowledges support from UKRI Future Leaders Fellowship MR/Y015517/1. This work was made possible by funding from the French National Research Agency (grant ANR-21-CE31-0026, project MBH\_waves) and from the Centre National d’Etudes Spatiales.  We thank Matteo Bonetti for sharing the three-body scattering data used in this work.

Disclosure of Delegation to Generative AI

The authors declare the use of generative AI in the research and writing process. According to the GAIDeT taxonomy (2026), the following tasks were delegated to GAI tools under full human supervision:

- Literature search and systematization
- Code optimization
- Data analysis
- Visualization
- Generating parts of the manuscript text 
- Proofreading and editing
- Adapting and adjusting emotional tone

The GAI tool used was: ChatGPT GPT-6 Astra, Claude Opus 5.5.
Responsibility for the final manuscript lies entirely with the authors.
GAI tools are not listed as authors and do not bear responsibility for the final outcomes.
Declaration submitted by: Kunyang Li

\section*{Data Availability}

The data underlying this article will be shared on reasonable request to the corresponding author.

\bibliographystyle{mnras}

\clearpage
\appendix
\clearpage

\section{Bonetti Outcome Probability Heatmaps}
\label{app:heatmaps}

\begin{figure*}
\centering
\includegraphics[width=\textwidth,height=0.85\textheight,keepaspectratio]{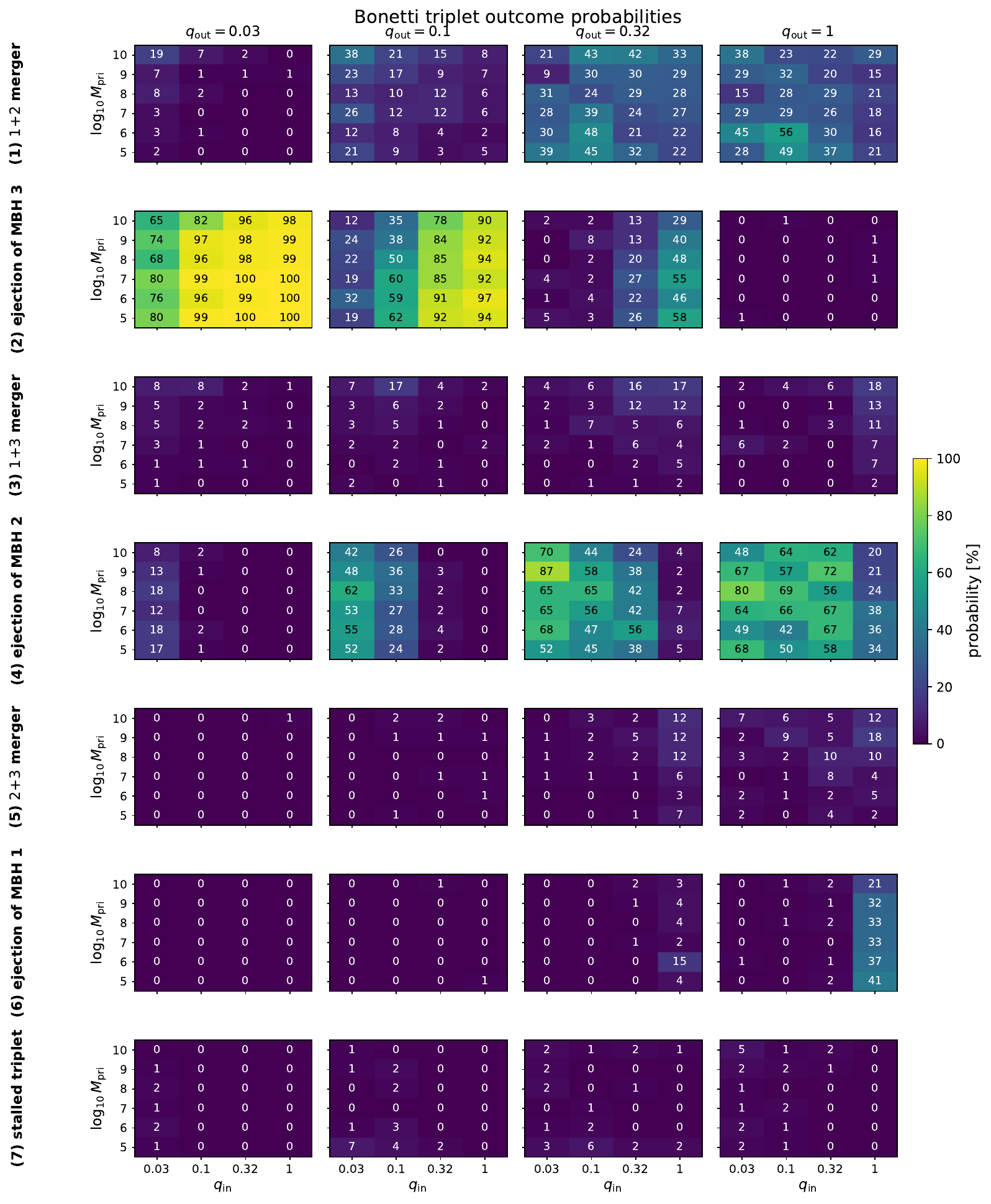}
\caption{Bonetti triplet outcome probabilities used by RAMCOAL, shown as two-dimensional maps. Each row corresponds to one of the seven triplet channels, which are ($1$) MBH1--MBH2 merger, ($2$) ejection of MBH3, ($3$) MBH1--MBH3 merger, ($4$) ejection of MBH2, ($5$) MBH2--MBH3 merger, ($6$) ejection of MBH1, and ($7$) stalled triplet. The four columns show the four tabulated bins of the outer mass ratio $q_{\rm out}=M_3/(M_1+M_2)$. Within each panel the probability (in \%, shown by the colour scale and printed in each cell) is mapped against the primary mass $\log_{10}(M_{\rm pri}/M_\odot)$ (decade bins, labelled by their upper edge) and the inner mass ratio $q_{\rm in}=M_2/M_1$. These probabilities supply the triplet outcome weights adopted by RAMCOAL.}
\label{fig:bonettiheatmap3d}
\end{figure*}

Figure~\ref{fig:bonettiheatmap3d} presents the full set of triplet outcome probabilities used to construct the RAMCOAL lookup table. We define $q_{\rm in}=M_2/M_1$ for the original inner binary and $q_{\rm out}=M_3/(M_1+M_2)$ for the incoming third MBH. The grid spans six primary-mass bins and four bins in each of $q_{\rm in}$ and $q_{\rm out}$. For every channel the probability is shown as a two-dimensional map in $(\log_{10}M_{\rm pri},\,q_{\rm in})$, with the four $q_{\rm out}$ bins placed side by side and the value printed in each cell so that all sampled probabilities can be read directly.
\clearpage

\section{COSCOAL-15 simulation setup}
\label{app:cosmo}

COSCOAL-15 follows a periodic box of side $14.7\,h^{-1}\,{\rm Mpc}$ (approximately 21\,Mpc) from $z=50$ to $z=2.3$. It assumes a $\Lambda$CDM cosmology with matter and baryon density parameters $\Omega_{\rm m}=0.281$ and $\Omega_{\rm b}=0.045$, dark-energy density parameter $\Omega_\Lambda=0.719$, matter power-spectrum normalization $\sigma_8=0.829$, Hubble constant $H_0=69.7\,{\rm km\,s^{-1}\,Mpc^{-1}}$, and primordial spectral index $n_{\rm s}=0.971$.

The simulation contains $256^3$ dark-matter particles with a uniform particle mass of $1.8\times10^7\,{\rm M_\odot}$ and a base mesh of $256^3$ cells. Cells are refined or derefined according to whether their mass exceeds or falls below eight times the dark-matter particle mass. Refinement is allowed up to level~16, corresponding to a spatial resolution of approximately 300\,pc. The MBH resolution sphere has a physical radius of approximately $R_{\rm sp}=1.3$\,kpc.

COSCOAL-15 uses a Schmidt star-formation law with a constant
efficiency per free-fall time $\varepsilon_{\rm ff}=0.02$, requiring
$n_{\rm H}>0.015\,{\rm cm^{-3}}$ and a gas overdensity above 200
relative to the mean cosmic baryon density. Its polytropic
pressure floor has $T_0/\mu=10^4$\,K at the formation threshold
and $\gamma_{\rm eff}=4/3$. The isolated tests instead use
$n_{\rm H}>0.125\,{\rm cm^{-3}}$ and a gravo-turbulent efficiency
that depends on the local Mach number and virial parameter,
with a normalization of 0.5 rather than a fixed efficiency
per free-fall time. Their pressure floor has $T_0/\mu=10$\,K
and $\gamma_{\rm eff}=1$.

Both configurations use mechanical core-collapse supernova
feedback, with $10^{51}$\,erg per supernova and a 10-Myr delay,
but differ in their stellar-population parameters. COSCOAL-15
adopts a mass-return fraction $\eta_{\rm SN}=0.20$, a mean
progenitor mass of $10\,{\rm M_\odot}$, and a metal-yield
parameter $y=0.10$. The isolated tests use a Kroupa initial
mass function, giving $\eta_{\rm SN}\simeq0.21$ and a mean
progenitor mass of $19\,{\rm M_\odot}$, with $y=0.05$.

COSCOAL-15 uses the galaxy-formation and MBH physics described in Section~\ref{sec:ramses}, with MBH seed masses of $10^5\,{\rm M_\odot}$. MBH seeding in COSCOAL-15 requires the local gas density to
exceed the threshold corresponding to
$n_{\rm H}=0.015\,{\rm cm^{-3}}$, and the local stellar density
to exceed the same mass-density threshold. The gas must also
be Jeans unstable on a scale of $4\Delta x$. No seed is formed
within 30 comoving kpc of an existing MBH. A uniform ultraviolet background is included, with a reionisation redshift of $z_{\rm reion}=10$. Unresolved MBH pairs and multiplets are evolved with the full RAMCOAL model. The computational performance of the simulation is discussed in Section~\ref{sec:cosmo}.
\end{document}